\pdfoutput=1
\documentclass[sigconf, nonacm]{acmart}
\usepackage{amsmath}
\usepackage{mathtools}
\usepackage{graphicx} % can better scale and rotate graphics than graphic package
\usepackage{url}
\usepackage{enumitem} % remove spaces inside 'itemize' [nolistsep]

\usepackage{stmaryrd} % for symbol \llceil
\usepackage{upgreek} % upright Greek letters; together with package "bm" -> upright bold letters
\usepackage{bbding} % for the \CheckmarkBold symbol

\usepackage{listings} % to get sql etc nicely formatted (source code listings)
\usepackage{tabularx} % table environment that allows to span a whole line
\usepackage{booktabs} % \toprule
\usepackage{multirow}
\usepackage{hhline} % cannot be overwritten by cellcolor
\usepackage{diagbox}
\usepackage{pgfplotstable} % powerful numerical table macros

\usepackage{xcolor,colortbl}    % !!! for some reason, overleaf needs xcolor separately, although that should already have been called by colortbl (10/2019)

\usepackage{tikz} % drawing figures
\usetikzlibrary{matrix}
\usetikzlibrary{calc}
\usetikzlibrary{math}
\usetikzlibrary{arrows.meta,positioning}
\usetikzlibrary{intersections, pgfplots.fillbetween}

\usepackage{easybmat} % \begin{BMAT} alternative syntax for math tabulars and also includes the command addpath which allows to draw a path between matrix elements.

\usepackage{adjustbox}
\def\eox{\unskip\kern 10pt{\unitlength1pt\linethickness{.4pt}$\diamondsuit${}}} % "\eox" command for end of example

\usepackage{rotating}		% for rotated text in tables, \begin{turn}{75}Accuracy\end{turn}

\usepackage{xspace} % for macros

\usepackage[bottom]{footmisc}   % for footnotes

\usepackage{subcaption} % recommended new package for subfigures (package subfig deprecated)
\newcommand{\hide}[1]{}

\usepackage[ruled,noend,linesnumbered]{algorithm2e} % boxruled, linesnumbered,ruled

\usepackage{setspace} % for setting line spacing

\DontPrintSemicolon     % does not print ";" at end of each line
\SetNlSty{}{}{}                % style of line numbers "\Setnlsty{small}{}{:}"
\SetAlgoInsideSkip{smallskip}   % distance between caption rule and beginning of algorithm
\SetAlFnt{\small}			% size of actual algorithm text\scriptsize\sf\rm
\SetAlCapFnt{\small}		% size of name of algorithm
\SetAlCapNameFnt{\small}

\hypersetup{                                                            % for both, DVI and PDF
	pdfpagemode=UseNone,               % pdfpagemode=UseOutlines
    pdfstartview=Fit,
    pdfpagelayout=SinglePage,       % pdfpagelayout=SinglePage
    bookmarks=false,
    bookmarksnumbered=true,
    bookmarksopen=false,             % bookmarkstree expanded
    pdftex=true,
    unicode,
    colorlinks=true,       % false: boxed links; true: colored links
    linkcolor=blue,          % color of internal links (change box color with linkbordercolor)
    citecolor=cyan,  % color of links to bibliography
    filecolor=magenta,      % color of file links
     urlcolor=blue           % color of external links
}

\usepackage[capitalise,nameinlink]{cleveref} %nameinlink makes whole word colored

\usepackage{aliascnt}  	
\newaliascnt{corollary}{theorem}

\aliascntresetthe{corollary}

\newaliascnt{example}{theorem}
\newtheorem{example}[example]{Example}
\aliascntresetthe{example}

\newaliascnt{definition}{theorem}
\newtheorem{definition}[definition]{Definition}
\aliascntresetthe{definition}

\newaliascnt{proposition}{theorem}

\aliascntresetthe{proposition}

\newaliascnt{lemma}{theorem}

\aliascntresetthe{lemma}

\newaliascnt{conjecture}{theorem}

\aliascntresetthe{conjecture}

\newtheorem{questionW}{Question}
\newtheorem{resultW}{Result}

{\end{itshape}
}
\AtEndPreamble{%
    \hypersetup{colorlinks,
      linkcolor=purple,
      citecolor=blue,
      urlcolor=ACMDarkBlue,
      filecolor=ACMDarkBlue}}

\usepackage{tcolorbox}		% also allows to have shaded backgrounds around environments
\tcbuselibrary{breakable,skins}		% allow to break boxes, allow to use "enhanced jigsaw"

\tcbset{examplestyle/.style={
		enhanced jigsaw,	% if box is broken, don't show border at continuation
		colback=blue!08,	% !20	gray!10
		colframe=blue!08,	% !40
		arc=2mm,
		boxrule=0pt,		% 0pt
		left=1mm,
		right=1mm,
		left skip= 0mm,  % breaks ACM template unless specified for specific style (4/2019)
		right skip= 0mm, % breaks ACM template unless specified for specific style (4/2019)
		top=1mm,		% 0mm, -1mm
		bottom=1mm,		% problem in package with eq at end of env. just add: \tcbset{bottom=2mm} \tcbset{bottom=0mm}
		breakable,		% allows page break
		parbox = false,		% restores normal text behavior. ELSE: paragraphs are formatted slightly different as the main text. DOWNSIDE: unwanted side effects
		before={\par\pagebreak[0]\vspace{1mm}\parindent=0pt},		
		after={\par\pagebreak[0]\vspace{1mm}\parindent=0pt},				
		bottomrule = 0mm,
		boxsep = 0mm,					% common padding of hlengthi between the text content and the frame of the box, added to left, right, top, ...
		topsep at break=0pt,			% Additional vertical space at the top of middle and last parts in a break sequence
		bottomsep at break=0pt,			% Additional vertical space at the first and middle parts in a break sequence		
		pad at break=0mm,
		pad before break=1mm,		
		pad after break=1mm,		
		bottomrule at break=0mm,
		toprule at break=0mm,		
		}}

\usepackage{footnote}

\tcolorboxenvironment{example}{examplestyle}
\makeatletter
\DeclareRobustCommand*\uell{\mathpalette\@uell\relax}
\newcommand*\@uell[2]{
  \setbox0=\hbox{$#1\ell$}
  \setbox1=\hbox{\rotatebox{10}{$#1\ell$}}
  \dimen0=\wd0 \advance\dimen0 by -\wd1 \divide\dimen0 by 2
  \mathord{\lower 0.1ex \hbox{\kern\dimen0\unhbox1\kern\dimen0}}
}
\makeatother

\usepackage{marginnote}

\renewcommand{\epsilon}{\varepsilon} % nicer epsilon symbol
\newcommand{\sql}[1]{\textup{\textsf{\small#1}}}

\usepackage{scalerel}
\usepackage{tikz}
\usetikzlibrary{svg.path}

\definecolor{orcidlogocol}{HTML}{A6CE39}
\tikzset{
  orcidlogo/.pic={
    \fill[orcidlogocol] svg{M256,128c0,70.7-57.3,128-128,128C57.3,256,0,198.7,0,128C0,57.3,57.3,0,128,0C198.7,0,256,57.3,256,128z};
    \fill[white] svg{M86.3,186.2H70.9V79.1h15.4v48.4V186.2z}
                 svg{M108.9,79.1h41.6c39.6,0,57,28.3,57,53.6c0,27.5-21.5,53.6-56.8,53.6h-41.8V79.1z M124.3,172.4h24.5c34.9,0,42.9-26.5,42.9-39.7c0-21.5-13.7-39.7-43.7-39.7h-23.7V172.4z}
                 svg{M88.7,56.8c0,5.5-4.5,10.1-10.1,10.1c-5.6,0-10.1-4.6-10.1-10.1c0-5.6,4.5-10.1,10.1-10.1C84.2,46.7,88.7,51.3,88.7,56.8z};
  }
}

\RequirePackage{etoolbox}
\DeclareRobustCommand\orcidicon[1]{\href{https://orcid.org/#1}{\mbox{\scalerel*{
\begin{tikzpicture}[yscale=-1,transform shape]
\pic{orcidlogo};
\end{tikzpicture}
}{|}}}} % includes various useful macros (keep macros separate from main content files)

\newcommand\vldbdoi{XX.XX/XXX.XX}
\newcommand\vldbpages{XXX-XXX}
 \newcommand\vldbvolume{XX}
 \newcommand\vldbissue{XX}
 \newcommand\vldbyear{2023}
\newcommand\vldbauthors{\authors}
\newcommand\vldbtitle{\shorttitle} 
\newcommand\vldbavailabilityurl{https://github.com/shraga89/ExplainDaV}
\newcommand\vldbpagestyle{plain} 
\usepackage{tabularx}
\usepackage{tikz}
\usepackage{tikz-qtree}
\usepackage{wrapfig}
\usepackage{pifont}% http://ctan.org/pkg/pifont
\usepackage{graphicx}
\usepackage{ifpdf}
\ifpdf
\usepackage[bookmarks=false]{hyperref}
\else

\fi
\newcommand*{\TechReport}{}

\newcommand{\revision}[1]{#1}
\newcommand{\shepherd}[1]{#1}
\newcommand{\lhda}{$L\Delta_{A}$}
\newcommand{\lhca}{$L\nabla_{A}$}
\newcommand{\lhdr}{$L\Delta_{r}$}
\newcommand{\lhcr}{$L\nabla_{r}$}
\newcommand{\rhda}{$R\Delta_{A}$}
\newcommand{\rhca}{$R\nabla_{A}$}
\newcommand{\rhdr}{$R\Delta_{r}$}
\newcommand{\rhcr}{$R\nabla_{r}$}
\newcommand{\lhdaD}{L\Delta_{A}}
\newcommand{\lhcaD}{L\nabla_{A}}

\newcommand{\rhdaD}{R\Delta_{A}}

\setcopyright{acmcopyright}
\copyrightyear{2022}
\acmYear{2022}
\acmDOI{}
\acmConference{}
\acmBooktitle{}
\acmPrice{15.00}
\acmISBN{978-1-4503-XXXX-X/18/06}

\begin{document}
\ifdefined\TechReport
\title{Explaining Dataset Changes for Semantic Data Versioning with \texttt{Explain-Da-V} (Technical Report)}
\else
\title{Explaining Dataset Changes for Semantic Data Versioning with \texttt{Explain-Da-V}}
\fi

\author{Roee Shraga, Ren\'ee J. Miller}
\affiliation{%
  \institution{Northeastern University}
  \city{Boston, MA, USA}
  \country{}
}
\email{{r.shraga,miller}@northeastern.edu}

\begin{abstract}
In multi-user environments in which data science and analysis is collaborative, multiple versions of the same datasets are generated. While managing and storing data versions has received some attention in the research literature, the semantic nature of such changes has remained under-explored. In this work, we introduce \texttt{Explain-Da-V}, a framework aiming to explain changes between two given dataset versions. \texttt{Explain-Da-V} generates \emph{explanations} that use \emph{data transformations} to explain changes. We further introduce a set of measures that evaluate the validity, generalizability, and explainability of these explanations. We empirically show, using an adapted existing benchmark and a newly created benchmark, that \texttt{Explain-Da-V} generates better explanations than existing data transformation synthesis methods.
\end{abstract}

% \ifdefined\TechReport
% \else
% \input{cover_letter}
% \fi

\maketitle
\setcounter{page}{1}	% reset page numbering to 1
\pagestyle{\vldbpagestyle}
\begingroup\small\noindent\raggedright\textbf{PVLDB Reference Format:}\\
\vldbauthors. \vldbtitle. PVLDB, \vldbvolume(\vldbissue): \vldbpages, \vldbyear.\\
\href{https://doi.org/\vldbdoi}{doi:\vldbdoi}
\endgroup
\begingroup
\renewcommand\thefootnote{}\footnote{\noindent
This work is licensed under the Creative Commons BY-NC-ND 4.0 International License. Visit \url{https://creativecommons.org/licenses/by-nc-nd/4.0/} to view a copy of this license. For any use beyond those covered by this license, obtain permission by emailing \href{mailto:info@vldb.org}{info@vldb.org}. Copyright is held by the owner/author(s). Publication rights licensed to the VLDB Endowment. \\
\raggedright Proceedings of the VLDB Endowment, Vol. \vldbvolume, No. \vldbissue\ %
ISSN 2150-8097. \\
\href{https://doi.org/\vldbdoi}{doi:\vldbdoi} \\
}\addtocounter{footnote}{-1}\endgroup

\ifdefempty{\vldbavailabilityurl}{}{
\vspace{.3cm}
\begingroup\small\noindent\raggedright\textbf{PVLDB Artifact Availability:}\\
The source code, data, and/or other artifacts have been made available at \url{\vldbavailabilityurl}.
\endgroup
}

\section{Introduction}\label{sec:intro}

Data is one of the most important ingredients in any decision making process. The amount and size of data is growing and datasets are being reused for multiple analyses. Data may be stored in different systems (e.g., data lakes~\cite{2019_nargesian_data_lake_management}), vary in their formats, and may or may not contain metadata. 
Data projects often involve multiple users that work on datasets conjointly or independently, creating different data versions. Accordingly, \emph{data versioning} becomes an important ingredient in data management~\cite{DBLP:conf/cidr/BhardwajBCDEMP15}. Nevertheless, even if versions are well managed~\cite{bhattacherjee2015principles}, the documentation 
%rjm is usually 
may be superficial, e.g., embedded in filenames, which can be very inadequate. In addition, the collaboration itself 
%rjm is usually 
may not be structured or properly managed and each user may perform different, often \emph{undocumented} processing steps on data~\cite{kery2017variolite,kery2018story,kery2019towards,hohman2020understanding,zhang2020data}. For example, some users may clean the data by removing rows or columns if they have duplicated or missing information. Other users extract features, transforming the current data to create new columns. 

Current tools have limited data versioning support
%rjm versions of data
~\cite{kery2017variolite}. Generally speaking, data, as opposed to code, 
%are much less 
may be less
documented~\cite{kery2019towards,zhang2020data} and 
%their 
data changes, even if documented, are rarely accompanied by useful descriptions, making it difficult to %distinguish and 
understand them~\cite{hohman2020understanding}. Within a close collaboration group, a notebook containing transformation code may be shared, but between organizations this is rarely done.  Consider, for example, the many versions of important datasets shared on open data portals~\cite{us_open_data,canada_open_data,uk_open_data} where transformations are generally not shared.
%Changes in data are typically not documented~\cite{kery2019towards,zhang2020data} and, even if they do, they are rarely given useful names, making it difficult to distinguish and understand the versions~\cite{hohman2020understanding}. 
The lack of sufficient version documentation results in reduced reproducibility and trust among users using the data~\cite{kery2018story,zhang2020data}. %Finally, w
While managing and storing data versions has received attention in literature~\cite{bhattacherjee2015principles,DBLP:conf/cidr/BhardwajBCDEMP15,huang2017orpheusdb, yilmaz2018datadiff,schule2021tardisdb}, the semantic nature of such changes has remained under-explored. We motivate our work using the following example.

\begin{figure}[h]
	\centering
	\begin{subfigure}[b]{0.3\textwidth}
		\includegraphics[width=\textwidth,trim=0 0 30 30]{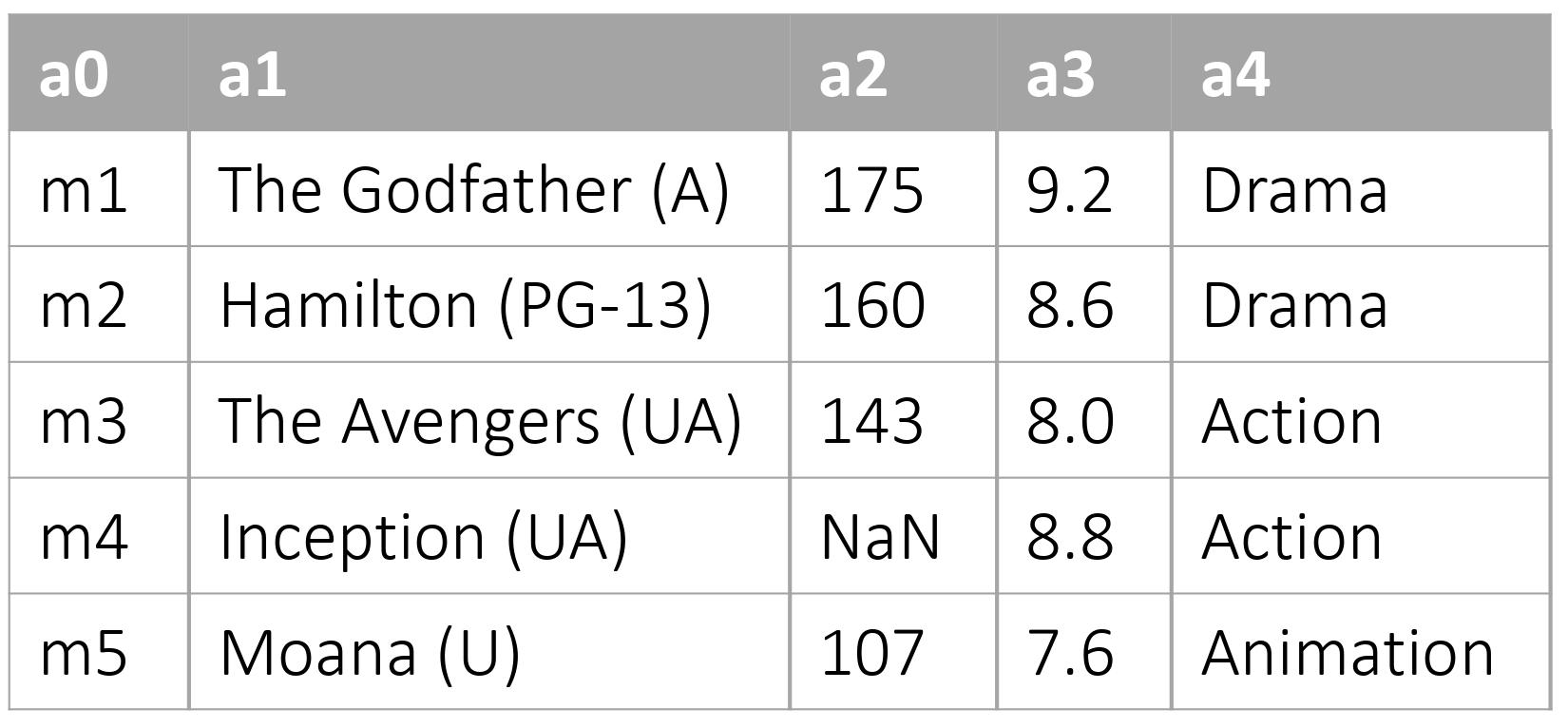}
	 	\caption{Dataset version created by \textsc{UserA}}
		\label{fig:example_a}
	\end{subfigure}
	\hfill
	\begin{subfigure}[b]{0.4\textwidth}
		\includegraphics[width=\textwidth,trim=0 0 30 30]{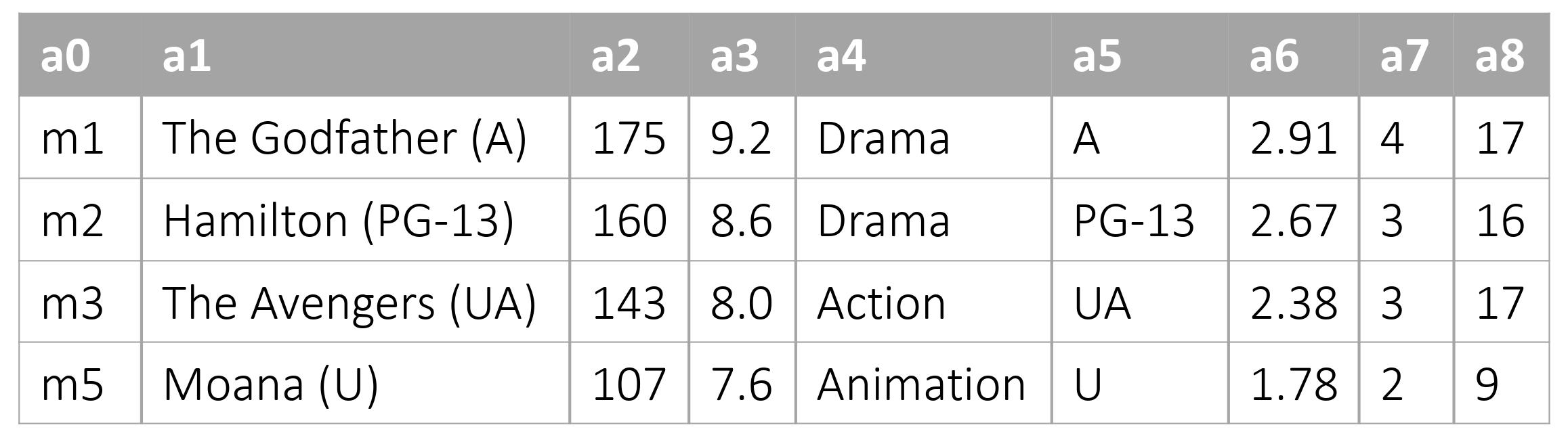}
		\caption{Dataset version created by \textsc{UserB}}
		\label{fig:example_b}
	\end{subfigure}
	\caption{Example dataset versions about movies created by two users. Attribute names are provided in Example~\ref{example:intro}.}
	\label{fig:example}
\end{figure}

\begin{example}\label{example:intro}
	Figure~\ref{fig:example} presents two dataset versions about movies. %\rs{should I provide the column names for the mutual columns (a1-a4)?} 
	We discard the column names from the figure to illustrate a realistic (data lake) scenario. For readability, %we provide the column names for Table~\ref{fig:example_a} in the caption. 
	\texttt{a0} is a tuple id, \texttt{a1} represents the movie title, \texttt{a2} measures the movie runtime in minutes, \texttt{a3} assigns a rating to the movie, and \texttt{a4} provides the genre of the movie. For convenience of presentation, lets assume that the table on the bottom (Figure~\ref{fig:example_b}) was created by \textsc{UserB} as a derivation of the table on the top (Figure~\ref{fig:example_a}) that was created by \textsc{UserA}. Even properly naming the tables, e.g., Table~\ref{fig:example_a} as \texttt{data1\_v1.csv} and Table~\ref{fig:example_b} as \texttt{data1\_v2.csv}, or knowing that Table~\ref{fig:example_b} is derived from Table~\ref{fig:example_a}~\cite{DBLP:conf/cidr/BhardwajBCDEMP15}, 
	%rjm cannot really 
	does not help \textsc{UserA} to get a \emph{semantic} understanding of what \textsc{UserB} changed in the table or, more importantly, what data processing steps  \textsc{UserB} has performed. % in the process.    
\end{example}

Example~\ref{example:intro} illustrates the need for a semantic understanding of a new dataset version. Aiming to fill this gap, this work provides the setup and new solution to \emph{explain} the semantic changes between two dataset versions. Specifically, our goal is to \emph{automatically} explain (in a simple user friendly way) the steps leading from one version of dataset to the other. For example, \textit{how was column \texttt{a6} in Figure~\ref{fig:example_b} created?} or \textit{why was the fourth row in Figure~\ref{fig:example_a} deleted?} Note that the changer’s intent, which is subjective, cannot be truly reverse engineered. Our objective is to provide the other user an accurate explanation, e.g., a set of functions, that describes the changes. Following this goal, we return to our motivating example.
%, e.g., as a function, that describes the changes. Following this goal, we now return to our motivating example.

\begin{figure}[h]
	\centering
% 	\vspace{-.1in}
	\includegraphics[width=.45\textwidth,trim=0 0 30 30]{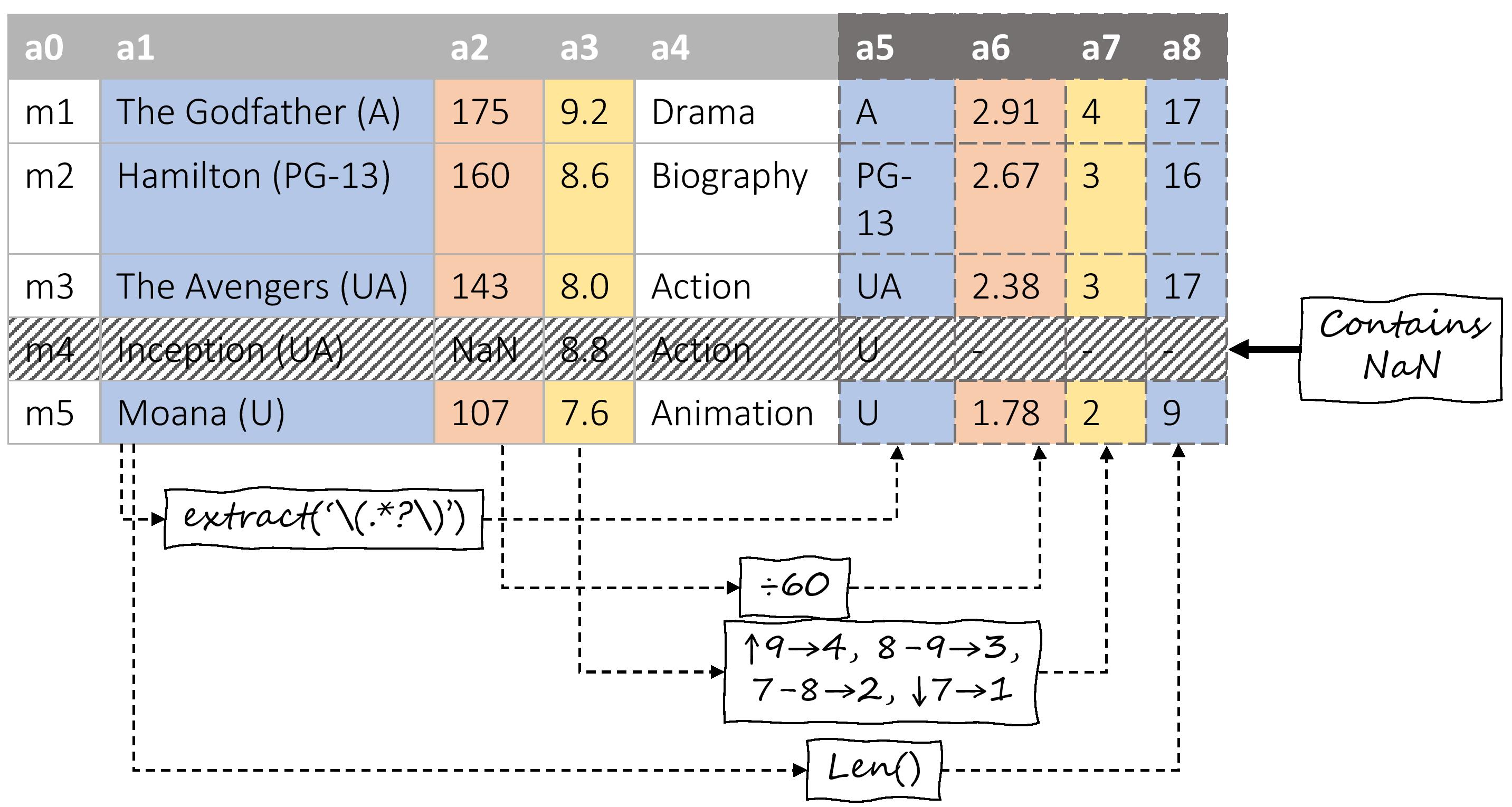}
% 	\caption{An interpretation of the changes between the dataset versions given in Figure~\ref{fig:example}. The columns are colored based on their origin (e.g., \texttt{a5} is blue because it  originates from the blue \texttt{a1}) and annotated column transformations are given at the bottom. The annotated row transformation is given on the right, in this case removing a row, which is also illustrated by diagonal stripes over the row.}
	\caption{An interpretation of the changes between the dataset versions given in Figure~\ref{fig:example}. The columns are colored based on their origin (e.g., \texttt{a5} is blue because it  originates from the blue \texttt{a1}) and annotated column transformations are given at the bottom. The annotated row transformation is given on the right, in this case removing a row, which is also illustrated by diagonal stripes over the row.}
	\label{fig:example_explained}
% 	\vspace{-.1in}
\end{figure}

\addtocounter{example}{-1}
\begin{example}[cont.]
	Figure~\ref{fig:example_explained} illustrates an annotated version of Figure~\ref{fig:example_b}, 
	%such that \textsc{UserA} will be able to understand the changes made by \textsc{UserB}.
	that explains the changes. In other words, %we aim to reverse engineer 
	Figure~\ref{fig:example_explained} reverse engineers the changes made by \textsc{UserB} in a way that a user can understand. Specifically, \textsc{UserB} cleaned the rows that contain \texttt{NaN} values (in this case \texttt{m4}) and extracted numerical features. The 
	%rj, certificate ?
	certification of the movie, given in parenthesis in \texttt{a1}, was extracted to create \texttt{a5} and the column \texttt{a6} converts the units of \texttt{a2}, the runtime of the movie, from minutes to hours. Since the range of movie ratings (\texttt{a3}) is limited, \textsc{UserB} also %decided to 
	discretized the values to create four rating classes %\roee{(it may be more intresting to use A, B, C and D instead of 1-4)} 
	in column \texttt{a7}. Aiming to examine the effect of title length (an effect found for paper citations~\cite{deng2015papers}) within the domain of movies, \textsc{UserB} added column \texttt{a8} that provides the length of titles from \texttt{a1}.   
\end{example}

As illustrated in Example~\ref{example:intro}, there are a variety of possible transformations (e.g., multiplying/dividing the values of a numeric column, e.g., \texttt{a3}, by a constant), potentially creating an infinite possible number of changes to a dataset. These changes can be vertical (changing columns) or horizontal (changing rows), they can add information (adding columns/rows) or remove information (removing columns/rows) and they can involve different data types, e.g., textual to numeric (\texttt{a1} to \texttt{a8}) or numeric to categorical (\texttt{a3} to \texttt{a7}). In addition, a user may also change a cell in the table, e.g., replacing the \texttt{NaN} value in row \texttt{m4} by 146.25 (the mean value of the other values in the column), or perform a full-table operation, e.g., transposing the table. Transformation discovery methods are used for multiple data management tasks including fuzzy joins~\cite{autojoin}, data wrangling~\cite{bogatuPFK19}, entity consolidation~\cite{DengTAEI0MOS019} and more%~\cite{foofah,autotransform,clx,ozmenEA21,tde,blinkfill,datamaran,dataxformer,flushfill,SinghG12,robustfill} 
~\cite{foofah,autotransform,dataxformer,flushfill} 
mainly focusing on textual (text-to-text) transformations and consider the transformed values (rather than the transformation itself). Our method mainly focuses on data versioning, for which, the transformations themselves, as a means of explaining changes among different dataset versions, is the main interest. Our resolved transformations also %are not limited to text and 
cover transformations that involve, among others, numeric transformations. The term ``explanation'' became quite common recently and may be associated with multiple meanings. %~\cite{glavic2021trends}.
For example, both El Gebaly et al.~\cite{el2014interpretable} and Kim et al.~\cite{kim2020summarizing} use data summaries as explanations. Explain3D~\cite{wang2019explain}, which shares a similar context to ours, explains dataset disagreements with syntactic provenance-based and value-based modification mappings. In this work the main component of an explanation is a transformation that explains change. %In particular, t
This paper makes the following contributions.
%The transformations are used for ``explanations'', a term that became quite common recently and may be associated with multiple meanings~\cite{glavic2021trends}. For example, both El Gebaly et al.~\cite{el2014interpretable} and Kim et al.~\cite{kim2020summarizing} use data summaries as explanations. Explain3D~\cite{wang2019explain}, which shares a similar context to ours, explains dataset disagreements with syntactic provenance-based and value-based modifications mappings between the datasets. We explicitly define explanation in \cref{def:explain}. In particular, this paper makes the following contributions.

% \renee{Is there some other way of characterizing these methods rather than by data type?  Are they all semantic version discovery methods but just with limited data types?}  While different transformation discovery methods have been explored before~\cite{foofah,autojoin,autopandas,autotransform,clx,ozmenEA21,bogatuPFK19,tde,DengTAEI0MOS019,blinkfill,datamaran,dataxformer,flushfill,SinghG12,robustfill}, most research focuses mainly on textual (text-to-text) transformations and considers the transformed values (rather than the transformation itself). 
% We, on the other hand, aim to resolve transformations between multiple data types (textual, numeric and categorical) and are mainly interested in the transformations themselves as a means of explaining changes among different dataset versions. In particular, this paper makes the following contributions.

\noindent(1) \textbf{Semantic Data Versioning Definition:} we define and solve a novel problem of semantic data versioning 
	%rjm in data lakes by zooming in on 
	by explaining the changes between two dataset versions.\\
\noindent(2) \textbf{Vertical and Horizontal Data Transformation Resolution Across Different Data Types:} we present a solution to the problem of semantic data versioning that examines both vertical (adding/removing columns) and horizontal (adding/removing rows) transformations that involve multiple data types.\\
\noindent(3) \textbf{Semantic Data Versioning Metrics:} we provide a set of evaluation measures to examine the quality of explanations in terms of validity, generalizablity, and explainability.\\
\noindent(4) \textbf{Semantic Data Versioning Benchmark:} we introduce a new data versioning benchmark composed of 5 version-sets including 342 different dataset versions representing a total of 1702 changes.\footnote{Code and benchmark are publicly available~\cite{gitURL}.}\\
%that represent a total of 1702 changes.\footnote{We will make code and benchmark publicly available upon acceptance. Main parts of the code and a benchmark sample are available~\cite{gitURL}.}\\
\noindent(5) \textbf{Empirical Evaluation:} our experiments show that \texttt{Explain-Da-V} performs better than multiple baselines on both our new version benchmark and on an existing data science pipeline benchmark~\cite{AutoPipelineRepo}.  We analyze the impact of different components of our solution on performance.

%In this paper we assume that the changes are known, internal (no external information was used) and that metadata may be unreliable (and thus only data values are used).
%\rjm{different systems
In this paper, we assume two tables are given where one is known to have been derived from the other (i.e., is a version it) %of the other) 
and we know a match between the attributes and tuples the two tables share. 
This work focuses on ``internal'' additions, deletions, or modifications (modeled as deletions followed by additions). External additions, e.g., finding joinable tables~\cite{2019_zhu_josie} and joining them with a table to create a new version, are reserved for future work.

\section{Related Work}
% \renee{we changed nothing in related work?  seems odd given the reviews.} \rs{todo: Add query explanation work.}

We are, to the best of our knowledge, the first to address the semantic aspect of data versioning. %However, 
%rjm several 
Yet,
related research exists ranging from %resolving \renee{does resolving mean discovery of?}  
synthesizing data transformations %from examples 
to exploring data %and schema 
change. 

%\vspace{.1cm}
%\textbf{Data Versioning in Data Lakes:}
% \subsection{Data Versioning in Data Lakes}\label{sec:related1}
% \textbf{dont emphasize data lakes}
\subsection{Data Versioning}\label{sec:related1}
% Traditional 
%rj works 
% research 
% on data versioning %for data lakes \renee{is it just of data lakes or enterprise as well?  maybe omit quantifer} 

Data versioning research mainly focus on developing version managers to decrease the need for storing many versions of large datasets~\cite{2019_nargesian_data_lake_management}. For example, DataHub provides a git-like interface to manage, store, recreate, and retrieve %and analyze 
versions using a directed version graph~\cite{DBLP:conf/cidr/BhardwajBCDEMP15}.  Follow-up research further studied the trade-off between recreation and storage in a principled way analyzing six different settings~\cite{bhattacherjee2015principles}. Recently, Sch{\"u}le et al. presented TardisDB~\cite{schule2021tardisdb}, an SQL extension to support version management. TardisDB uses named branches over tables, to monitor table versions and  track %(record) 
their modification history. In 
contrast, we focus on the semantic aspects of data versioning, zooming in 
%rjm of 
on
explaining the semantic differences between dataset versions. %Finally, 
%some research directions also explored 
Schema versioning has also been studied~\cite{roddick1995survey}. Although schemata may change over time~\cite{snodgrass2008validating}, %and these changes 
which provides semantic hints to %the changes in the 
data change, %, as we focus on data lakes, 
%rjm I changed the negative, it is not that we assume it is missing, rather we don't assume it is available!
%rjm we will assume metadata may be missing, inconsistent, or incomplete, 
%we do not assume metadata is always complete and unambiguous~\cite{2019_nargesian_data_lake_management}.
we assume metadata is not always complete and may be ambiguous~\cite{2019_nargesian_data_lake_management}.
Hence, we focus only on the versioning of the data itself.

% \renee{In the following, more strongly distinguish "finding" tables or discovery, from versioning.  This work assumes the finding part is done.}

% \rs{Remember to add for CR:}
\ifdefined\TechReport
Multiple methods find or discover related tables~\cite{2012_sarma_finding_related_tables,badSearchSurvey} (e.g., joinable~\cite{2019_zhu_josie} and unionable~\cite{2018_nargesian_tus,2023_khatiwada_santos}). While different versions of a table may be related and found using these methods, our work assumes that the discovery has already been done and aims at providing a semantic explanation for the differences between versions. 
\else
\fi
% Techniques for finding related tables can also be used to highlight differences and similarities between two dataset versions, which we assume is given in this work and require semantic explanation. 

% Also in the scope of data lakes, we can find multiple works that aim to find related tables~\cite{2012_sarma_finding_related_tables} (e.g., joinable~\cite{2019_zhu_josie} and unionable~\cite{2018_nargesian_tus}). In this work we address a special kind of related tables, namely a different version of the same table. Techniques for finding related tables can also be used to highlight differences and similarities between two dataset versions, which we assume is given in this work and require semantic explanation. 

%\vspace{.1cm}
%\textbf{Data Change, Difference, and Integration:}
\subsection{Data Change, Difference, and Integration}\label{sec:related2}
We assume that some match between the attributes and tuples of the versions is given. This assumption is rooted in many years of data integration research, exploring attribute matching (schema matching)~\cite{rahm2001survey,shraga2020adnev}, tuple matching (entity resolution)
%rjm I don't think we need footnote
%\footnote{also known as entity matching, deduplication, record linkage and more}
\cite{elmagarmid2006duplicate,li2020deep}, and others~\cite{bellatreche2013special,miller2018open,2022_khatiwada_alite}. Earlier works looked into change and copy detection in structured data~\cite{chawathe1996change,chawathe1997meaningful}, which was later extended also to semi-structured documents such as XML~\cite{nierman2002evaluating,cobena2002detecting,wang2003x}. 

% Acknowledging that data change, Bleifu{\ss} et al.~\cite{bleifuss2018exploring} envision systems and methods that can interactively explore such change. They present a model of what changed, where, when and how using what they call a
% \renee{what's a change-cube?  is this common terminology?  If not, define/explain it.}. \textbf{change to what they call..}
Acknowledging data change, Bleifu{\ss} et al.~\cite{bleifuss2018exploring} envision systems %and methods 
that can interactively explore such change. They present a model of what changed, where, when and how, using what they call a
change-cube to monitor the history of changes over time using methodologies such as time-series clustering~\cite{bornemann2018data}. %In follow-up research, t
%They presented 
DBChEx~\cite{bleifuss2019dbchex} %, which
is a tool to explore data and schema change using a set of exploration primitives. While similar in nature, this line of work focuses mainly on how to \emph{explore} change aiming to answer questions such as ``How many changes have there been in  recent minutes%, days or years
?" and ``How old are the entities in table Y? When were they last updated?"~\cite{bleifuss2018exploring,bleifuss2019dbchex}. %In contrast, o
Our work focuses on local changes between versions and
%, using the notation of~\cite{bleifuss2018exploring},
\emph{how} the changes were performed (which transformations were applied?), e.g., how did \textsc{UserB} create the table in Figure~\ref{fig:example_b} from Figure~\ref{fig:example_a}.  
%\renee{Something's wrong with last sentence.  How was table 1a created to get table 1b??  You mean how was 1b created from 1a?  If the latter, do they create transformations like we do?} \rs{"This" should be replaced with "Our" and, yes, 1b was created over 1a.} \rs{\textbf{Our} work focuses on local changes between...} 
Another related research area is explaining query answers~\cite{roy2015explaining, miao2019going}.  Given a query and a database, they explain query answers using the tuples in the given database, e.g., using provenance~\cite{cheney2009provenance}. We, in contrast, explain dataset changes using data transformations. While changes can be thought of as a set of queries, explanations for queries that involve non-trivial transformations (e.g., adding \texttt{a5} in \cref{example:intro}) cannot be explained only using the tuples.

%\vspace{.1cm}
%\textbf{Data-Transformation-By-Example}
\subsection{Data Transformation By Example}\label{sec:related3}
The final related line of 
%rjm works 
research
we cover aims to automatically transform data. Largely, given input and output tables (datasets) or their subsets (examples), the goal of such approaches is to find a transformation (program) such that if it is applied over the input we get the output. It is worth noting that earlier work has referred to this problem as query reverse engineering~\cite{tran2014query,orvalho2020squares}, which is roughly the same idea, i.e., finding 
%rjm the 
a
query that generates the output using the input. This line of work can be divided into two main groups. 

The first group is rooted in a paradigm called programming-by-example (PBE)~\cite{foofah,blinkfill,autojoin,clx,flushfill,bogatuPFK19,SinghG12,datamaran}, where the goal is to synthesize a program that manipulates a given input to get a given output. 
To do so, methods design different search spaces (operators to be applied over the input) and apply different search algorithms. For example, Foofah~\cite{foofah} creates a search space using operators such as drop (delete a column) and split (separate a column by some delimiter) and search the space using A* heuristic search. Clx~\cite{clx} also introduces string patterns such as regular expressions to the search space and tokenizations. Data Diff~\cite{sutton2018data} applies a search approach to ``patch'' transformations, summarizing distribution changes, %. for data wrangling and 
%Sutton et al.~\cite{sutton2018data} also offer 
including one numeric patch (operator) supporting linear transformations with pre-defined (randomly selected) parameters. %Finally, 
Muller et al. describes differences between relational databases with what they call ``update distance''~\cite{muller2006describing} using a similar searching approach. Finally, Bogatu et al. introduced functional dependencies to navigate the search space~\cite{bogatuPFK19}, which we also use in our work. 
%In an earlier work Muller et al. also look at describing differences between databases with what they call ``update distance''~\cite{muller2006describing} using a similar approach.}

The second group 
%rjm of works 
focuses on creating transformation repositories from
%for 
external sources such as Web Forms, Knowledge Bases~\cite{dataxformer,ozmenEA21}, GitHub and Stackoverflow~\cite{tde,autotransform}. Transform Data by Example (TDE)~\cite{tde}, instead of searching through a space of pre-defined possible operations, 
%treats the problem as 
creates a search engine where transformation functions are crawled from GitHub and Stackoverflow. %\renee{what does "in what follows" refer to?}
%In what follows, i
Instead of applying heuristic search such as A*, TDE 
%tried to 
ranks candidate functions to find  relevant 
%one to execute. 
functions. TDE was later extended to allow transformation search based on patterns~\cite{autotransform}. DataXFormer~\cite{dataxformer} and Proteus~\cite{ozmenEA21} create a repository of tables from which desired output values can be extracted.

%DataXFormer~\cite{dataxformer} creates a repository of tables from which desired output values can be extracted. Proteus~\cite{ozmenEA21} enhances DataXFormer to also support the combination of multiple tables to find output values.

A %nother 
 similar line of research revolves around resolving data preparation and analysis transformations~\cite{autopandas,yan2020auto,yang2021auto}. AutoPandas~\cite{autopandas} %, which we use as a baseline in our experiments
focuses on %the 
Pandas %python 
library~\cite{Pandas} and %, similar to the above methods, 
aims to synthesize a program using pandas functions. Auto-pipeline~\cite{yang2021auto} extends the ``by-example" paradigm to ``by-target", %which basically 
meaning that the output the user provides is not necessarily aligned with the input and can require table-reshaping operations (e.g., group by). Auto-pipeline comes in two variations, namely, search (which is equivalent, yet extended to what is described above) and deep reinforcement learning. The former can be a candidate baseline for our approach. %\footnote{Reproducible code of Auto-pipeline is not publicaly available, so, in our experiments, we reproduced its functionality using Foofah to allow a quantitative comparison.} 
The latter requires training data which we assume does not exist in our setting. 

% \renee{This last paragraph makes this work sound like a delta over TBE.  Let's recast this to say something like "In contract to TBE and query reverse engineering, we do not focus on matching input to output, rather, we introduce new notions of validity, generalizability and explainability to guide the discovery of transformations between versions...  " } 
%\renee{We may need to come back to this at the end once we've got the paper more fully written. I think we can say this without making this work sound "delta". }

% Our work focuses on semantic data versioning, which in part includes resolving data transformations. Different from other approaches, we focus on the transformation itself from an explainability 
%\renee{do you mean "point of view" or "standpoint"?}
% point-of-view to support data versioning, rather than providing the appropriate output for some new input examples. 
In contrast to PBE and query reverse engineering, we do not focus on matching input to output.
Rather than looking only at success rates (is the transformation valid), our search is guided by the principle of creating valid, generalizable, and explainable transformations.
%Rather, we introduce new notions of validity, generalizability, and explainability to guide the discovery of transformations between versions. 
In our approach, these transformations can be multi-dimensional (adding/removing attributes/tuples) and address multiple data-types (e.g., numeric, categorial, and text transformations). While our string-based transformation resolution is based on an extended Foofah, we also support numeric transformations using explainable machine-learning algorithms to \emph{fit the appropriate transformation} rather than searching a very large space of possible transformations. 
%We further extend our choice of PBE 
We go beyond Foofah to support text-to-numeric transformations (e.g., measuring the length of a string)
%and in the \texttt{a2} to \texttt{a9} transformation in Example~\ref{example:intro}) 
and text cleaning operations (e.g., stopword removal and lemmatization). 
%rjm putting some of this back in...
%Finally, from an evaluation perspective, rather than looking only at success rates as in previous works, we look at the validity, generalizablity and explainability of the generated transformation. 
%Importantly, rather than looking only at success rates (is the transformation valid), our search is guided by the principle of creating valid, generalizable, and explainable transformations.

%\roee{remember saying something about date data type in related work}
%
%\roee{\begin{itemize}
%		\item data versioning in data lakes
%		\item data transformations by example
%		\item data diff (e.g., xml diffs)
%		\item unionable/joinable/related - other relationships between tables in a data lake
%\end{itemize}}
% \vspace{-.05in}
\section{Semantic Data Versioning}\label{sec:prelim}
%In this section we define the problem of Semantic Data Versioning and formally define the problem we address in this paper.

% In this section, we provide the preliminaries and problem definition of semantic data versioning.
% We now provide preliminaries and problem definition. %we address in this paper.%definition of semantic data versioning.

% We now provide preliminary %\rjm{
% notation.
% %}
% %rjm the problem definition is no longer in this section
% %and the problem definition.

% \subsection{Preliminaries}\label{sec:prelim}
%For the scope of this work we assume that 
A dataset is 
%given 
denoted by a table $T$, composed of a set of attributes $T_{A} = \{A_1, \dots, A_n\}$ and tuples $T_{r} = \{r_1, \dots, r_m\}$.
%such that 
Each tuple is defined as $r_i = \langle r_{i0}, r_{i1}, \dots, r_{in}\rangle$, such that $r_{i0}$ is the tuple identifier and $r_{ij}$ ($j\neq 0$) is a value assigned to the attribute $A_j$ in the tuple $r_i$. 
% \renee{I would add to this that each tuple has a unique identifier $id_i$ (you pick notation).  Perhaps explain that tuples in $T$ and $T'$ with the same identifier are assumed to represent the same real work entity or more generally have been "resolved"?    Are you assuming that $r_i$ in $T$ and $r_i$ in $T'$ share the same identifier?}
%Data lakes are multi-user environments, where 
%In a data lake setting, more than one user has access to some table $T$. In what follows, 
% A table may have multiple versions. The versions may be a result of multiple data processing phases. 
%For example, when data cleaning users usually removes rows or columns if they have duplicated information or missing information. When feature extracting, users usually transform the current data to create new columns. 
% \renee{I suggest removing the next three sentences since we do not do anything with multiple versions.}
% A table may have multiple versions, which may be a result of multiple data processing phases. 
% Dataset versions %of the same tables 
% can be derived sequentially (users working on consecutive versions of the table) or in-parallel (users working on a shared table independently). 
% %Either the former or the latter, 
% Regardless, in a real-world scenario these versions are rarely documented properly~\cite{kery2017variolite,kery2018story,kery2019towards,hohman2020understanding,zhang2020data}.
% %rjm cut above.
Often we may have two datasets and know one was derived from the other but the actual transformation code or documentation has been lost~\cite{kery2017variolite,kery2018story,kery2019towards,hohman2020understanding,zhang2020data}.
In what follows, we address the problem of explaining the changes between two dataset versions.

%Formally, we assume that there exist some initial table from which multiple versions can be derived in ??? (users working on consecutive versions of the table) or in parallel (users working on a shared table independently). Either the former or the latter, in a real-world scenario these versions are rarely documented properly.

%Managing different versions of a table was explored from a ... point of view. For example, .... However, the semantic perspective, that is monitoring the actual changes that created diffrent versions was yet to be explored. In this work, we focus on the perspective of two versions of a table such that one was derived from the other, for which we are required to ``explain" the changes.  

%Formally, we assume that there exists some \emph{root table} $\mathcal{T}$, from which all other tables were derived.

%$\{\langle r_{11}, r_{12}, \dots, r_{1m}\rangle, \dots, \langle r_{n1}, r_{n2}, \dots, r_{nm}\rangle\}$ such that

%Let $\mathcal{T}$ be a \emph{root table} and $T$, $T^{\prime}$ two \emph{derived tables} rooted at $\mathcal{T}$.

%We now describe the formal problem definition. 

% \begin{table}[h]
% 	\caption{Basic Notations.}\label{tab:notaions}
% 	\begin{tabular}{|c|c|} 
% 		\bottomrule
% 		\textbf{Notation} & \textbf{Meaning} \\\toprule
% 		$T$ & Left-hand dataset \\\midrule
% 		$T^{\prime}$ & Right-hand (revised) dataset \\\midrule
% 		$T_{A}$ & The attribute set of dataset $T$\\\midrule
% 		$T_{r}$ & The tuple set of dataset $T$ \\\bottomrule
% 	\end{tabular}
% \end{table}

Given two dataset versions, $T$ and $T^{\prime}$, we assume the latter, wlog, is a \emph{derived table}, i.e., a user changed the table $T$ and as a result obtained the table $T^{\prime}$ with $T^{\prime}_{A} = \{A^{\prime}_1, \dots, A^{\prime}_n\}$ and tuples $T^{\prime}_{r} = \{r^{\prime}_1, \dots, r^{\prime}_m\}$. %Formally, let $T$ be a table with attributes $T_{A}  = \{A_1, A_2, \dots, A_N\}$ and tuples $T_{r} = \{r_1, r_2, \dots, r_M\}$ and $T^{\prime}$ be a table derived with attributes $T^{\prime}_{A} = \{A^{\prime}_1, A^{\prime}_2, \dots, A^{\prime}_n\}$ and tuples $T^{\prime}_{r} = \{r^{\prime}_1, r^{\prime}_2, \dots, r^{\prime}_m\}$. 
%rjm As motivated in Section~\ref{sec:related2}, %in this work, 
We %shall 
assume that an alignment between $T_{A}$ and $T^{\prime}_{A}$ (attribute-match, denoted $\Sigma_{A}$) is given and that tuples in $T$ and $T'$ with the same identifier ($r_{0i}$ and $r^{\prime}_{0i}$) are assumed to represent the same real world entity. %to create a tuple-match, denoted  $\Sigma_{r}$. 
An attribute $A_i\in T$ (or $A^{\prime}_j\in T^{\prime}$) is considered unmatched if it does not appear in $\Sigma_A$. %Similarly, 
%the 
A record $r_i\in T$ (or $r^{\prime}_j\in T^{\prime}$) is considered unmatched 
%rjm you haven't defined \sigma_r
%it does not appear in $\Sigma_r$.}
if there is no record in $T^{\prime}$ (respectively, $T$) with the same %tuple 
identifier.

Given an attribute-match $\Sigma_{A}$, %and a tuple-match $\Sigma_{r}$, 
we define the changes between the two dataset versions to be explained using a three symbols notation. The first refers to whether the dataset is the left-hand one (L) or the right-hand (revised) one (R), the second to whether it is the matched ($\nabla$) or unmatched (unmatched is also called delta ($\Delta$)), and the third refers to attributes (A) or tuples (r). Specifically, \lhda (left-hand delta attributes) and \lhca (left-hand 
%rjm consistent 
matched attributes) are the set of unmatched (delta) and matched (consistent) attributes in $T$, respectively. Similarly, \rhda{} and \rhca{} are the unmatched and matched attributes in $T^{\prime}$. Using these sets, we create \emph{projected} tuples. Let $r_{j}$ be a tuple of table $T$, the projected tuple is given by $\pi_{\lhcaD}[r_{j}]$, projecting out non-matching attributes. Given such projected tuples, we can define similar sets for tuples, namely \lhdr (left-hand delta tuples), \rhdr (right-hand delta tuples), \lhcr (left-hand consistent tuples) and \rhcr (right-hand consistent tuples). We summarize 
%rjm notation is a collective noun typically used in singular
%these notations 
this notation in Table~\ref{tab:notaions}. %\added{For clarity, note that the notation is composed of three symbols. The first refers to whether the dataset is the left-hand one (L) or the right-hand (revised) one (R), the second to whether it is the matched ($\nabla$) or unmatched ($\Delta$) part, and the third refers to attributes (A)/tuples (r).}
Intuitively, we are interested in explaining the deltas between the datasets, i.e., \lhda, \rhda, \lhdr, and \rhdr.

\begin{example}\label{example:changes}
	Given the dataset versions in Figures~\ref{fig:example_a} ($T$) and~\ref{fig:example_b} ($T^{\prime}$), the attribute-match is simply given by aligning the columns headers (e.g., \texttt{a2} $\leftrightarrow$ \texttt{a2}). The tuple ids are given under \texttt{a0} (e.g., \texttt{m1} $\leftrightarrow$ \texttt{m1}).
	%The tuple-match is given by the tuple ids under attribute \texttt{a1} (e.g., \texttt{m1} $\leftrightarrow$ \texttt{m1}). %rjm In what follows, t
	The following are the change sets:
	%can be derived 
	\lhda = $\emptyset$ (no removed columns), \rhda = \{\texttt{a5}, \texttt{a6}, \texttt{a7}, \texttt{a8}\} (four added attributes), \lhdr = \{\texttt{m4}\} (one removed tuple), and \rhdr = $\emptyset$ (no added tuples).%, for which Example~\ref{exp:explanations} provides a possible set of explanations.
\end{example}

\subsection{Change Explanations}\label{sec:explanations}
We use the term \emph{explanation} to refer to a user friendly way to interpret a change between two relations. 
%rjm added -- give the big picture first
Intuitively, an explanation is a transformation $\mathcal{P}$ from an origin $\mathcal{O}$ to a goal $\mathcal{G}$.  
%rjm end
Formally, an explanation $\mathcal{E}$ is defined with respect to a \emph{goal} $\mathcal{G}$ with a name $\mathcal{G}_{name}$ and an associated relation $\mathcal{G}_{relation}$ it represents. 
%An explanation $\mathcal{E}$ is composed of an \emph{origin} $\mathcal{O}$ and a transformation $\mathcal{P}$. 
As the goal, the origin is also associated with a name ($\mathcal{O}_{name}$) and a relation ($\mathcal{O}_{relation}$). A transformation $\mathcal{P}$ is an expression that transforms the origin relation $\mathcal{O}_{relation}$ into the goal relation $\mathcal{G}_{relation}$. The origin relation may also be empty. A formal definition is as follows

% \renee{Presumably a goal can be empty too, e.g., delete attribute.  I realize this is not that useful but possible, right?} \rs{Yes. The goal can be empty if nothing changed, as you said - not useful. When deleting an attribute, the goal won't be empty, the goal would be the deleted attribute (in $T$) and we would have to find the solution in $T^{\prime}$, e.g., in contains another attribute that is a duplicate and this is why the attribute was deleted.}

\begin{definition}[Explanation ($\mathcal{E}$)]\label{def:explain}
	Let $\mathcal{G}$ be a goal. An \emph{explanation} $\mathcal{E}_{\mathcal{G}} = (\mathcal{O}, \mathcal{P})$ of $\mathcal{G}$ is composed of an origin $\mathcal{O}$ and a transformation $\mathcal{P}$, such that $\mathcal{G}_{relation} = \mathcal{P}(\mathcal{O}_{relation})$.
\end{definition}

% \renee{Hmm, I like this but seems like it belongs more in related work section?} \rs{Originally, I wanted to put it there, what do you think would be the right position for this paragraph within the related work section?}

% \added{The term ``explanation'' became quite common recently and may be associated with multiple meanings~\cite{glavic2021trends}. For example, both El Gebaly et al.~\cite{el2014interpretable} and Kim et al.~\cite{kim2020summarizing} use data summaries as explanations. Explain3D~\cite{wang2019explain}, which shares a similar context to ours, explains dataset disagreements with syntactic provenance-based and value-based modifications mappings between the datasets. In this work, whenever we refer to an ``explanation'', we refer to the one defined in \cref{def:explain}.}

\subsection{Explaining Dataset Changes}\label{sec:changesets}

We focus on two \emph{orientations} of explanations, namely, \emph{vertical} explanations %$\overrightarrow{\mathcal{E}}$ 
and \emph{horizontal} explanations. %$\downarrow\mathcal{E}$. 
%In addition, we also 
We further distinguish between \emph{removal} %explanations %$\mathcal{E}^{-}$ 
and \emph{addition} explanations.  Modifications can be modeled as a removal followed by an addition. %$\mathcal{E}^{+}$. 
%The explanation types differ in the type of relations that the origin and the goal represent.
Explanation types differ in the type of relations that the origin and the goal represent.

The goal ($\mathcal{G}_{relation}$) and origin ($\mathcal{O}_{relation}$) relations are defined with respect to versions $T$ or $T^{\prime}$. Specifically, the relations $\mathcal{O}_{relation}$ and $\mathcal{G}_{relation}$ can be a projection (subset of attributes) or selection (subset of tuples) over either $T$ or $T^{\prime}$. In vertical explanations (adding or removing attributes), the associated goal and origin names ($\mathcal{G}_{name}$ and $\mathcal{O}_{name}$) are the projected attributes. For horizontal explanations (adding or removing tuples), $\mathcal{G}_{name}$ and $\mathcal{O}_{name}$ correspond to the set of tuple ids in the subset. When clear from context, we refer to the goal and the origin by their names.

% \begin{example}
%     Recall Figure~\ref{fig:example_explained} and the dataset versions in Figures~\ref{fig:example_a} ($T$) and~\ref{fig:example_b} ($T^{\prime}$). An example explanation for a goal $\mathcal{G} = (a6, \pi_{a6}[T^{\prime}])$ is composed of an origin $\mathcal{O} = (a2, \pi_{a2}[T])$ and a transformation $\mathcal{P} = \pi_{a2}[T] \div 60$, which is tuple-based, i.e., divide each tuple in $\pi_{a2}[T]$ by 60. When clear from context, we denote this explanation as $\mathcal{E}_{a6} =(a2, a2 \div 60)$. %$\overrightarrow{\mathcal{E}^{+}}_{a7} =(a3, a3 \div 60)$. 
% \end{example}

\ifdefined\TechReport
\begin{table}[t]
	\caption{Notations used in the paper. The changes \lhda, \lhca, \lhdr, and \lhcr{} are defined wrt the left-hand table $T$. The right-hand notation %\lhda, \lhca, \lhdr and \lhcr 
	can be obtained by replacing $T$ with $T^{\prime}$ below.}\label{tab:notaions}
	\begin{tabular}{|c|c|c|} 
		\bottomrule
		&\textbf{Notation} & \textbf{Meaning} \\\toprule
		\parbox[t]{2mm}{\multirow{4}{*}{\rotatebox[origin=c]{90}{Basic}}} & $T$ & Left-hand dataset \\\cline{2-3}
		&$T^{\prime}$ & Right-hand (revised) dataset \\\cline{2-3}
		&$T_{A}$ & The attribute set of dataset $T$\\\cline{2-3}
		&$T_{r}$ & The tuple set of dataset $T$ \\\midrule
% 		\parbox[t]{2mm}{\multirow{4}{*}{\rotatebox[origin=c]{90}{Changes}}} & \lhda & The set of unmatched attributes in T\\
		\parbox[t]{2mm}{\multirow{8}{*}{\rotatebox[origin=c]{90}{Changes}}} & \lhda & The set of unmatched attributes in T\\
		& & $\{A_{i}: A_{i}\in T_{A} \wedge \nexists A'_{j}\in T': (A_{i}, A'_j)\in \Sigma_A\}$\\\cline{2-3}
		&\lhca & The set of matched attributes in T \\
		& & $T_{A}\setminus$ \lhda \\\cline{2-3}
        % &\lhdr & $\{\pi_{\lhcaD}[r_{j}]: r_{j}\in T_{r} \wedge \nexists r'_i \in T: (r_i, r'_i)\in \Sigma_r\}$ \\\cline{2-3}
        &\lhdr & The set of unmatched tuples in T \\
        & & $\{\pi_{\lhcaD}[r_{j}]: r_{j}\in T_{r} \wedge \nexists r'_i \in T: r_{0i} = r^{\prime}_{0i}\}$ \\\cline{2-3}
		&\lhdr & The set of matched tuples in T \\
		& & $\{\pi_{\lhcaD}[r_{j}]: r_{j}\in T_{r}\}\setminus $\lhdr \\\bottomrule
	\end{tabular}
\end{table}
\else
\begin{table}[t]
  \caption{Notations used in the paper. The changes \lhda, \lhca, \lhdr, and \lhcr{} are defined wrt the left-hand table $T$. The right-hand notation %\lhda, \lhca, \lhdr and \lhcr 
	can be obtained by replacing $T$ with $T^{\prime}$ below.}\label{tab:notaions}
	\vspace{-.075in}
\scalebox{.625}{\begin{tabular}{|l|cl|cl|}
\bottomrule
&\textbf{Notation}                    & \textbf{Meaning}                           & \textbf{Notation}                   & \textbf{Meaning}                \\ \toprule
\parbox[t]{2mm}{\multirow{2}{*}{\rotatebox[origin=c]{90}{\textbf{basic}}}} & $T$ & Left-hand dataset    & $T^{\prime}$                         & Right-hand (revised) dataset                  \\
&$T_{A}$ & The attribute set of dataset $T$ & $T_{r}$ & The tuple set of dataset $T$ \\\hline  
\parbox[t]{2mm}{\multirow{4}{*}{\rotatebox[origin=c]{90}{\textbf{changes}}}} &\lhda & Unmatched attributes in T  &\lhca & Matched attributes in T\\
&& $\{A_{i}: A_{i}\in T_{A} \wedge \nexists A'_{j}\in T': (A_{i}, A'_j)\in \Sigma_A\}$ & &$T_{A}\setminus$ \lhda \\
&\lhdr & Unmatched tuples in T & \lhcr & Matched tuples in T\\
&& $\{\pi_{\lhcaD}[r_{j}]: r_{j}\in T_{r} \wedge \nexists r'_i \in T: r_{0i} = r^{\prime}_{0i}\}$ & 
& $\{\pi_{\lhcaD}[r_{j}]: r_{j}\in T_{r}\}\setminus $\lhdr\\\bottomrule
\end{tabular}}%
\end{table}

\fi

\begin{example}
    Recall the versions in Figures~\ref{fig:example_a} ($T$) and~\ref{fig:example_b} ($T^{\prime}$). An example explanation for a goal $\mathcal{G} = (a6, \pi_{a6}[T^{\prime}])$ is composed of an origin $\mathcal{O} = (a2, \pi_{a2}[T])$ and a transformation $\mathcal{P} = \pi_{a2}[T] \div 60$, which is tuple-based, i.e., divide each tuple in $\pi_{a2}[T]$ by 60. When clear from context, we denote this explanation as $\mathcal{E}_{a6} =(a2, a2 \div 60)$. %$\overrightarrow{\mathcal{E}^{+}}_{a7} =(a3, a3 \div 60)$. 
\end{example}
Recalling the change sets, we aim to find vertical addition explanations for \lhda, vertical removal explanations for \rhda, horizontal addition explanations for \lhdr, and horizontal removal explanations for \rhdr.  An explicit problem definition, that relies on the quality of explanations is provided in \cref{sec:problem}.
%the explicit problem addressed in this paper is as follows.

% \begin{definition}\label{def:problem}
% 	Let $T$ and $T^{\prime}$ be dataset versions and \lhda, \rhda, \lhdr, and \rhdr be the left-hand delta attributes, right-hand delta attributes, left-hand delta tuples, and right-hand delta tuples. We aim to find vertical addition explanations to \lhda, vertical removal explanations to \rhda, horizontal addition explanations to \lhdr and horizontal removal explanations to \rhdr.
% \end{definition}

% \begin{definition}\label{def:problem}
% 	Let $T$ and $T^{\prime}$ be dataset versions and \lhda, \rhda, \lhdr, and \rhdr be the left-hand delta attributes, right-hand delta attributes, left-hand delta tuples, and right-hand delta tuples. We aim to find vertical addition explanations ($\overrightarrow{\mathcal{E}}^{+}$) to \lhda, vertical removal explanations ($\overrightarrow{\mathcal{E}}^{-}$) to \rhda, horizontal addition explanations ($\downarrow\mathcal{E}^{+}$) to \lhdr and horizontal removal explanations ($\downarrow\mathcal{E}^{-}$) to \rhdr.
% \end{definition}

\begin{example}\label{exp:explanations}
% 	Recall Figure~\ref{fig:example_explained} that annotates the changes between the dataset versions of Figure~\ref{fig:example} and \lhda, \rhda, \lhdr, and \rhdr, defined in Example~\ref{example:changes}. A possible set of explanations to explain the changes is as follows
Recall the versions of Figure~\ref{fig:example}, annotated changes in Figure~\ref{fig:example_explained}, and \lhda, \rhda, \lhdr, and \rhdr, defined in Example~\ref{example:changes}. A possible set of explanations to explain the changes is as follows
	\begin{itemize}
% 		\item[$\mathbf{\overrightarrow{\mathcal{E}^{+}}_{a6}}$] $= (a2,$ \texttt{extract(a2, `(.*?)')}$)$ 
% 		\item[$\mathbf{\overrightarrow{\mathcal{E}^{+}}_{a7}}$] $= (a3, a3 \div 60)$  
% 		\item[$\mathbf{\overrightarrow{\mathcal{E}^{+}}_{a8}}$] $= \left(a4, \begin{cases}
% 					4, & \text{if } 9\leq a4\\
% 					3, & \text{if } 8\leq a4 < 9\\
% 					2, & \text{if } 7\leq a4 < 8\\
% 					1, & \text{otherwise}
% 					\end{cases}\right)$
% 		\item[$\mathbf{\overrightarrow{\mathcal{E}^{+}}_{a9}}$] $= (a2,$ \texttt{len(a2)}$)$ 
% 		\item[$\mathbf{\downarrow\mathcal{E}^{-}_{m4}}$] $= (\emptyset,$ \texttt{has\_NaN}$)$
\item[$\mathbf{\mathcal{E}_{a5}}$] $= (a1,$ \texttt{extract(a1, `(.*?)')}$)$ 
		\item[$\mathbf{\mathcal{E}_{a6}}$] $= (a2, a2 \div 60)$  
		\item[$\mathbf{\mathcal{E}_{a7}}$] $= \left(a3, \begin{cases}
					4, & \text{if } 9\leq a3\\
					3, & \text{if } 8\leq a3 < 9\\
					2, & \text{if } 7\leq a3 < 8\\
					1, & \text{otherwise}
					\end{cases}\right)$
		\item[$\mathbf{\mathcal{E}_{a8}}$] $= (a1,$ \texttt{len(a1)}$)$ 
		\item[$\mathbf{\mathcal{E}_{m4}}$] $= (\emptyset,$ \texttt{has\_NaN}$)$
	\end{itemize}
Most of the explanations are self-explanatory (as they should be). Interesting cases are $\mathcal{E}_{a5}$, %$\overrightarrow{\mathcal{E}^{+}}_{a6}$
%rjm , for which we use some programming language 
%rjm notation 
%jm to express that a value, given in parenthesis, is extracted from the column \texttt{a1}. 
that extracts value in parenthesis.  Another example is the 
%rj horizontable 
horizontal
explanation $\mathcal{E}_{m4}$, %$\downarrow\mathcal{E}^{-}_{m4}$
for which the origin is an empty set and 
%rjm added
the tuple
was removed due to a \texttt{NaN} (null equivalent) value. 
%rjm We use \texttt{has\_NaN} for convenience, in practice, the transformation is one that removes tuples with NaN values from T. 
Although the goal is $m4$, the transformation we find is one that removes $m4$ and no other tuples.

%The term \texttt{has\_NaN} is used for convenience, but in practice the transformation is one that removes tuples with NaN (null equivalent) values from T. While that although the goal is $m4$, the transformation we find is one that removes $m4$ and no other tuples.
%Notice that although the goal is $m4$, the transformation we find is one that removes $m4$ and no other tuples.
% $m4=\begin{cases}
% \emptyset, & \text{if } NaN\in m4\\
% m4, & \text{otherwise}
% \end{cases}$
%$m4=\begin{cases}
%m4, & \text{if } NaN\in m4\\
%\emptyset, & \text{otherwise}
%\end{cases}$
% Notice that although the goal is $m4$, the transformation we find is one that removes $m4$ and no other tuples.
\end{example}

\noindent\texttt{Explain-Da-V} is a data-driven\footnote{Data-driven reflects that we only use data values (we do not use meta-data).} method composed of four parts corresponding to adding/removing attributes/tuples.
 We first describe our core explanation methods (\cref{sec:methods}), which are then utilized to explain vertical %changes
 (\cref{section:vertical}) and horizontal changes (\cref{section:horizontal}).

\revision{\section{Core Semantic Explanation Methods}\label{sec:methods}}

\revision{Our core explanation methods rely on fitting an appropriate explanation methodology to data types we find in the origin $\mathcal{O}$ and the goal $\mathcal{G}$. Rather than the traditional database attribute types (strings, integers, floats, etc.), given the nature of our analysis, we look into ML feature types~\cite{shah2021towards}. %Specifically, w
We focus on three main types, namely, \emph{Numeric}, \emph{Categorical} and \emph{Textual} (mixed types are considered textual), which characterizes the core changes \texttt{Explain-Da-V} covers.\footnote{\texttt{Explain-Da-V} can be easily extended to support additional types such as dates.} 
%rjm repetitious
%Specifically, \texttt{Explain-Da-V} works iteratively and looks into cases where the goal is either numeric, categorical or textual.
Aiming to resolve a high variety of changes, we develop methods that are built on top of different types of origin sets 
%rjm (e.g., including numeric, categorical and textual values) 
using multiple approaches that exploit the type of change. Accordingly, \texttt{Explain-Da-V} can be applied over any pair of versions, regardless of how far apart the versions are (meaning how many transformations have been applied).  
%rjm as long as the changes are known and internal (see \cref{sec:intro}).} 
%\rjm{
If an explanation is not found for a specific change, it is declared idiopathic (unexplained).} 
%}

\revision{Note that among the different changes, the vertical additions are the most common and complex and, thus, the presented methods mostly address such a scenario. Specifically, for the presentation of methods, we assume the goal as a single attribute %\rjm{
(a right-hand attribute to be explained) %} 
with its data values. Also, given a goal, finding its origin is not straight forward. %\rjm{
For the moment, assume the origin is the original left-hand table $T$. %} 
We discuss a method to ``find'' an origin, given a goal, in \cref{sec:origin}.} 

%Rather than the traditional database attribute types (strings, integers, floats, etc.), given the nature of our analysis, we look into ML feature types~\cite{shah2021towards}. %Specifically, w
% We focus on three main types, namely, \emph{Numeric}, \emph{Categorical} and \emph{Textual} (mixed types are considered textual).\footnote{\texttt{Explain-Da-V} can be easily extended to support additional types such as dates.}

% \revision{We now cover the main explanation methods we use as a part of \texttt{Explain-Da-V}. Note that among the different changes, the vertical additions are the most common and complex and, thus, the presented methods mostly address such a scenario. Specifically, for the presentation of methods, we instantiate the goal as a single attribute with its data values. Also, given a goal, finding its origin is not straight forward. We discuss a method to ``find'' an origin, given a goal, in \cref{sec:origin}.} 

\revision{\subsection{Numeric Change Explanations}\label{sec:num_methods}}
% In this case, both the origin and goal are numeric. 
\revision{Whenever we need to explain a numeric goal using an origin that contains numeric data, w}e position the problem as \emph{regression} in which the origin relation tuples are treated as independent variables and the goal relation tuples as dependent variables. Aiming at explainable transformations, we build on top of linear regression~\cite{burkart2021survey}. To reduce model complexity and prevent over-fitting~\cite{thrampoulidis2015regularized}, we experiment with Lasso and Ridge regularization. 

\begin{example}\label{example:num2num0}
	%One example of numeric-to-numeric 
	A numeric transformation is given in Figure~\ref{fig:example_b}, where explaining \texttt{a6} can be resolved by fitting a regressor $\frac{1}{60}\cdot a2$.
\end{example}

Not all numeric transformations can be covered by a linear function. Accordingly, to allow richer, more flexible, numeric transformations, we \emph{extend} the feature space (i.e., the origin) by generating additional features. \revision{Note that while these extensions are motivated by commonly used data science and engineering operations~\cite{kuhn2019feature}, they do not (and cannot) cover every possible transformation.}

\vspace{.05cm}
\noindent\textbf{Polynomial Regression and Inter-relation Features:}
To explain polynomial transformations, we generate additional polynomial features~\cite{edwards2002alternatives} over $\mathcal{O}$. %Formally, g
Given a predefined degree $d$, the polynomial extension of $\mathcal{O}$ is given by $poly(\mathcal{O})$ whose attributes correspond to $\{A_{i}^{2}, \dots, A_{i}^{d}, \forall A_{i}\in \mathcal{O}_{name}\}$. The %composed 
extended origin relation is created on a tuple level by applying the associated operation. For example, the attribute $A_{i}^{2}$ of the tuple $r_{j}$ in the extended relation would get the value in attribute $A_i$ squared, meaning $\pi_{A_{i}}[r_{j}]^{2}$. We also introduce feature inter-relation, that is, multiplication and division between different attribute values in $\mathcal{O}$. Note that addition and subtraction are already supported when using linear regression. %Formally, t
The inter-relation extension of $\mathcal{O}$, $inter(\mathcal{O})$ corresponds to $\{A_{i}\cdot A_{j}, A_{i}\div A_{j}\dots, \forall A_{i}, A_{j}\in \mathcal{O} \wedge A_{i} \neq A_{j}\}$. Also here the transformations are done on the tuple level, e.g., the attribute $A_{i}\cdot A_{j}$ of the tuple $r_{j}$ would get the value $\pi_{A_{i}}[r_{j}]\cdot\pi_{A_{j}}[r_{j}]$. The extensions can also be applied consecutively, e.g., $inter(poly(\mathcal{O}))$ to create attributes such as $A_{i}\div A_{j}^{2}$ to resolve, for example, the BMI formula ($kg\div m^{2}$). \ifdefined\TechReport For example, recall Figure~\ref{fig:example} and let $\mathcal{O}_{name} = \{a_2, a_3\}$, the extended features using $poly(inter(\mathcal{O}))$ with $d=2$ would be:

\begin{figure}[h]
	\centering
	\includegraphics[width=.45\textwidth,trim=0 0 30 30]{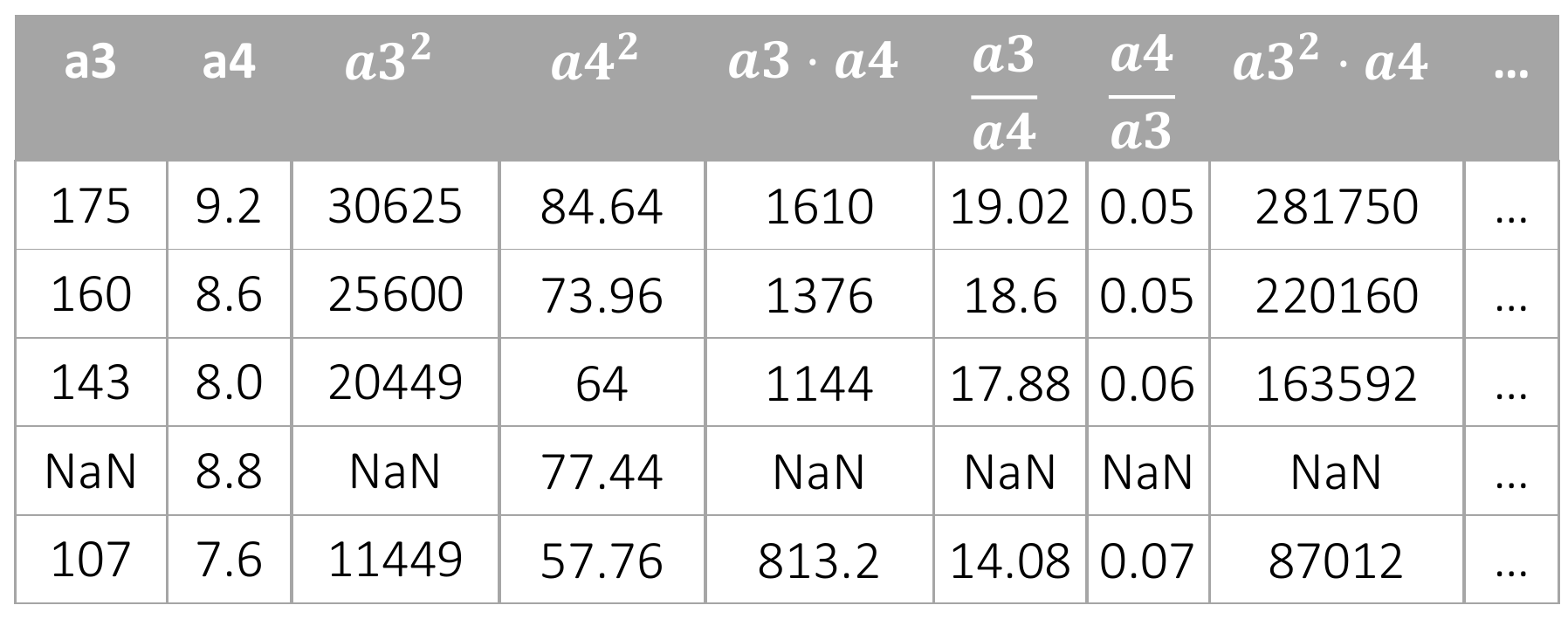}
	\caption{Polynomial and Inter-relation extensions over \texttt{a3} and \texttt{a4} Figure~\ref{fig:example_a}.}
	\label{fig:poly_inter_ext}
\end{figure}
\else
% \begin{figure}[h]
% 	\centering
% % 	\includegraphics[width=.45\textwidth]{./figures/poly_example}
% 	\includegraphics[width=.4\textwidth]{./figures/poly_example_SMALL}
% % 	\caption{Polynomial and Inter-relation extensions over \texttt{a3} and \texttt{a4} Figure~\ref{fig:example_a}.}
% \caption{Poly. and Inter. extensions over \texttt{a3} and \texttt{a4} Figure~\ref{fig:example_a}.}
% 	\label{fig:poly_inter_ext}
% \end{figure}
\fi

\vspace{.05cm}
\noindent\textbf{Mathematical Transformations:} When generating new features over numeric data, it is also common to use mathematical operations~\cite{brook2018applied}. Specifically, to support this 
% \rjm{type of}
type of explanation we generate a math extension of $\mathcal{O}$, $math(\mathcal{O})$, with the attributes $\{log(A_{i}), sqrt(A_{i}), reciprocal(A_{i}), exp(A_{i}), \dots, \forall A_{i}\in \mathcal{O}\}$, where $sqrt(A_{i}) = \sqrt{A_{i}}$, $reciprocal(A_{i}) = A_{i}^{-1}$, and $exp(A_{i}) = e^{A_{i}}$. %$\mathcal{O}_{math} = math(\mathcal{O}) = \{log(A_{i}), sqrt(A_{i}), reciprocal(A_{i}), exp(A_{i}) \dots, \forall A_{i}\in \mathcal{O}\}$, where $sqrt(A_{i}) = \sqrt{A_{i}}$, $reciprocal(A_{i}) = A_{i}^{-1}$, and $exp(A_{i}) = e^{A_{i}}$. 
%As above, t
The transformations are %done on a 
tuple-based, e.g., the %new 
feature $log(A_{i})$ of the tuple $r_{j}$ in the extended relation would get the value $log(\pi_{A_{i}}[r_{j}])$. \ifdefined\TechReport For illustration, recall our running example, the extended features using $math(\{a_2, a_3\})$ would be:
%For illustration, recall our running example and let $\mathcal{O} = \{a_3, a_4\}$, the extended features using $math(\mathcal{O})$ would be:

\begin{figure}[h]
	\centering
	\includegraphics[width=.45\textwidth,trim=0 0 30 30]{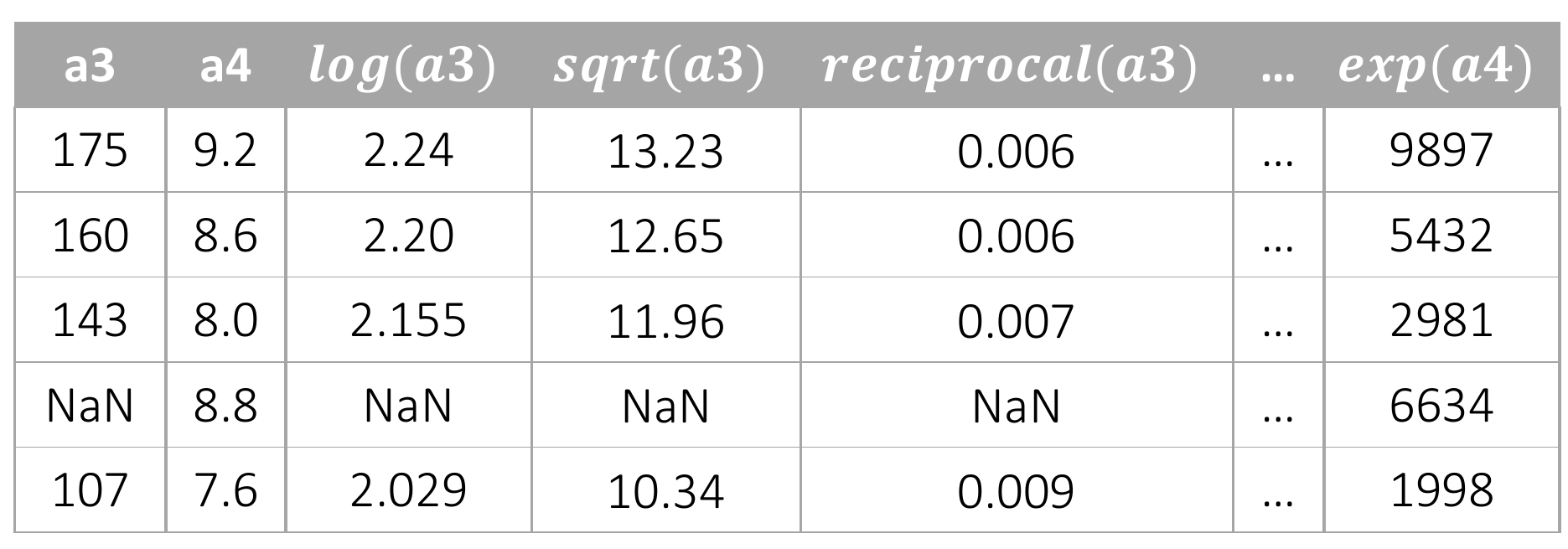}
	\caption{Math extension over \texttt{a3} and \texttt{a4}  Figure~\ref{fig:example_a}.}
	\label{fig:math_ext}
\end{figure}
\else
% \begin{figure}[h]
% 	\centering
% 	\includegraphics[width=.4\textwidth]{./figures/math_example_SMALL}
% 	\caption{Math extension over \texttt{a3} and \texttt{a4}  Figure~\ref{fig:example_a}.}
% 	\label{fig:math_ext}
% \end{figure}
\fi
% \vspace{.05cm}
\vspace{.05cm}
\noindent\textbf{Global Aggregations:} We also generate aggregate features. This extended set is especially important when looking at one of the most common transformations in machine learning, that is, (value) normalization~\cite{zheng2018feature}. 
%rjm Similar to 
%As above, w
We introduce an extension of $\mathcal{O}$, $agg(\mathcal{O})$, with the attributes $\{sum(A_{i}), mean(A_{i}), max(A_{i}), min(A_{i}) \dots, \forall A_{i}\in$  $\mathcal{O}\}$. Here, although assigned on a tuple level, the extended values are computed over all the values in the attribute. For example, the feature $sum(A_{i})$ of the tuple $r_{j}$ would get the value $\sum_{r_{ik}\in \pi_{A_{i}(T)}}r_{ik}$. Then, if a user applies a sum normalization over $A_{i}$, the feature-set $agg(inter(\mathcal{O}))$ that includes the feature $A_{i}\div sum(A_{i})$ would be able to resolve and explain this added attribute. Similarly, in the case of min-max normalization~\cite{zheng2018feature} the additional features of $min(A_{i})$ and $max(A_{i})$ can be used to generate accurate explanations. Another example is the common collaborative filtering transformation of subtracting the mean value ($A_{i} - mean(A_{i})$)~\cite{marlin2004collaborative}. %Another example that can be solved using $agg(\mathcal{O})$ is the common collaborative filtering transformation of subtracting the mean value ($A_{i} - mean(A_{i})$)~\cite{marlin2004collaborative}. %For illustration, r
\ifdefined\TechReport Recalling our running example and let $\mathcal{O} = \{a_2, a_3\}$, the extended features using $agg(\mathcal{O})$ would be:

\begin{figure}[h]
	\centering
    \includegraphics[width=.45\textwidth,trim=0 0 30 30]{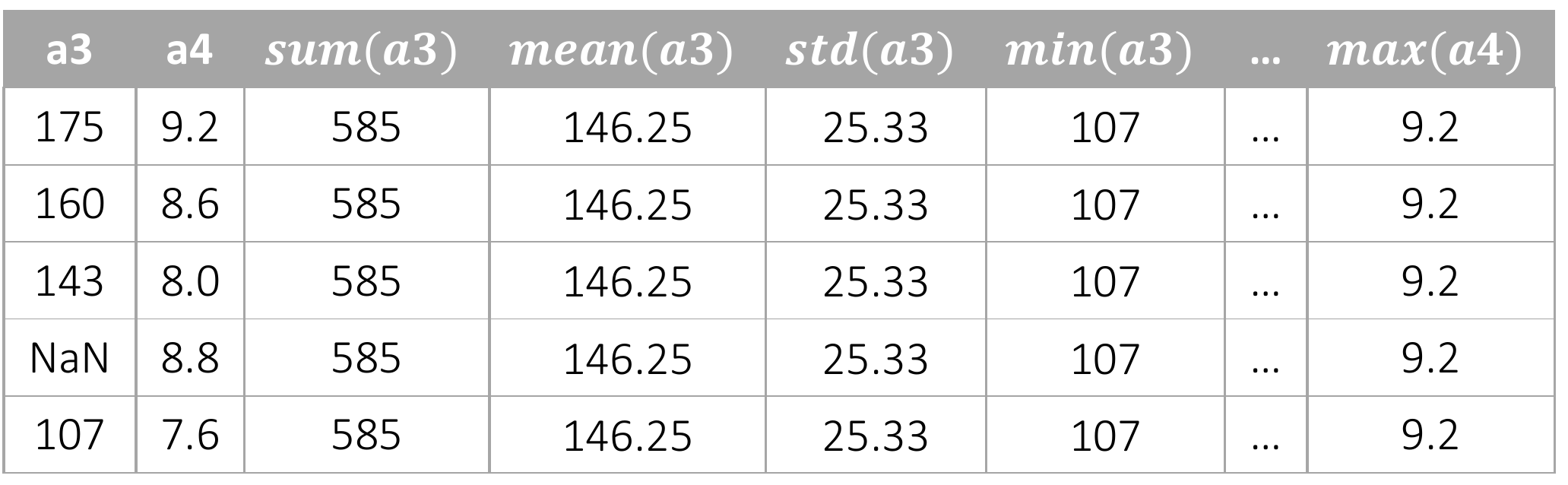}
	\caption{Aggregation extension over \texttt{a3} and \texttt{a4}  Figure~\ref{fig:example_a}.}
	\label{fig:agg_ext}
\end{figure}
\else
% \begin{figure}[h]
% 	\centering
%     \includegraphics[width=.4\textwidth]{./figures/agg_example_SMALL}
% 	\caption{Aggregation extension over \texttt{a3} and \texttt{a4}  Figure~\ref{fig:example_a}.}
% 	\label{fig:agg_ext}
% \end{figure}
\fi
% \vspace{.05cm}

An extended feature-set is used to fit a %transformation 
regressor that %can 
assigns %different 
coefficients for %the 
extended feature-set. %in the set. %in the selected feature-set. 
A perfect regressor would be able to deal %with 
one all-inclusive feature-set (i.e., $poly(inter(\dots(\mathcal{O})\cup inter(poly(\dots(\mathcal{O}), \dots$); yet, since it is not realistic to expect that (from an explainable regressor), we apply each set independently (power-set of the extensions) and generate multiple ``possible'' explanations. Section~\ref{sec:search} addresses the issue of choosing among them.
%We address the issue of choosing one among them in Section~\ref{sec:search}.

Note that the transformation is ``learned'' (fitted) only based on the two dataset versions $T$ and $T^{\prime}$ and \textbf{no additional training data} is required. \revision{\cref{sec:col_addition} discusses the main application of numeric change explanations and illustrates it using multiple examples.}

% \vspace{-.1in}
\revision{\subsection{Categorical Change Explanations}\label{sec:cat_methods}}
% When the origin is numeric and the goal is categorical,
\revision{Aiming to explain a categorical goal using an origin relation that contains numeric,}
%or categorical data,} 
we position the problem as \emph{classification}, in which the tuples of the origin relation are treated as explanatory variables and the goal relation tuples are used as the output class labels. The output class can be binary (e.g., is movie longer than two hours) or multi-class (e.g., \texttt{a7}, Example~\ref{example:intro}). Aiming at explainability, we focus on decision trees~\cite{burkart2021survey}. \revision{Note that decision trees cover only explanations that can be represented as disjunction of conjunctions~\cite{mitchell1997decision}. Each path from the root to a leaf corresponds to a conjunction and the tree %itself 
is the disjunction of these conjunctions.}

\begin{wrapfigure}[7]{r}{0.225\textwidth}
    \raggedright
    \vspace{-.05in}
    \scalebox{0.75}{\begin{tikzpicture}[every tree node/.style={draw},
	level distance=.8cm,sibling distance=0.1cm,
	edge from parent path={(\tikzparentnode) -- (\tikzchildnode)}]
	\revision{\Tree
	[.\texttt{a3>9}
	\edge node[auto=right] {$False$};
	[.\texttt{a3<7}
	\edge node[midway,left] {$True$};
	[.\textbf{a7=1} ]
	\edge node[midway,right] {$False$};
	[.\texttt{a3>8}
	\edge node[midway,left] {$True$};
	[.\textbf{a7=3} ]
	\edge node[midway,right] {$False$};
	[.\textbf{a7=2} ]
	]
	]
	\edge node[auto=left] {$True$};
	[.\textbf{a7=4}
	]
	]}
	\vspace{-.25in}
	\end{tikzpicture}}
\end{wrapfigure}
\begin{example}\label{example:categorical0}
	\revision{Figure~\ref{fig:example_b} provides an example %of numeric-to-
	a categorical transformation, namely, \texttt{a7}, which can be resolved with the help of the following decision tree.}  
\end{example}

\revision{\subsection{Textual Change Explanations}\label{sec:tex_methods}}
When the origin and/or goal are textual, we follow the PBE approach (see Section~\ref{sec:related3}), using a search-based solution. Specifically, we adopt an existing framework called Foofah~\cite{foofah}. The %adopted 
PBE solution is composed of designing a space of possible operators and a search algorithm. The search algorithm (A*, following Foofah) navigates the space of operators using a heuristic function (based on dissimilarity of tables) that estimates the cost of any proposed partial solution. The space is %further 
pruned %by a set of rules 
to boost search speed~\cite{foofah}.

% \vspace{.05cm}
\noindent\subsubsection{\revision{Textual-to-Textual}}\label{sec:text2text} %\renee{The next sentence is out of place now, we haven't really explained this yet, right?  Perhaps remove this sentence?}. Recall that in addition to finding valid transformations (as done in PBE systems like Foofah), an important contribution of our work is to find transformations that are explainable and generalizable.  
%rjmIn addition, 
% To a
Addressing %the 
data versioning, %management problem, 
we extend the traditional PBE operators to include operators the cover frequent text-processing steps~\cite{vijayarani2015preprocessing}, including text lowering, lemmatization, removal of special charterers (e.g., punctuations and numeric values) and tokens (e.g., stop-words and html tags). All implemented additional operators for foofah are given in our repository~\cite{FoofahExt}. 
%\renee{Is there anyway to make this stronger?  Anything novel in the design of this search space?  Should we at least say "unlike Foofah we consider textual-to-numeric and textual-to-categorical?  But if we can say something more about search I think that would help with novelty/ad-hoc comments.}. In addition to textual-to-textual transformations, we use this framework also for textual-to-numeric and textual-to-categorical transformations by designing a search space that suits this broader domain

Unlike Foofah, we also consider textual-to-numeric and textual-to-categorical. Note that although in recent years transformer-based models have become a standard way to extract (latent) features from text, traditional feature engineering over text, that is, extracting manual numeric and categorical features from textual values, is still an important ingredient in NLP~\cite{da2019fine,kovaleva2019revealing,chernyavskiy2021transformers} and in other research disciplines such as HCI~\cite{cao2021my} and information management~\cite{gkikas2022text}. 

% \vspace{.05cm}
\noindent\subsubsection{\revision{Textual-to-Numeric}}\label{sec:text2num} The search space defined for resolving this kind of transformation includes a meta-operation that counts the occurrences of some pattern $pat$ in a value (\texttt{count}$_{pat}(r_{ij})$), where $r_{ij}$ is a value in the table, see Section~\ref{sec:prelim}).
%rjm  of some pattern in a value. 
Using this operation we can define operations such as \texttt{number\_of\_words} = \texttt{count}$_{\text{`␣'}}(r_{ij})$ and \texttt{number\_of\_questions} = \texttt{count}$_{\text{`?'}}(r_{ij})$. We also cover counting a pre-defined set of stop-words and punctuation marks.

% \vspace{.05cm}
\noindent\subsubsection{\revision{Textual-to-Categorical}}\label{sec:text2cat} %As above, 
% We define a similar meta-operation for the existence of a pattern $pat$ (\texttt{contains}$_{pat}(r_{ij})$), which is used to generate operations such as \texttt{contains\_multiple\_lines} = \texttt{contains}$_{`\backslash n '}(r_{ij})$.
We define a similar meta-operation for a pattern existence (\texttt{contains}$_{pat}(r_{ij})$), which is used to generate operations such as \texttt{contains\_percent} = \texttt{contains}$_{`\%'}(r_{ij})$.

\begin{example}\label{example:num2cat0}
	Figure~\ref{fig:example_b} provides an example of textual-to-textual transformation, namely, \texttt{a5}, which can be resolved with the help of the foofah environment in its original implementation. Specifically, if we consider \texttt{a1} as an origin, foofah would consider $\pi_{a1}[T]$ as input examples and $\pi_{a5}[T^{\prime}]$ as corresponding output examples (goal in our terms) and synthesize a tuple-based data transformation program 
	\sql{(1) t = split(t, 0, `(')}, \sql{(2) t = split(t, 1, `)')}, \sql{(3) t = drop(t, 0)}, \sql{(4) t = drop(t, 2)}\\
	\noindent the tuple value \texttt{Moana (U)}, for example, would be transformed as follows \texttt{[Moana , U)]} $\rightarrow$\texttt{[Moana , U, ]} $\rightarrow$ \texttt{[U, ]} $\rightarrow$ \texttt{U}.
	
	Resolving \texttt{a8} requires \texttt{Explain-Da-V}'s
	extensions that includes textual-to-numeric transformations (\texttt{len()}).
\end{example}

%rjm \revision{Obviously, also the extensions we introduce to Foofah cannot cover all possible textual changes, especially when it comes to idiopathic text transformations.}
% \revision{Note that Foofah aims to synthesize transformations using as little input (example tuples) as possible, and is often able to do so using just one or two examples. Our goal in contrast is to generate an explanation that correctly explains a full table transformation (a dataset version). Hence, our tremendous expansion of the search space beyond text-to-text transformations plays a critical role.  
% }

% \renee{anything we can say about changes to the search strategy which you allude to above?}

\revision{Note that Foofah aims to synthesize transformations using a given set of example tuples, and is often able to do so using just a few examples. Our goal, in contrast, is to generate an explanation that correctly explains a full table transformation (a dataset version). Hence, our tremendous expansion of the search space beyond text-to-text transformations plays a critical role. Moreover, \cref{sec:origin} introduces a technique that prunes the search space in this context.  
}

\subsection{\revision{Categorical-encoding Change Explanations}}\label{sec:encode_methods}

% \revision{Whenever dealing with mixed types, that also contain some textual values%for a machine learning task
% , these values may be encoded.} 
\revision{Whenever dealing with mixed types, textual and categorical values may be encoded.} A common encoding approach, which we use here, is one-hot-encoding~\cite{rodriguez2018beyond}. Let $A_{i}\in \mathcal{O}$ be a textual/categorical attribute in the origin. One-hot-encoding of this attribute generates an additional attribute for each unique value (or category) in $A_{i}$ and assigns a value of 1 to each tuple that corresponds to this value (category). This addition, not only allows the resolution of this common encoding scheme, but also a richer representation that can be used to resolve other types of encoding (e.g., ordinal encoding) and additional transformations. \revision{Such encodings are also commonly used in data preparation for machine learning~\cite{brownlee2022data}. Recalling \cref{example:intro}, if a user aims to predict the rating of a movie, extracting features such as \texttt{Is\_Drama} or \texttt{Is\_Action} can be beneficial for learning.}

% \rs{ADD TO RESPONSE}

%\revision{%\rjm{
% Such encodings are commonly used before training a classifier, e.g., we may extract features such as {\tt contains a dollar sign} or {\tt contains a dash} to create a version of a dataset on which a schema matching algorithm can be applied.}
%}

% \renee{R3 has a question about this should we explain more?  feel free to change my example or create a full example in an example environment} 

\subsection{\revision{Reshaping Change Explanations}}\label{sec:reshape_methods}

% \renee{is this blue above meant to mean the whole subsection is new or just the subsection division?}
% \rs{Just new division. The text is taken from a different place in the previous version.}

Generally group-by is a table-reshaping operation~\cite{yang2021auto}, i.e., a natural attribute-match and tuple-match do not exist. However, when it comes to feature engineering, group-by can also be used to generate aggregated features based on some other attribute. The latter is addressed %rjm similarly 
in a manner that is similar
to numeric change explanations.
% numeric-to-numeric. %Let $A_{i}\in \mathcal{O}$ be a textual value and $A_{j}\in T_{A}$ be a numeric value. 
An extended origin, similar to Section~\ref{sec:num_methods}, would be created for each numeric attribute $A_{j}\in T_{A}$ with respect to each textual/categorical attribute $A_{i}\in \mathcal{O}$ independently or conjointly (grouping by multiple attributes). We use an SQL syntax for clarity. 
\begin{wrapfigure}{r}{0.225\textwidth}
    \raggedright
    \vspace{-.05in}
    \noindent\fbox{%
    \parbox{.225\textwidth}{\sql{SELECT $A_{i}$, mean($A_{j}$), max($A_{j}$)...}\\
    \sql{FROM T}\\
    \sql{GROUP BY $A_{i}$}}}
    \vspace{-.05in}
    \caption{Group by Query}
    \label{fig:groupby}
\end{wrapfigure}
A helper query (Figure~\ref{fig:groupby}) can be used to generate the extended group-by features of the numeric attribute $A_{j}\in T_{A}$ with respect to a textual/categorical attribute $A_{i}\in \mathcal{O}$.

% \renee{Why not group by categorical attributes?  I think this ties into the origin.  Remember we haven't told the reader how to compute an origin yet.}

If more than one numeric attribute exists, it will be added to the GROUP BY and SELECT clauses.
Using this helper query, by joining it with the origin, we obtain the additional possible attributes. %For example, if the attribute \texttt{a4} of Figure~\ref{fig:example_a} is found in an origin, 
\ifdefined\TechReport Consider, for example, the attribute \texttt{a4} of Figure~\ref{fig:example_a}. The additional aggregation features would be generated over the numeric attributes \texttt{a2} and \texttt{a3} and joined into the table, as illustrated in \cref{fig:groupby_ext}. Among these, we can find the mean \texttt{a2} (runtime) and \texttt{a3} (rating) per \texttt{a4} (genre), supporting such explanations.

\begin{figure}[h]
	\centering
	\includegraphics[width=.45\textwidth,trim=0 0 30 30]{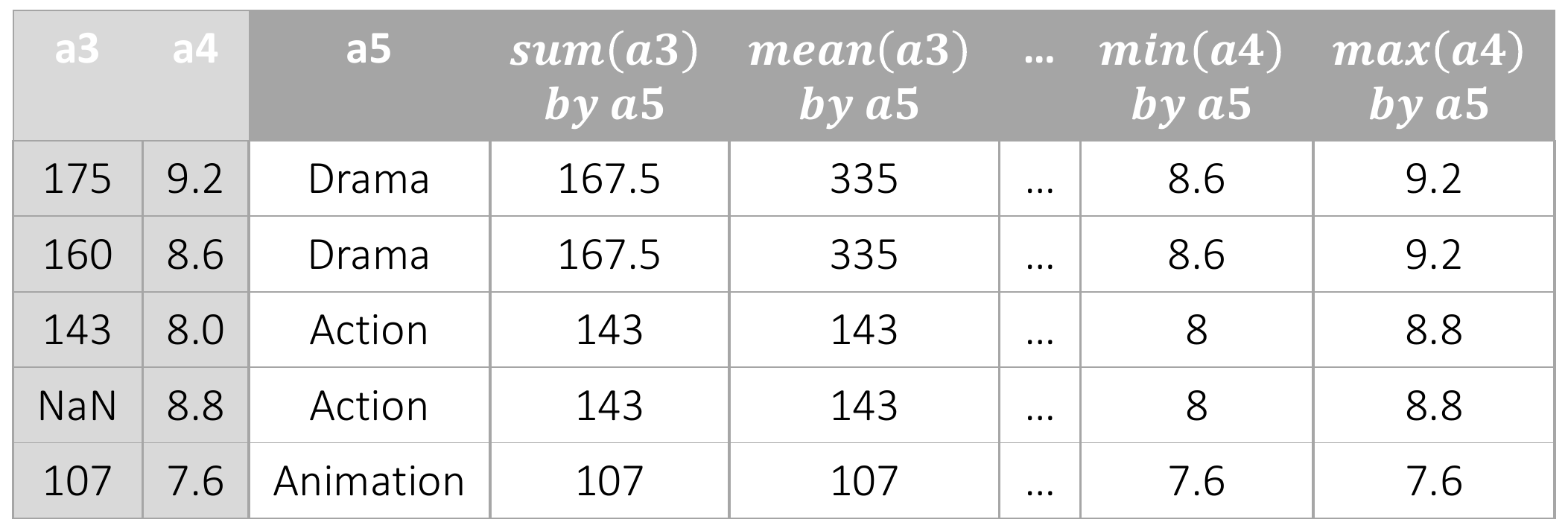}
	\caption{Groupby feature extension over \texttt{a3} and \texttt{a4} Figure~\ref{fig:example_a}.}
	\label{fig:groupby_ext}
\end{figure}
\else
% \begin{figure}[h]
% 	\centering
% 	\includegraphics[width=.4\textwidth]{./figures/groupby_example_SMALL}
% 	\caption{Groupby feature extension over \texttt{a3} and \texttt{a4} Figure~\ref{fig:example_a}.}
% 	\label{fig:groupby_ext}
% \end{figure}
\fi

% \renee{Not sure I understand.  First there are two sentences back-to-back starting with for example, and second are these added to feature set or group by?}
% \noindent For example, the mean \texttt{a2} (runtime) and \texttt{a3} (rating) per \texttt{a4} (genre) %will be
% are added to the feature set to support resolving such %transformation
% explanations.  

We also consider the reshaping scenario introduced in previous work, e.g.,~\cite{yang2021auto}. Reshaping is often considered as %one of the 
a possible operation %s 
that can be applied over a table throughout the search (see Section~\ref{sec:related3}). %As an alternative, we introduce a data-driven approach. 
We introduce an alternative data-driven approach. Specifically, reshaping is associated only with attribute-match 
%without a 
%\rjm{
when there is no %} 
tuple-match. We explicitly reshape the table using the query of Figure~\ref{fig:groupby} and fit a regressor over it. \revision{%In its current form
Currently, \texttt{Explain-Da-V} does not support other reshaping transformations such as transpose and pivot, which we intend to explore in future work.}

%\revision{Except the aforementioned group-by reshaping transformation, \texttt{Explain-Da-V} mainly considers }

% \renee{Roee what do you think about moving "finding origin here and talking about the origin for all types of changes - I know it is most important for attribute additions, but it would look more systematic to describe for all.  And placing it here we can argue both why we needed to expand the search space greatly over Foofah but also how we use the origin to tame this search space.}

\subsection{\revision{Finding the Origin}}\label{sec:origin}
\revision{In our discussion so far, we have assumed that the origin for the transformations $\mathcal{P}$ is the original relation $T$.
However, we can make our search more efficient if we can determine that for a specific goal, the origin is only a portion of $T$.
%In addition to methods that aim to find the transformation $\mathcal{P}$, the origin is an important ingredient as well. 
Accordingly, we aim to \emph{``find the origin''}. Different from some related literature where the input-output scope is clear (see Section~\ref{sec:related3}), in data versioning we do not know what was used to derive a specifical goal. Hence, our approach first searches for an origin for the transformation.}

A na\"ive solution to use all available data values. For example, in the context of adding attributes, using all attributes of $T$ as an origin, i.e., $\mathcal{O} = (T_{A}, T)$. The problem with this method is twofold. First, unrelated attributes may serve as noise when aiming to find a proper transformation for the goal. For example, referring back to \cref{example:intro}, aiming to resolve \texttt{a5} using all attributes (\texttt{a1}-\texttt{a4}) presents a much larger search space than aiming to resolve it using \texttt{a2}. A second issue has to do with the data types. \revision{Using a numeric change explanation (\cref{sec:num_methods}) may be more beneficial than a textual change explanation (\cref{sec:tex_methods}). For example, using \texttt{a2} as an origin to explain \texttt{a5} instead of using  \texttt{a1} to \texttt{a4}.}

When examining the creation of a new attribute from existing attributes%\footnote{\rs{should we remove this from here since we say that in the intro?} This work focuses on ``internal'' additions.  External additions, for example finding joinable tables~\cite{2019_zhu_josie} and joining them into $T$, are reserved for future work.}
, we observe a side effect of creating a \emph{functional dependency} between the origin and the new attribute (goal). For example, if two movies have the same runtime in minutes (\texttt{a2}), they will have the same runtime in hours (\texttt{a5}), which, by definition constructs a functional dependency between \texttt{a2} and \texttt{a5}. 
Accordingly, we use a functional dependency discovery algorithm~\cite{2015_papenbrock_fdep_comparison} to find the origin.\footnote{In our experiments we use a discovery algorithm called FDEP~\cite{1999_flach_database_dependency_discovery}.} 
%rjm shortening
%The dependencies we are interested in are the ones 
We find dependencies in which our goal is the dependent set and the discovered determinant is used as the origin. Note that there can be more than one attribute set that determines the goal and accordingly multiple origins may be generated.

We analyze all determining attribute sets by considering each one of them as a candidate origin. %In what follows, 
Accordingly, multiple explanations may be generated for a %single 
goal. %In Section~\ref{sec:search}, we describe how we choose among them. In our implementation we rank the determinants by size and cardinality and, if desired, an early stop condition can be introduced based on the size or quality of the discovered transformation.  
Section~\ref{sec:search} describes how we choose among them. Specifically, we rank the determinants by size and cardinality and, if desired, an early stop condition can be introduced based on the size or quality of the discovered transformation. 

\begin{example}\label{example:origin}
  Recall %the tables from 
  Figure~\ref{fig:example} and consider attribute \texttt{a7} as a goal. Since the example tables %in the example 
  are small, any combination of attributes in \{\texttt{a1}, \texttt{a2}, \texttt{a3}, \texttt{a4}\} can be considered as an origin. If no high quality explanations are found for singleton attributes, the algorithm can consider combinations of attributes. In a larger real example, only a few attributes or combinations of them may be an origin.%In a larger real example, only a few of the attributes may be an origin or even just some combinations of attributes.
\end{example}

\section{Explaining Vertical Changes}\label{section:vertical}
% \section{Explaining Horizontal Dataset Changes}\label{section:vertical}
% \renee{try to get titles to one line when posible:  Explaining Horizontal Changes?}
\revision{%\rjm{
With our arsenal of explanation methods, we now consider how to use them to explain changes. We begin with attribute additions,} %}
%now look at attribute transformation and begin with the more common type of adding attributes, 
after which, we describe our approach to handling attribute removal.

%\subsection{Addition Explanations ($\overrightarrow{\mathcal{E}x}^{+}$ for RHDA)}
\subsection{Addition Explanations for \rhda}\label{sec:col_addition}
%\renee{are we using both RHDA and \rhda?}

\begin{figure}[h]
	\centering
% 	\vspace{-.05in}
	\includegraphics[width=.4\textwidth,trim=0 0 30 30]{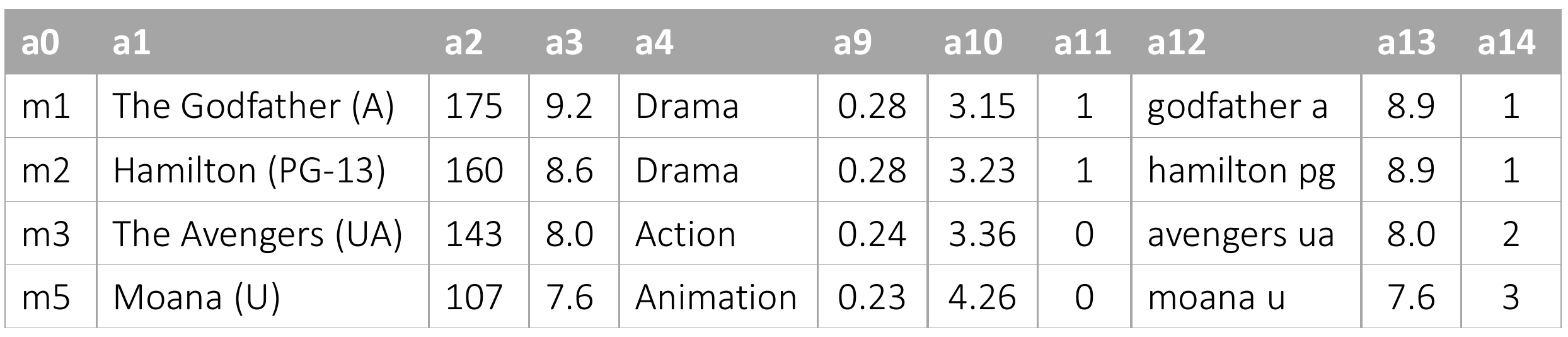}
	\vspace{-.025in}
	\caption{Dataset version created by \textsc{UserC} over Figure~\ref{fig:example_a}.}
	\label{fig:example_c}
    % \vspace{-.05in}
\end{figure}

Adding an attribute is a very common operation in data science, mainly revolving around data preparation and feature engineering for machine learning (ML)~\cite{zheng2018feature}. Added attributes are usually a result of applying some transformation over the existing data. %, which is a rich research direction in the last few years (see Section~\ref{sec:related3}). 
% Before diving into details on how to resolve and explain transformations, we are first interested in \emph{``finding the origin''}. In other words, different from some related literature where the input-output scope is clear (see Section~\ref{sec:related3}), in data versioning we do not know what was used to create a new attribute. %\rjm{
% Hence, our approach first searches for an origin for the transformation. %}.  
% After an origin is found, we turn our efforts to resolve the transformation that, when applied to the origin relation (input), generates the desired goal relation (output). To do so, we aim to understand how data types (both origin and goal) should be treated. Figure~\ref{fig:framework} provides a sketch of the approach (left and middle parts) and highlights its main novelties (right part). 
%rjm:  I think the framework requires a few more words..
\revision{We first find the origin (\cref{sec:origin}) an then utilize the core explanation methods (\cref{sec:methods}) to find transformations that, when applied to the origin relation, generate the desired goal relation.} 
\ifdefined\TechReport
Figure~\ref{fig:framework} provides a sketch of the approach (left and middle parts) and highlights its main novelties (right part) in the context of adding attributes. 
\else
\fi
% \vspace{.1cm}
% \noindent\textbf{Methodology:} when resolving added attributes, \texttt{Explain-Da-V} works iteratively. In each iteration, we look at an attribute $A_{i}\in \rhdaD$ as a goal ($\mathcal{G} = (A_{i}, \pi_{A_{i}}[T^{\prime}])$), find its origin ($\mathcal{O}$) as a projection over 
% %rjm $T^{\prime}$ 
% $T$ and use it to find a transformation $\mathcal{P}$ generating %an explanation $\overrightarrow{\mathcal{E}^{+}}_{A_{i}} = (\mathcal{O}, \mathcal{P})$. 
% %rjm if we get rid of the overrightarrow...
% a vertical explanation $\mathcal{E}_{A_{i}} = (\mathcal{O}, \mathcal{P})$. %${\mathcal{E}^{+}_{A_{i}} = (\mathcal{O}, \mathcal{P})$. 
\ifdefined\TechReport
\begin{figure}[h]
	\centering
	\includegraphics[width=.4\textwidth,trim=0 0 30 30]{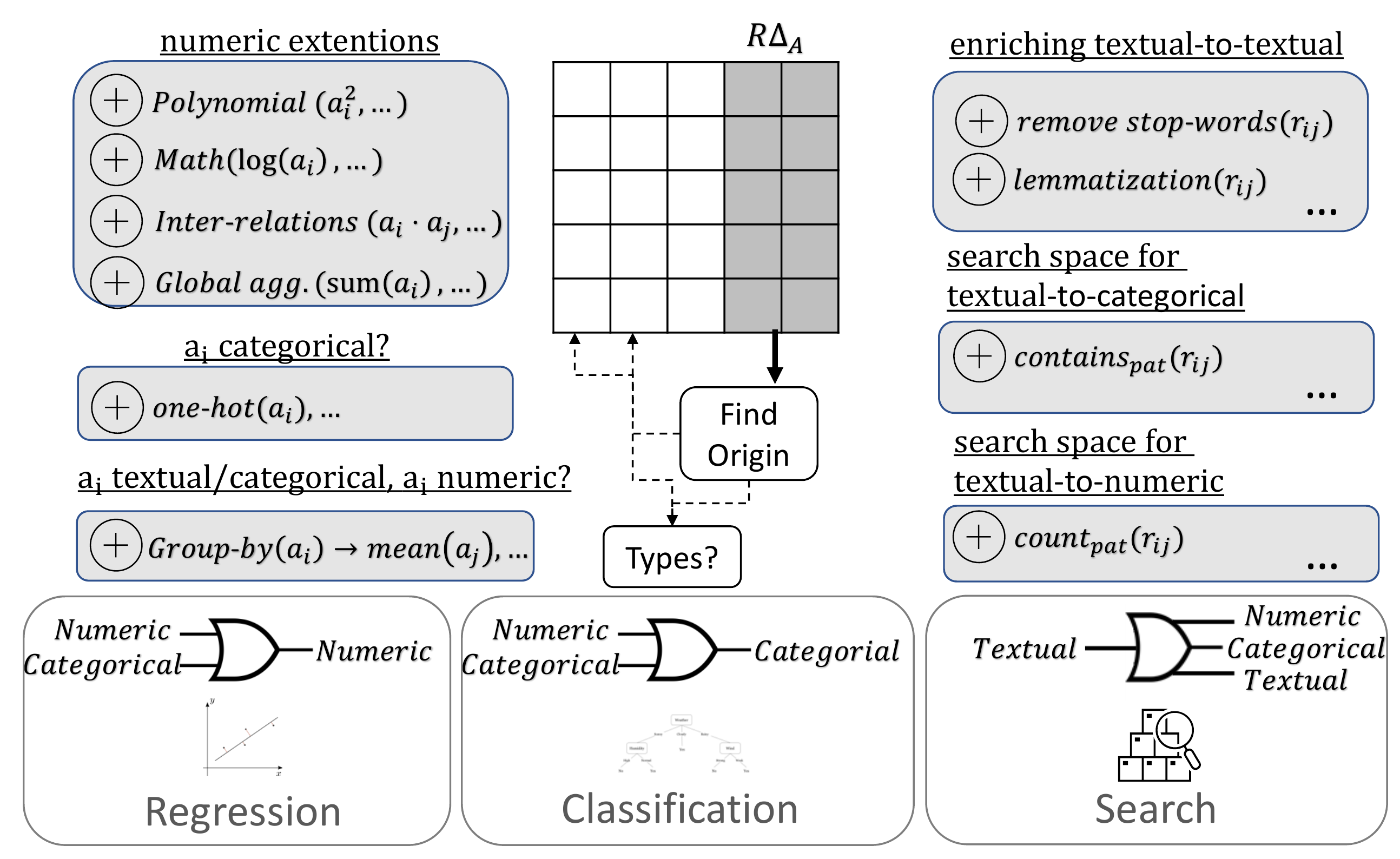}
% 	\caption{\texttt{Explain-Da-V} 
% 	%rjm I prefer one line captions when possibl
% 	%approach to explain 
% 	explanation of attribute addition. }
\caption{Explaining attribute additions}
	\label{fig:framework}
% \vspace{-.05in}
\end{figure}
\else
\fi
\ifdefined\TechReport
\revision{
\begin{algorithm}[h]
\caption{Explaining Attribute Additions}\label{alg:vertical}
\setstretch{0.9} % sets line height
\small
\SetAlgoLined
\LinesNumbered
\textbf{Input}: A set of attributes to-be-explained $\rhdaD$\\ \label{line:input}
\textbf{Output}: A set of explanations for each attribute $\mathcal{E}_{\rhdaD}$ \\ \label{line:output}
$\mathcal{E}_{\rhdaD}$ := $\emptyset$\; 
\For{$A_{i}\in \rhdaD$}{\label{line:forbegin}
    $\mathcal{E}_{A_{i}}$ := $\emptyset$; %\;
    $\mathcal{O}_{A_{i}}$ := \emph{find origin} (\cref{sec:origin})\label{line:origin}\;
    \For{$\mathcal{O}\in \mathcal{O}_{A_{i}}$}{\label{line:innerforbegin}
        \If {$A_{i}$ is numeric}{
            \If {$\mathcal{O}$ is numeric}{\label{line:originnum}
                $\mathcal{E}^{\mathcal{O}}_{A_{i}}\gets$ Numeric Explanations (\cref{sec:num_methods})\;\label{line:numtonum}
            }
            \Else{
               $\mathcal{E}^{\mathcal{O}}_{text}\gets$ Textual Explanations (\cref{sec:tex_methods})\;\label{line:text2num}
                $\mathcal{E}^{\mathcal{O}}_{encode}\gets$ Encoding Explanations (\cref{sec:encode_methods})\;\label{line:numencode}
                $\mathcal{E}^{\mathcal{O}}_{reshape}\gets$ Reshaping Explanations (\cref{sec:reshape_methods})\;\label{line:numreshape}
                $\mathcal{E}^{\mathcal{O}}_{A_{i}}\gets \mathcal{E}^{\mathcal{O}}_{encode}\cup\mathcal{E}^{\mathcal{O}}_{reshape}$\;
            }
        }
        \ElseIf {$A_{i}$ is categorical}{
            \If {$\mathcal{O}$ is numeric}{\label{line:originnum2}
                $\mathcal{E}^{\mathcal{O}}_{A_{i}}\gets$ Categorical Explanations (\cref{sec:cat_methods})\;\label{line:numtocat}
            }
            \Else{
               $\mathcal{E}^{\mathcal{O}}_{text}\gets$ Textual Explanations (\cref{sec:tex_methods})\;\label{line:text2cat}
                $\mathcal{E}^{\mathcal{O}}_{encode}\gets$ Encoding Explanations (\cref{sec:encode_methods})\;\label{line:encodecat}
                $\mathcal{E}^{\mathcal{O}}_{reshape}\gets$ Reshaping Explanations (\cref{sec:reshape_methods})\;\label{line:reshapecat}
                $\mathcal{E}^{\mathcal{O}}_{A_{i}}\gets \mathcal{E}^{\mathcal{O}}_{encode}\cup\mathcal{E}^{\mathcal{O}}_{reshape}$\;
            }
        }
        \ElseIf {$A_{i}$ is textual}{
            $\mathcal{E}^{\mathcal{O}}_{A_{i}}\gets$ Textual Explanations (\cref{sec:tex_methods})\;\label{line:text}
        }
    }\label{line:innerforend}
    $\mathcal{E}_{\rhdaD}\gets \mathcal{E}_{\rhdaD}\cup\mathcal{E}_{A_{i}}$\;
}\label{line:forend}
\textbf{Return}:$\mathcal{E}_{\rhdaD}$
\end{algorithm}
}

\revision{\texttt{Explain-Da-V} attribute addition explanation algorithm is provided in \cref{alg:vertical}. \texttt{Explain-Da-V} works iteratively, aiming to resolve each added attribute (i.e., the goal $\mathcal{G} = (A_{i}, \pi_{A_{i}}[T^{\prime}])$) at a time (Lines~\ref{line:forbegin}-\ref{line:forend}). After a set of possible origins $\mathcal{O}_{A_{i}}$ is found (Line~\ref{line:origin}), we utilize the core explanation methods to generate a set of explanations $\mathcal{E}^{\mathcal{O}}_{A_{i}}$ for each origin (Lines~\ref{line:innerforbegin}-\ref{line:innerforend}). For example, if a numeric origin is found for a numeric goal (Line~\ref{line:numtonum}), \texttt{Explain-Da-V} uses the numeric explanation method (\cref{sec:num_methods}). Note that multiple explanations are generated for each target, for which we introduce a search strategy in \cref{sec:search}. We illustrate it over Figure~\ref{fig:example_c}, which provides an additional version of the table in Figure~\ref{fig:example_a}.}

\else
\revision{\texttt{Explain-Da-V} attribute addition explanation is iterative, aiming to resolve each added attribute (i.e., the goal $\mathcal{G} = (A_{i}, \pi_{A_{i}}[T^{\prime}])$) at a time. First a set of possible origins for $A_{i}$ is found following \cref{sec:origin}. Then, \texttt{Explain-Da-V} uses the core explanation methods (\cref{sec:methods}) in a case-based manner (according to the types of the origin and the goal) to generate explanations. For example, if a numeric origin is found for a numeric goal, \texttt{Explain-Da-V} uses a numeric explanation method (\cref{sec:num_methods}). Multiple explanations may be generated for each goal (e.g., due to multiple origins), for which we introduce a search strategy in \cref{sec:search}. A full algorithm is given in a technical report~\cite{technical_report}. We illustrate it over Figure~\ref{fig:example_c}, which provides an additional version of the table in Figure~\ref{fig:example_a}. A more detailed example can be found in a technical report~\cite{technical_report}.}
\fi

% \rs{\textbf{UPDATE!}}
\ifdefined\TechReport
\begin{example}\label{example:full}
    \revision{Among the new added attributes \texttt{a9} and \texttt{a10} are numeric (Lines~\ref{line:originnum}). The attribute \texttt{a9} is a (sum) normalization of the values in \texttt{a3} (normalized rating). \texttt{Explain-Da-V} would first find its origins (Line~\ref{line:origin}). As in the case of \texttt{a6} (see Example~\ref{example:origin}), also \texttt{a9} can be determined by multiple attribute sets. Among the possible origins, consider $\mathcal{O}=$\texttt{a3}. Since both \texttt{a3} and \texttt{a9} are numeric, \texttt{Explain-Da-V} uses numeric change explanations (Line~\ref{line:numtonum}). Zooming in on the method (\cref{sec:num_methods}), a baseline explanation will be generated by fitting a regressor over $\mathcal{O}$. Then, different extensions will be applied over the origin, each will be associated with an explanation by fitting a regressor. Among the generated extensions we will find $agg(inter(\mathcal{O}))$, that contains the feature \texttt{a3}$\div$\texttt{sum(a3)} over which a transformation \texttt{a9}=\texttt{a3}$\div$\texttt{sum(a3)} will be fitted to generate an explanation $\mathcal{E}_{a9}$ $= (a3, a3 \div sum(a3))$ that will be added to $\mathcal{E}^{\mathcal{O}}_{a9}$. Similarly, \texttt{a10} is determined by multiple attribute sets. Among them, consider $\mathcal{O}=$\{\texttt{a2}, \texttt{a3}\}, over which the following explanation will be generated $\mathcal{E}_{a10}$ $= (\{a2,a3\}, 60\cdot a3 \div a2)$ (rating per hour) and added to $\mathcal{E}^{\mathcal{O}}_{a10}$.}
\end{example}
    \begin{wrapfigure}[5]{r}{0.15\textwidth}
    \raggedright
    \vspace{-.05in}
    \begin{tikzpicture}[every tree node/.style={draw},
    	level distance=0.6cm,sibling distance=0.5cm,
    	edge from parent path={(\tikzparentnode) -- (\tikzchildnode)}]
    	\Tree
    	[.a3>8
    	\edge node[auto=right] {$True$};
    	[.\textbf{a11=1}
    	]
    	\edge node[auto=left] {$False$};
    	[.\textbf{a11=0}
    	]
    	]
    \end{tikzpicture}
\end{wrapfigure}

\addtocounter{example}{-1}
\begin{example}[cont.]
    \revision{Next we look at \texttt{a11}, which is categorical, and for example, consider the origin $\mathcal{O}=$\texttt{a3}. Accordingly, \texttt{Explain-Da-V} uses categorical change explanations (Line~\ref{line:numtocat}) and might fit the following decision tree to explain \texttt{a11}.
    \noindent Note that \texttt{a11} might been created by applying \texttt{a3}>8.5 which differs from \texttt{Explain-Da-V}'s data-driven explanation. Our goal however is to provide an accurate explanation which both are. Interestingly, if we consider \texttt{a2} as an origin we can derive a similar decision tree rooted at $a2 > 150$. These issues refer to the \emph{generalizability} of explanations which will be discussed in Section~\ref{sec:expeval}.}
    
    \revision{The added attribute \texttt{a12} is textual and focuses on text cleaning. Consider an origin $\mathcal{O}=$\texttt{a1}, \texttt{Explain-Da-V} will execute Line~\ref{line:text} and specifically \cref{sec:text2text} which describes text-to-text transformations. The resolved transformation includes removing punctuation marks (e.g., `(') and numeric values (`13') and lowering the text. Note that such a transformation requires our extensions to Foofah and would not be accurately resolved using the original implementation of Foofah~\cite{foofah}. A textual-to-categorical (\cref{sec:text2cat}) example would be to use `contains\_multiple\_words?', in this case the tuples \texttt{m1} and \texttt{m3} would get the value 1 and \texttt{m2} and \texttt{m5} the value 0.}  
    
    \revision{Next, consider \texttt{a13}, which is numeric and among possible origins, consider $\mathcal{O}=$\{\texttt{a3,a4}\}, which is mixed. \texttt{Explain-Da-V} would turn to encoding (Line~\ref{line:encode}) and reshaping (Line~\ref{line:reshape}). Consider the latter and note that new generated features are illustrated in Figure~\ref{fig:groupby_ext}. \texttt{Explain-Da-V} will generate an explanation using the transformation $1\cdot(mean(a3)\text{ }by\text{ }a4)$ with represents a group by \texttt{a4} and computing the mean of \texttt{a3} (mean rating by genre).} 
    
    \revision{Finally, attribute \texttt{a14} is categorical and consider, \texttt{a4} as an origin. In addition to applying trying to find textual explanations (Line~\ref{line:text2cat}), \texttt{Explain-Da-V} would also turn to encoding (Line~\ref{line:encodecat}) and reshaping (Line~\ref{line:reshapecat}) explanations. Consider the former and note that \texttt{a14} is an ordinal encoding of attribute \texttt{a4} (Drama$\rightarrow$1, Action$\rightarrow$2, Animation$\rightarrow$3). The three encoded attributes, namely \texttt{is\_Drama}, \texttt{is\_Action?} and \texttt{is\_Animation?} are used to resolve \texttt{a14}. Note that in a real-world scenario, an attribute like \texttt{a14} would not necessarily be recognized as a categorical (e.g., high cardinality or misclassification as a numeric value). In this case \texttt{Explain-Da-V} would turn to Line~\ref{line:numencode} resulting in the transformation $1\cdot$\texttt{is\_Drama?} + $2\cdot$\texttt{is\_Action?} + $3\cdot$\texttt{is\_Animation?}.}
\end{example}
\else
\begin{example}\label{example:full}
    \revision{Among the new added attributes \texttt{a9} and \texttt{a10} are numeric. The attribute \texttt{a9} is a (sum) normalization of the values in \texttt{a3} (normalized rating). \texttt{Explain-Da-V} would first find its origins (\cref{sec:origin}). As in the case of \texttt{a6} (see Example~\ref{example:origin}), also \texttt{a9} can be determined by multiple attribute sets. Among the possible origins, consider \texttt{a3}. Since both \texttt{a3} and \texttt{a9} are numeric, \texttt{Explain-Da-V} uses numeric change explanations (\cref{sec:num_methods}). A baseline explanation will be generated by fitting a regressor over \texttt{a3}. Then, different extensions will be applied over the origin, each will be associated with an explanation by fitting a regressor. Among the generated extensions we will find $agg(inter(\mathcal{O}))$, that contains the feature \texttt{a3}$\div$\texttt{sum(a3)} over which a transformation \texttt{a9}=\texttt{a3}$\div$\texttt{sum(a3)} will be fitted to generate an explanation $\mathcal{E}_{a9}$ $= (a3, a3 \div sum(a3))$. Similarly, consider \{\texttt{a2}, \texttt{a3}\} as an origin for \texttt{a10}, the following explanation will be generated $\mathcal{E}_{a10}$ $= (\{a2,a3\}, 60\cdot a3 \div a2)$ (rating per hour).}
    %Similarly, \texttt{a10} is determined by multiple attribute sets (origins). Among them, consider \{\texttt{a2}, \texttt{a3}\}, over which the following explanation will be generated $\mathcal{E}_{a10}$ $= (\{a2,a3\}, 60\cdot a3 \div a2)$ (rating per hour).}
\end{example}
    \begin{wrapfigure}[5]{r}{0.15\textwidth}
    \raggedright
    % \vspace{-.1in}
    \begin{tikzpicture}[every tree node/.style={draw},
    	level distance=0.6cm,sibling distance=0.5cm,
    	edge from parent path={(\tikzparentnode) -- (\tikzchildnode)}]
    	\Tree
    	[.a3>8
    	\edge node[auto=right] {$True$};
    	[.\textbf{a11=1}
    	]
    	\edge node[auto=left] {$False$};
    	[.\textbf{a11=0}
    	]
    	]
    \end{tikzpicture}
    \vspace{-.15in}
\end{wrapfigure}

\addtocounter{example}{-1}
\begin{example}[cont.]
    \revision{Attribute \texttt{a11} is categorical and consider, for example, the numeric origin \texttt{a3}. Accordingly, \texttt{Explain-Da-V} uses categorical change explanations (\cref{sec:cat_methods}) and might fit the following decision tree to explain \texttt{a11}.
    \noindent Note that \texttt{a11} might been created by applying \texttt{a3}>8.5 which differs from \texttt{Explain-Da-V}'s data-driven explanation. Our goal however is to provide an accurate explanation which both are. Interestingly, if we consider \texttt{a2} as an origin we can derive a similar decision tree rooted at $a2 > 150$. These issues refer to the explanations generalizability, which is discussed in Section~\ref{sec:expeval}.}
    
    \revision{Attribute \texttt{a12} is textual and focuses on text cleaning. \texttt{Explain-Da-V} executes \cref{sec:text2text} and using, for example, \texttt{a1} as an origin, the resolved transformation removes punctuation marks (e.g., `(') and numeric values (`13') and lowers the text. Note that such a transformation requires our extensions to Foofah and would not be accurately resolved using the original Foofah~\cite{foofah}.}  
    
    % \revision{Attribute \texttt{a12} is textual and focuses on text cleaning. Consider an origin \texttt{a1} (textual). \texttt{Explain-Da-V} executes \cref{sec:text2text} and the resolved transformation includes removing punctuation marks (e.g., `(') and numeric values (`13') and lowering the text. Note that such a transformation requires our extensions to Foofah and would not be accurately resolved using the original implementation of Foofah~\cite{foofah}. A textual-to-categorical (\cref{sec:text2cat}) example would be to use `contains\_multiple\_words?', in this case the tuples \texttt{m1} and \texttt{m3} would get the value 1 and \texttt{m2} and \texttt{m5} the value 0.}  
    
    \revision{Attribute \texttt{a13} is numeric and among possible origins, consider \{\texttt{a3,a4}\}, which is mixed. \texttt{Explain-Da-V} would turn to encoding (\cref{sec:encode_methods}) and reshaping (Line~\ref{sec:reshape_methods}). Consider the latter, %and note that new generated features are illustrated in Figure~\ref{fig:groupby_ext}. 
    \texttt{Explain-Da-V} will generate an explanation using the transformation $1\cdot(mean(a3)\text{ }by\text{ }a4)$ which represents a group by \texttt{a4} and computing the mean of \texttt{a3} (mean rating by genre).} 
    
    \revision{Finally, attribute \texttt{a14} is categorical and consider, \texttt{a4} as an origin. In addition to applying 
    %rjm trying to find 
    textual explanations (\cref{sec:text2cat}), \texttt{Explain-Da-V} would also turn to encoding (\cref{sec:encode_methods}) and reshaping (\cref{sec:reshape_methods}) explanations. Consider the former and note that \texttt{a14} is an ordinal encoding of attribute \texttt{a4} (Drama$\rightarrow$1, Action$\rightarrow$2, Animation$\rightarrow$3). The three encoded attributes, namely \texttt{is\_Drama}, \texttt{is\_Action?} and \texttt{is\_Animation?} are used to resolve \texttt{a14}. Note that in a real-world scenario, an attribute like \texttt{a14} would not necessarily be recognized as a categorical (e.g., if it has high cardinality it could be  misclassified as a numeric value). In this case \texttt{Explain-Da-V} would turn to \cref{sec:num_methods} resulting in the transformation $1\cdot$\texttt{is\_Drama?} + $2\cdot$\texttt{is\_Action?} + $3\cdot$\texttt{is\_Animation?}.}
\end{example}
\fi

\subsection{Removal Explanations for \lhda}\label{sec:col_removal}
Removing attributes is less common and usually include superficial transformations. We treat each attribute in \lhda separately as a goal. We cover two main types of explanations for removal reflecting data cleaning (removing duplicated and noisy attributes). 

First, we examine a \emph{table-independent} attribute removal, which in our terms reflects an empty origin ($\mathcal{O} = \emptyset$). Specifically, we use a threshold to decide whether a attribute was removed because it has too many (above a threshold) missing (NaN) values.\footnote{The threshold can be treated as an hyper-parameter or a user-provided input.} In this case an explanation for a removed attribute $A_{i}\in \lhdaD$ will be in the form of %rjm $\mathcal{E}x_{A_{i}} = (\emptyset,
$\mathcal{E}_{A_{i}} = (\emptyset,
\text{`contains missing information'})$. Formally, the `contains missing information' can be defined as $A_{i} = \begin{cases}
\emptyset, & \text{if ratio of NaN values} > \alpha\\
A_{i}, & \text{otherwise}
\end{cases}$,\\ 
where $\alpha\in [0,1]$ is some threshold.

As a second case, we look into duplicated information. A trivial explanation can be provided for an identical attribute in $T^{\prime}$. Given a goal $A_{i}$, the origin is some attribute $A^{\prime}_{j}\in T^{\prime}$ such that $A^{\prime}_{j}\notin T$ %\renee{this sentence doesn't make sense, A'j can't be both in T' and not in T'} \rs{typo- fixed} 
and $\pi_{A_{i}}[T]=\pi_{A^{\prime}_{j}}[T^{\prime}]$ (full overlap of values). A natural extension of finding duplications is looking into similarities between attributes. %In this work, w
We look into two types of similarities, measuring the overlap between attributes and if there is a one-to-one dependency between them. Overlap is measured and, if it meets some threshold, an explanation is generated using the overlapping attribute $A^{\prime}_{j}$ as the origin and an `overlaps with $A^{\prime}_{j}$' transformation, which is defined similar to above. We also check if some attribute in $T^{\prime}$ determines (using a similar methodology as described in Section~\ref{sec:origin}) $A_{i}\in \lhdaD$. Obviously many other measures of similarity exist, which we intend to explore in future work. Finally, note that sometimes attribute removal can be idiopathic, i.e., the user simply removed an attribute because they are not interested in some parts of the data.  
% \renee{do you want to say that sometime attribute removal can be idiopathic (unexplainable)?  I may just not be interested in some parts of the data?} \rs{ADD!}

%\subsection{Vertical Explanations}
% \section{Explaining Vertical Dataset Changes}\label{section:horizontal}
\section{Explaining Horizontal Changes}\label{section:horizontal}
% We now look at horizontal explanations, beginning with the common data cleaning operation of tuple removal, after which we discuss adding tuples. 
We begin with the common data cleaning operation of tuple removal, after which we discuss adding tuples. 

\subsection{Removal Explanations for \lhdr}\label{sec:row_removal}
Tuple removal is a very common operation in data preparation, which mainly revolves around cleaning data. Our examination begins iteratively by looking into each removed tuple in \lhdr\ independently. This may, for example, result in the horizontal explanation $\mathcal{E}_{m4}$ from \cref{exp:explanations}. %For all remaining unexplained 
%\renee{you mean unmatched?} \rs{No. "Unexplained" meaning an independent data cleaning explanation was not found.} 
% For all remaining unexplained tuples, we aim to resolve if some \emph{predicate} was applied to remove them conjointly. 
Finally, we explore if a \emph{predicate} was applied to remove them all remaining (unexplained) tuples conjointly. 

%Similar to 
As in Section~\ref{sec:col_removal}, we aim to find tuples that were removed collectively in a table-independent manner due to missing values (NaNs), see the \texttt{m4} explanation in Example~\ref{exp:explanations}. \revision{For table-dependent explanations, we find duplicated tuples, which is a common result of data cleaning using entity resolution
%rjm, which is a rich and active research
~\cite{elmagarmid2006duplicate,li2020deep}. We focus only on identical tuple removal. Note that this strict requirement can be relaxed and any entity resolution technique, e.g., using declarative rules~\cite{BFKPT16}, can be used to find duplicated-tuple removal explanations.} Specifically, if a duplicated tuple $r^{\prime}_{j}\in T^{\prime}$ is found for a goal tuple $r_{i}\in$ \lhdr, we create a horizontal explanation of the form
$\mathcal{E}_{r_{i}} = (\emptyset, \text{duplicated of } r^{\prime}_{j})$. 
This transformation can be expressed as follows $r_{i} = \begin{cases}
\emptyset, & \exists r^{\prime}_{j}\in T^{\prime} s.t. r_{i} =  r^{\prime}_{j}\\
r_{i}, & \text{otherwise}
\end{cases}$\\ % and there's a rich and active research on how to find duplicated tuples~\cite{elmagarmid2006duplicate,li2020deep} including using declarative rules~\cite{BFKPT16}, see Section~\ref{sec:related2}. 
% For table-dependent explanations, we aim to find duplicated tuples. Specifically, if a duplicated tuple $r^{\prime}_{j}\in T^{\prime}$ is found for a goal tuple $r_{i}\in$ \lhdr, we create a horizontal explanation of the form %$\downarrow\mathcal{E}^{-}_{r_{i}} = (\emptyset, \text{duplicated of } r^{\prime}_{j})$. 
% $\mathcal{E}_{r_{i}} = (\emptyset, \text{duplicated of } r^{\prime}_{j})$. 
% %rjm what is a "formal transformation" ? %Similar to above a formal transformation 
% This transformation can be expressed as follows $r_{i} = \begin{cases}
% \emptyset, & \exists r^{\prime}_{j}\in T^{\prime} s.t. r_{i} =  r^{\prime}_{j}\\
% r_{i}, & \text{otherwise}
% \end{cases}$\\ Clearly, the strict requirement of equal tuples can be relaxed and there's a rich and active research on how to find duplicated tuples~\cite{elmagarmid2006duplicate,li2020deep} including using declarative rules~\cite{BFKPT16}, see Section~\ref{sec:related2}. %In this work, w
% We focus only on identical tuple removal and note that any entity matching technique can be used to find %and explain the removal of 
% duplicated tuples removal explanations. %Finally, we also 
%rjm reverse engineer 
% \rjm{use}
% use outlier detection to serve as explanations for removed tuples. %\rs{remember to talk about the conflict in the experiments}
% Specifically, we use the Z-method and IQR-method~\cite{chandola2009anomaly}. 
Finally, outlier detectors (Z-method and IQR-method~\cite{chandola2009anomaly,ting2018outlier}) also serve as explanations for removed tuples.

Not all tuples can be explained independently, thus, for all unexplained 
%\renee{unmatched? for consistency} \rs{see above. i do mean unexplained and not unmatched. Unmatched refers to all the tuples that needs to be explained and unexplained refers to the ones for which an independent explanation was not found.} 
tuples, \lhdr$_{unexplained}$, we aim to find a joint explanation in the form of a predicate. Given a set of unexplained tuples, \revision{we use a categorical explanation method (\cref{sec:cat_methods}) to find a joint explanation. Similar to \cref{sec:col_addition}, in case the origin has mixed types, a decision tree is applied also over encoded (using categorical-encoding change explanation, \cref{sec:encode_methods}) attributes in the table.}
%we use a decision tree to find a joint explanation (similar to Section~\ref{sec:numtocat}). The decision tree is applied over the numeric and categorical (using one-hot-encoding) attributes in the table. 

%\renee{Why is one-hot-encoding important here?}

% \subsubsection{Predicate Resolution} 

% \begin{figure}[h]
% 	\centering
% 	\includegraphics[width=.35\textwidth]{./figures/example_table_d}
% % 	\caption{Dataset version created by \textsc{UserD} as a Derivative of the Dataset in Figure~\ref{fig:example_a}.}
% 	\caption{Dataset version created by \textsc{UserD} over Figure~\ref{fig:example_a}.}
% 	\label{fig:example_d}
% \end{figure}

\begin{figure}[h]
% \vspace{-.1in}
	\begin{subfigure}[b]{0.2\textwidth}
	\centering
	\includegraphics[width=\textwidth,trim=0 0 30 30]{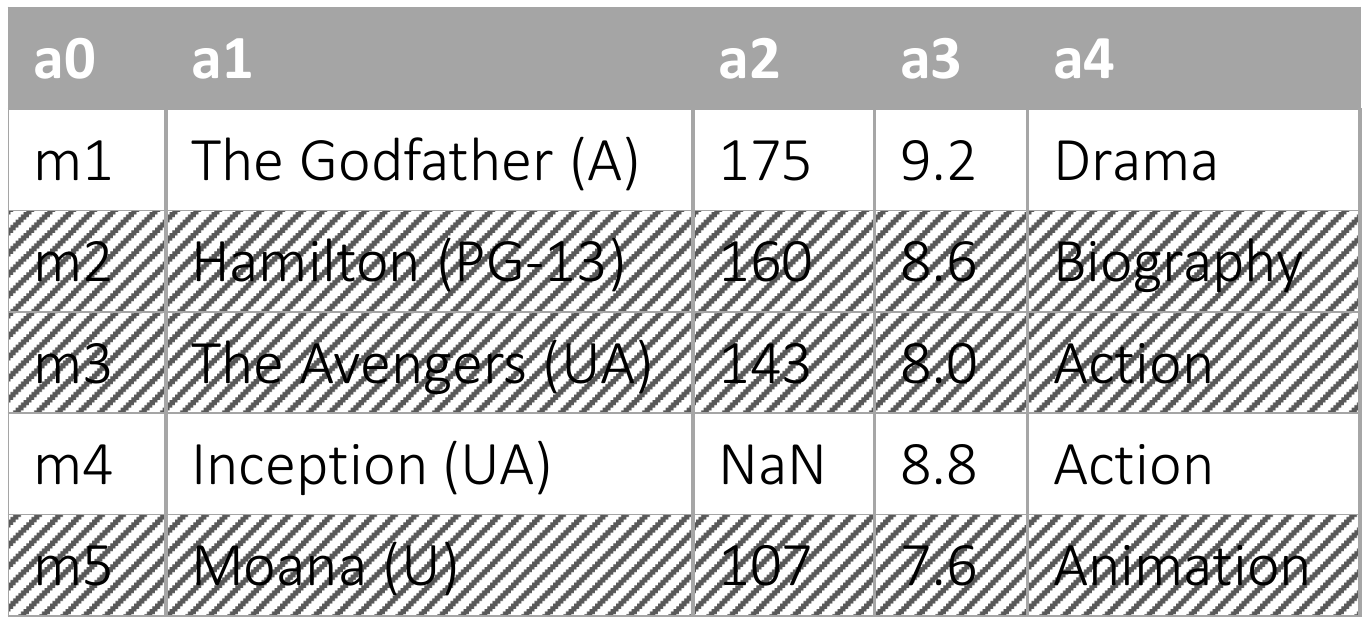}
	\caption{Dataset version created by \textsc{UserD} over Figure~\ref{fig:example_a}.}
	\label{fig:example_d}
	\end{subfigure}
	\hfill
    \begin{subfigure}[b]{0.2\textwidth}
    \centering
    \resizebox{.95\textwidth}{0.6\textwidth}{\begin{tikzpicture}[every tree node/.style={draw},
	level distance=1cm,sibling distance=0.1cm,
	edge from parent path={(\tikzparentnode) -- (\tikzchildnode)}]
	\Tree
	[.a3>8.5
	\edge node[auto=right] {$True$};
	[.\texttt{is\_Drama?}
	\edge node[midway,left] {$True$};
	[.\textbf{maintain} ]
	\edge node[midway,right] {$False$};
	[.\texttt{is\_Action?}
	\edge node[midway,left] {$True$};
	[.\textbf{maintain} ]
	\edge node[midway,right] {$False$};
	[.\textbf{remove} ]
	]
	]
	\edge node[auto=left] {$False$};
	[.\textbf{remove}
	]
	]
	\end{tikzpicture}}
	\caption{Tree}
	\label{fig:tree}
     \end{subfigure}
     \vspace{-.025in}
     \caption{\textsc{UserD} version of Figure~\ref{fig:example_a} and its explanation.}
% \vspace{-.1in}
\end{figure}
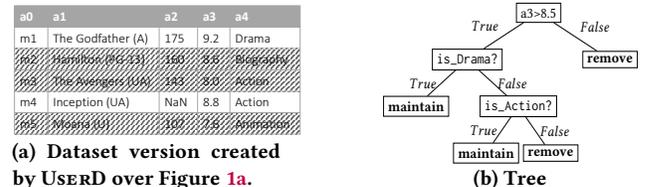

\begin{example}\label{example:row_remove}
	In the example of Figure~\ref{fig:example_b} we present a simple example of tuple removal due to NaN value. Figure~\ref{fig:example_d} provides an example of applying a predicate over the table. To resolve this predicate, \texttt{Explain-Da-V} will first add the one-hot-encoded features corresponding to \texttt{a4} (\texttt{is\_Drama?}, \texttt{is\_Action?} and \texttt{is\_Animation?}), then, using a decision tree, it will try to resolve the predicate. The decision tree in Figure~\ref{fig:tree} will be generated.
% 	\begin{center}
% 	 \begin{tikzpicture}[every tree node/.style={draw},
% 	level distance=1cm,sibling distance=0.1cm,
% 	edge from parent path={(\tikzparentnode) -- (\tikzchildnode)}]
% 	\Tree
% 	[.a4>8.5
% 	\edge node[auto=right] {$True$};
% 	[.\texttt{is\_Drama?}
% 	\edge node[midway,left] {$True$};
% 	[.\textbf{maintain} ]
% 	\edge node[midway,right] {$False$};
% 	[.\texttt{is\_Action?}
% 	\edge node[midway,left] {$True$};
% 	[.\textbf{maintain} ]
% 	\edge node[midway,right] {$False$};
% 	[.\textbf{remove} ]
% 	]
% 	]
% 	\edge node[auto=left] {$False$};
% 	[.\textbf{remove}
% 	]
% 	]
% 	\end{tikzpicture}
% 	\end{center}

\end{example}

% Not all predicates can be resolved using a decision tree. Specifically, decision trees represent disjunctions of conjunctions~\cite{mitchell1997decision}, where each path from the root to a leaf corresponds to a conjunction and the tree itself is the disjunction of these conjunctions.
%rjm shortening - I think this is redundant
%\revision{As mentioned in \cref{sec:cat_methods}, only predicates that can be represented as disjunctions of conjunctions can be resolved.}
% \textbf{what predicates cant be resolved??}

\subsection{Addition Explanations for \rhdr}
% \rjm{
The non-idiopathic addition of
% } 
%Adding
tuples
may be a result of over-sampling (bootstrapping). To detect such a transformation, we use a similar methodology as in Section~\ref{sec:row_removal}. Given an added tuple $r^{\prime}_{i}\in$ \rhdr, we aim to find a duplicated (equal or similar) tuple $r_{j}\in T$ %as an origin 
to create an explanation noting that the tuple has been bootstrapped.

\section{Evaluating Explanations}\label{sec:expeval}
%rjm shortening In this paper 
We aim to generate user friendly explanations that capture the semantics of changes. Specifically,
%rjm if an explanation captures the semantics of change, it means it
the explanation (transformation) can reproduce the change and generalize it beyond a specific setup. Aiming to assess such semantics, we now describe how we evaluate explanations.
% Specifically, our goal is to \emph{automatically} explain (in a simple user friendly way) the steps leading from one version of dataset to the other. For example, \textit{how was column \texttt{a6} in Figure~\ref{fig:example_b} created?} or \textit{why was the fourth row in Figure~\ref{fig:example_a} deleted?} Note that the changer’s intent, which is subjective, cannot be truly reverse engineered. Our objective is to provide the other user an accurate explanation, e.g., a set of functions, that describes the changes. Following this goal, we return to our motivating example.
%Sections~\ref{section:vertical} and~\ref{section:horizontal} describe the way \texttt{Explain-Da-V} resolves and explains the changes between two versions of a dataset. Note that s
Sometimes multiple explanations can be generated with respect to a change. Recall Example~\ref{example:full} in which we present two possible valid explanations for \texttt{a11}. The first decision tree explanation is rooted at $a3>8$ and the second is rooted at $a2>150$. Also a decision tree rooted at $a3>7.5$ is a possible (invalid) explanation. In what follows, an important question that needs to be asked is \emph{how to compare (and choose among) possible explanations?} 

Related work on data transformation (see Section~\ref{sec:related3}) %usually evaluates success rates, that is, whether the output was generated successfully by applying the transformation over the input. 
employ success rates that measure whether the output was generated successfully by applying the transformation over the input. 
Yang et al. also introduce a ranking measure (MRR) over possible transformations (pipelines in their terms), which still views the transformation as a whole~\cite{yang2021auto}. We claim that solely using such a 
%conclusive 
measure 
does not capture the true nature of the transformation, especially when evaluating an attribute-to-attribute (Section~\ref{sec:col_addition}) transformations. To provide a more fine-grained evaluation, we evaluate both the \emph{validity} and \emph{generalizablity} computed over the transformed values to 
%rjm evaluate -- evaluate used too many times
% \rjm{assess}
assess the coverage of the 
%rjm provided 
transformation. As we are interested in providing explainable solutions, we also use two \emph{explainability} dimensions,  \emph{conciseness} and \emph{concentration}.

%three dimensions of the explainability of the explanation, namely, \emph{breadth}, \emph{conciseness}, and \emph{concentration}.

\subsection{Explanation Validity and Generalizablity}\label{sec:valngen}
% As mentioned in Section~\ref{sec:col_addition},
Attribute (vertical) addition explanations are richer
%\rjm{than attribute removal or tuple transformations}
than %attribute 
removal or tuple transformations, so their evaluation is addressed accordingly. 

\noindent\textbf{Vertical Additions:} 
We separate this evaluation into validity (\emph{does the generated transformation recreate the goal using the origin?}) and generalizability (\emph{will the generated transformation be able to recreate a similar goal using a similar origin?}). Recall Definition~\ref{def:explain} and the notation of origin ($\mathcal{O}$), goal ($\mathcal{G}$), and transformation ($\mathcal{P}$). For simplicity, we denote the output of a transformation applied to an origin relation as $\mathcal{\hat{G}} = \mathcal{P}(\mathcal{O}_{relation})$. Validity is computed in a tuple-based manner over value-pairs $(\hat{r}_{ij}, r_{ij})$ such that $\hat{r}_{ij}\in\mathcal{\hat{G}}$ is a transformed value corresponding to a goal value $r_{ij}\in\mathcal{G}_{relation}$, i.e., $\hat{r}_{i0}=r_{i0}$ 
 ($r_{i0}$ is the tuple id so this means the tuples are matching, see Section~\ref{sec:prelim}).
Explanation validity is measured as follows:
%\vspace{-.025in}
\begin{equation}\label{eq:val}\small
	Val(\mathcal{E}_{\mathcal{G}}) = \frac{1}{|\mathcal{G}_{relation}|} \sum\limits_{\substack{\hat{r}_{ij}\in\mathcal{\hat{G}}, r_{ij} \in \mathcal{G}_{relation}: \\ s.t. \hat{r}_{i0}=r_{i0}}} \mathbb{I}(\hat{r}_{ij}=r_{ij})%\vspace{-.025in}
\end{equation}\noindent where $\mathbb{I}(\hat{r}_{ij}=r_{ij})$ is an indicator returning the value 1 if the transformed value equals to the corresponding goal value and 0 otherwise. %\renee{Is it 0 otherwise?  Or is some notion of similarity permitted?}. 
The validity can be viewed as a tuple-based success rate, i.e., the proportion of the tuples that were successfully transformed.% out of all tuples.

\begin{example}
\revision{Recall the vertical explanation ${\mathcal{E}}_{a9}$ $= (a3, a3 \div sum(a3))$ which was created for the attribute \texttt{a9} in Figure~\ref{fig:example_c} (see Example~\ref{example:num2num0}). Also consider an alternative vertical explanation ${\mathcal{E}^{\prime}}_{a9}$ $= (a3, a3 \div 33.4)$. 
Both explanations would have a validity score of $1$ as applying the corresponding transformation recreates \texttt{a9} perfectly.}
\end{example}

%The validity measure only looks at the given dataset versions $T$ and $T^{\prime}$. 
\revision{As illustrated in the example, validity only looks at the given dataset versions $T$ and $T^{\prime}$, which may result in overfitting (e.g., selecting $a3 \div 33.4$ over $a3 \div sum(a3)$). Aiming to measure such scenarios, we introduce generalizability, measuring the extent to which a solution can explain an equivalent set of versions. Specifically, generalizability can be measured if a pair of versions $\tilde{T}$ and $\tilde{T}^{\prime}$ exist such that $\tilde{T}^{\prime}$ was generated as a version of $\tilde{T}$ using the same transformations that were used to generate $T^{\prime}$ from $T$.} 
% We are also interested 
% %to measure 
% in measuring 
% the generalizability of the solution, that is, can the generated explanation be used to also explain 
% %rjm an equivalent 
% a different
% pair of versions $\tilde{T}$ and $\tilde{T}^{\prime}$ (if available) such that $\tilde{T}^{\prime}$ was generated as a version of $\tilde{T}$ using the same transformations that were used to generate the version $T^{\prime}$ from $T$. %? %The generalizability of an explanation $Gen(\mathcal{E}x_{\mathcal{G}}, \tilde{\mathcal{R}})$ is measured using an output set $\tilde{\mathcal{R}}$ which is created from $\tilde{T}$ and $\tilde{T}^{\prime}$ similar to above, which \texttt{Explain-Da-V} is not aware of. We illustrate the difference and importance of the two measures using the following example. 
Let $\mathcal{\tilde{O}}$ be the origin over $\tilde{T}$ and $\mathcal{\tilde{G}}$ the goal over $\tilde{T}^{\prime}$. The generalizability of an explanation $Gen(\mathcal{E}_{\mathcal{G}})$ is measured by applying $\mathcal{P}$ over $\mathcal{\tilde{O}}$ to generate $\mathcal{\hat{\tilde{G}}}$ and is computed as in Eq.~\ref{eq:val} over $\mathcal{\hat{\tilde{G}}}$ and $\mathcal{\tilde{G}}$. We illustrate the importance of generalizability using the following example. 

%Note that the values of $\mathcal{\tilde{G}}$ are unknown and a potential user would have to annotate if the values are transformed correctly.  %\renee{I don't understand this, how does the T/T' problem differ from the $\tilde{T}/\tilde{T'}$ problem? if manual annotation is not required in the former, why would you need it in the latter?  More explanation is needed here.  Below in example you don't talk about human annotation.} 
%rjm revision newly suggested 
%\rjm{
%We illustrate the difference and importance of the two measures using the following example.
% \ifdefined\TechReport
\begin{figure}[h]
% \vspace{-.05in}
	\centering
	\begin{subfigure}[b]{0.25\textwidth}
		\includegraphics[width=\textwidth,trim=0 0 30 30]{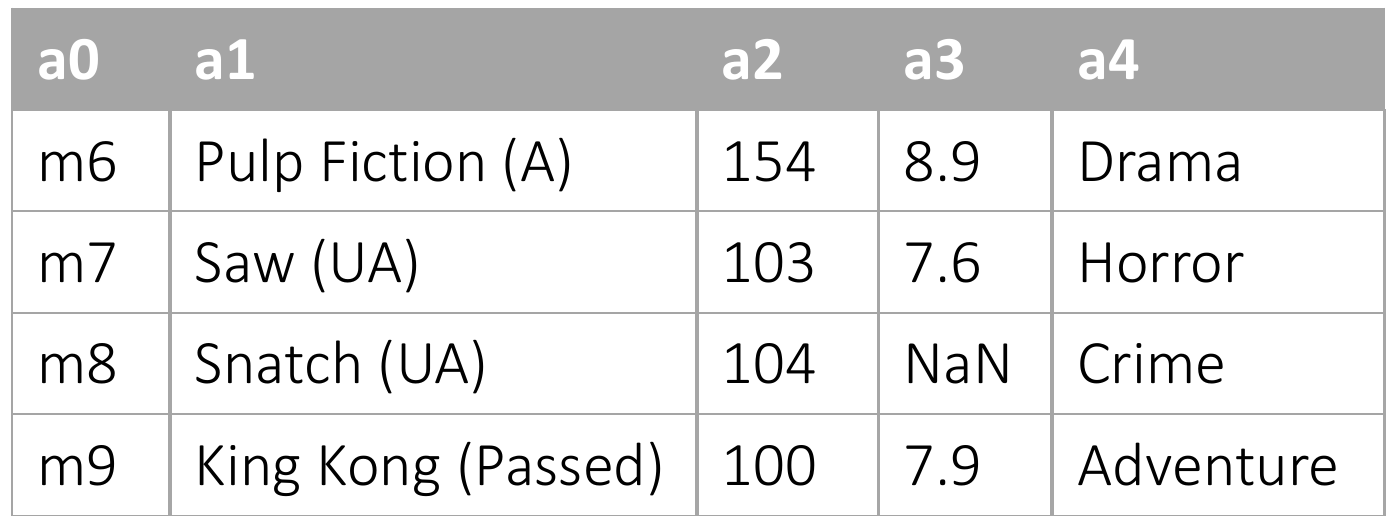}
% 	 	\caption{Dataset version created by \textsc{UserA}}
% 		\label{fig:example_a}
	\end{subfigure}
	\hfill
	\begin{subfigure}[b]{0.45\textwidth}
		\includegraphics[width=\textwidth,trim=0 0 30 30]{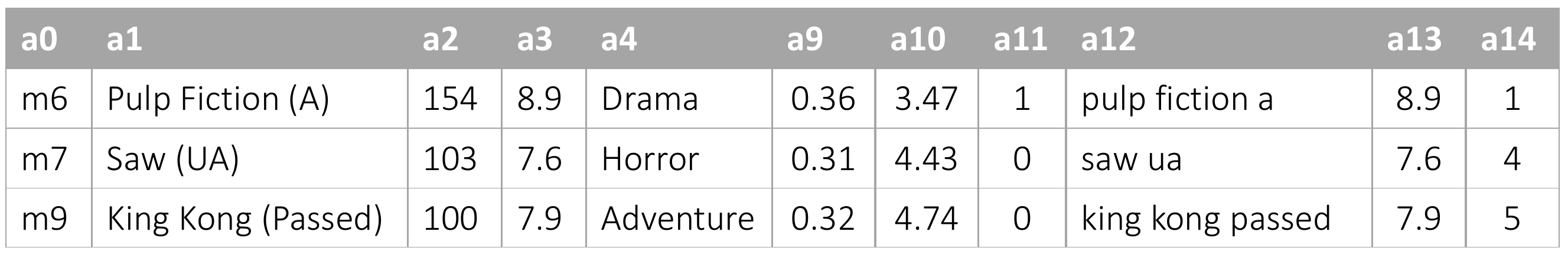}
% 		\caption{Dataset version created by \textsc{UserB}}
% 		\label{fig:example_b}
	\end{subfigure}
	\caption{\revision{Example versions for generalizability}}
	\label{fig:example_gen}
% 	\vspace{-.05in}
\end{figure}

\begin{example}\label{example:gen}

\revision{\cref{fig:example_gen} provides two dataset versions. The top table is similar to \cref{fig:example_a} and the bottom table corresponds to \cref{fig:example_c} such that the same transformations over \cref{fig:example_a} generates \cref{fig:example_c}.
%rjmwere applied here as well. 
Recall the explanations ${\mathcal{E}}_{a9}$ $= (a3, a3 \div sum(a3))$ and ${\mathcal{E}^{\prime}}_{a9}$ $= (a3, a3 \div 33.4)$. While these two are valid, using \cref{fig:example_gen}, we observe that ${\mathcal{E}}_{a9}$ is also generalizable while ${\mathcal{E}^{\prime}}_{a9}$ is not. Specifically, if we apply ${\mathcal{E}}_{a9}$  over \texttt{a3} in \cref{fig:example_gen} we obtain the values of \texttt{a9} in the bottom table. However, if we apply ${\mathcal{E}^{\prime}}_{a9}$, we obtain the values $0.27$, $0.23$, and $0.24$ for the records \texttt{m6}, \texttt{m7}, and \texttt{m9}, respectively, resulting in a $0$ generalizability.}
\end{example}
\noindent \shepherd{In practical settings, generalizability can be computed when the same changes are applied to multiple datasets, e.g., in a data pipeline such as ETL~\cite{vassiliadis2009survey}.} For our new benchmark (Section~\ref{sec:benchmark}), we generate an annotated \emph{hold-out} set, used to compute generalizability.

% When designing our new benchmark (Section~\ref{sec:benchmark}), we generate an annotated \emph{hold-out} set, which is not available for \texttt{Explain-Da-V} and used it to compute generalizability.

\noindent\textbf{Other Explanations:} For tuple removal, we apply a reconstruction methodology to evaluate validation and generalizability globally. We gather all %rjm what does unique mean here?
%unique 
generated explanations and apply them over $T$ and try to regenerate $T^{\prime}$. Then, we check the overlap between the removed tuples and the tuples that are not included in $T^{\prime}$. For example, if a \texttt{a3}>8 predicate was used to explain the removed tuples, the same predicate would be applied over $T$ and compared to $T^{\prime}$. This overlap, i.e., the proportion of tuples that were correctly removed using the explanations of \texttt{Explain-Da-V}, is used as the overall validity of tuple removal. The generalizability is measured similar to above using an additional 
%rjm ? (analogical) 
dataset version pair $\tilde{T}$ and $\tilde{T}^{\prime}$.

We also compute validity and generalizability for other explanations. Validation and generalizability of a removed attribute or added tuple are computed independently, i.e., a score of 1 is given if an attribute was removed correctly or a tuple was added correctly.
%For a removed attribute or added tuple, validation and generalizability are computed independently, i.e., a score of 1 is given if an attribute was removed correctly or a tuple was added correctly.
% \vspace{-.1in}
\revision{\subsection{Problem Definition}\label{sec:problem}}
\noindent\revision{We now formally state the problem of explaining data versions. Recall the change sets defined over the $T$ and $T^{\prime}$ (see \cref{sec:changesets}).}

% \revision{\begin{definition}\label{def:problem}
% Given $T^{\prime}$, a version of $T$ where the goals are \lhda (left-hand delta attributes), \rhda (right-hand delta attributes), \lhdr (left-hand delta tuples), and \rhdr (right-hand delta tuples). From a search space of possible explanations, the version explanation problem is to find, for each goal, a set of explanations with the highest validity.
%  \end{definition}}
 
\begin{definition}\label{def:problem}
Given $T^{\prime}$, a version of $T$ where the goals are \lhda (left-hand delta attributes), \rhda (right-hand delta attributes), \lhdr (left-hand delta tuples), and \rhdr (right-hand delta tuples). From a search space of possible explanations, the version explanation problem is to find, for each goal, a set of explanations with the highest validity.
 \end{definition}
 
% \noindent\revision{Hence, a solution to the version explanation problem is a set of explanations that composed come closest to producing $T^{\prime}$ from $T$. Note that with validity alone we may have ties (as in our examples where multiple explanations have validity 1).} 
\noindent\revision{A solution to the version explanation problem is a set of explanations that composed come closest to producing $T^{\prime}$ from $T$. Note that with validity alone we may have ties (as in our examples where multiple explanations have validity 1).} 
\shepherd{If we have multiple datasets (or a dataset holdout), we can use generalizability to pick among the multiple solutions. We may also relax \cref{def:problem} to find solutions whose validity is within some range of the best and then use generalizability to select among these candidate solutions.  In addition, we can use explainability (described next) in this selection and pick explanations that a user can better understand.}

\subsection{Explanation Explainability}\label{sec:explainability}
As motivated %in the introduction
above, we care about the explainability of the generated solution. 
We, again, mainly focus on the attribute addition transformations. Since we use different models to generate explanations (regressors, decision trees, and programs), we seek a common ground %when it comes 
to measure explainability. Inspired by Narayanan et al.~\cite{narayanan2018humans} and Lakkaraju et al.~\cite{lakkaraju2016interpretable},
%rjm citations are annotations on sentences and shouldn't be a grammatical part of the sentence
% where the focus is on %explaining 
who focus on decision sets, 
we introduce two explainability dimensions, namely \emph{conciseness}, and \emph{concentration}, that can be measured across different explanations types. 

% \subsubsection{Explainability Conciseness}\label{sec:breadth} 
% \vspace{.1in}
\noindent\textbf{Explainability Conciseness:} Studies show that the fewer the components in a model and the shorter it is, the easier it is for a user to understand it~\cite{cruz2015bayesian,poursabzi2021manipulating}. In what follows, we measure the conciseness of the transformation as the number of components ($N_{c}$) it holds. For regression models we use the \emph{number of coefficients}, for decision trees we use the \emph{number of nodes}, and for programs we use the \emph{number of implementation lines}.

 \begin{example}\label{example:breadth}
    % As a simple example, consider an $exp(A_{i})$ transformation. Obviously a desired explanation would use the $math(\cdot)$ extension (using \cref{sec:num_methods}) to generate an explanation $(A_{i}, exp(A_{i}))$ that, in addition to being valid and generalizable, obtains an explainability conciseness of 1 (a sole coefficient with the value 1). 
    Consider an $exp(A_{i})$ transformation. Obviously a desired explanation would use the $math(\cdot)$ extension (\cref{sec:num_methods}) to generate a valid and generalizable explanation $(A_{i}, exp(A_{i}))$ that obtains an explainability conciseness of 1 (a sole coefficient). 
    
    An alternative explanation would use a Taylor Series over the $poly(\cdot)$ extension to generate a valid and generalizable explanation $(A_{i}, 1 + A_{i} + \frac{A_{i}^2}{2} + \frac{A_{i}^3}{6} + \dots)$ with an explanability conciseness of $\frac{1}{d+1}$, where $d$ is the polynomial degree. Note that this case also highlights the trade-off between validity (or generalizability) and explainability. The bigger the selected degree, the higher the validity (and generalizability) and the lower the explainability. 
 \end{example}
 
% \subsubsection{Explainability Concentration}
% \vspace{.1in}
\noindent\textbf{Explainability Concentration:}
 While 
 %rjm a smaller 
 a more concise explanation is favorable, it should also contain as few components as possible~\cite{letham2015interpretable} (i.e., it should be as concentrated as possible). Specifically, since humans have a limited working memory, a solution that is grouped into 
 %less 
 fewer chunks of information is favorable~\cite{narayanan2018humans}. For example, a linear regression function is easier to understand than a polynomial regression with reciprocal and logarithmic transformations, even if the former is longer. For regressors, we count \emph{the extensions that were used} (e.g., polynomials and math operations). For decision trees, we count \emph{the number of internal nodes} that represent conditions and for programs we use \emph{the number of intermediate transformations}. Let $N_{g}$ be the number of chunks, the explanability concentration is then given as $1\div N_{g}$ such that a more concentrated transformation gets a higher score.
 
  \begin{example}
    % Recall Example~\ref{example:breadth} and the explanations $(A_{i}, exp(A_{i}))$ and $(A_{i}, 1 + A_{i} + \frac{A_{i}^2}{2} + \dots)$. The concentration of these explanations is 0.5 (1 extension, 1 degree) and $\frac{1}{d}$, respectively. %\rs{can we (should we) come up with an example where we have two transformations such that one is more concise that the other and the second is more elementary than the first.} \renee{yes if possible}
    The concentration of $(A_{i}, exp(A_{i}))$ and $(A_{i}, 1 + A_{i} + \frac{A_{i}^2}{2} + \dots)$ is 0.5 (1 extension, 1 degree) and $\frac{1}{d}$, respectively.
    
    To highlight the difference between conciseness and concentration consider, for example, $\mathcal{E}_{1} = A_1+A_2+5$ and $\mathcal{E}_{2} = log(A_1)+A_2^2\cdot A_1$. %$\mathcal{E}_{2} = log(A_1)+\frac{A_2^2}{A_1}$. 
    While $\mathcal{E}_{1}$ is less concise ($\frac{1}{3}$ vs $\frac{1}{2}$ of $\mathcal{E}_{2}$), it is more concentrated ($1$) than $\mathcal{E}_{2}$ ($\frac{1}{3}$) which involves two additional extensions.
 \end{example}
The \emph{total explanability} is a linear combination of conciseness and concentration that can be defined by a user or a system.\footnote{In our experiments we use a uniform combination.}

\subsection{On Choosing an Explanation}\label{sec:choosing}\label{sec:search}
\revision{\texttt{Explain-Da-V} works iteratively, aiming to find valid explanations for each detected change following \cref{section:vertical} and \cref{section:horizontal}. As mentioned above, for each goal, multiple explanations can be generated, for example, if there are multiple origins (\cref{sec:origin}) or we have more than one methodology to explain a transformation (e.g., different extensions in \cref{sec:num_methods}). \texttt{Explain-Da-V} chooses the most \emph{explainable valid} explanation for each goal.}
%rjmIn what follow, it 
% It can be applied regardless of how far apart the versions are. Zooming in on one explanation,} \texttt{Explain-Da-V} chooses the most \emph{explainable valid} solution. Recall that given a goal, multiple explanations can be generated, for example, if there are multiple origins (\cref{sec:origin}) or we have more than one methodology to explain a transformation (e.g., different extensions in \cref{sec:num_methods}).

%We begin by generating a set of explanations. Given a goal, multiple explanations can be generated, for example, if there are multiple origins (\cref{sec:origin}) or we have more than one methodology to explain a transformation (e.g., different extensions in \cref{sec:num_methods}).

% \renee{Keep the word "validity" for our precise definition.  So I would not use validity for mean squared error - instead say something like "Each explanation is derived to optimize some notion of error that is not always the same as our definition of validity."}

Each independent explanation is derived in a way that optimizes some notion of error within the respective context that is not always the same as our definition of validity. A regressor (\cref{sec:num_methods}) minimizes the mean squared errors, a PBE solution (\cref{sec:tex_methods}) directly optimizes accuracy via search and a decision tree (greedily) optimizes the split functions of nodes. Given a set of explanations, we choose one as follows. (1) Find the highest validity in the set. (2) If multiple explanations share this value, return the most explainable based on total explainability (see \cref{sec:explainability}).

% \begin{enumerate}
%     \item Find the highest validity in the set.
%     \item If multiple explanations share this value, return the most explainable based on total explainability (see \cref{sec:explainability}).
% \end{enumerate}

% \renee{This is now inconsistent with my revision of the definition to be threshold based.  I didn't see how to make a binary definition of validity work.  If you agree, make this threshold based.}

Note that generalizability can not be used for explanation selection
%rjm as we assume no access t  
unless we have access to a $\tilde{T}$ and $\tilde{T}^{\prime}$ (see Section~\ref{sec:valngen}). 

Potentially, there can be a large number of transformations. Dealing with this size, the explanations in a set are generated in a sorted order by the size and cardinally of their origin (see Section~\ref{sec:origin}). Similarly, among regression models the explanations are sorted by the amount of extensions that were applied (i.e., first, a model without extensions is considered). Accordingly, we introduce an \emph{early stop condition} such that if an explanation meets a predefined threshold of validity and explainability, it is returned and the search is stopped.\footnote{In our experiments the threshold was set to .95.} Empirically, almost 70\% of cases are terminated early.

%\revision{the user-defined validity threshold $\theta$ (see~\cref{def:problem}), it is returned and the search is stopped.\footnote{In our experiments the $\theta$ was set to .95.} Empirically, almost 70\% of cases are terminated early.}

% a predefined threshold of validity and explainability, it is returned and the search is stopped.\footnote{In our experiments the threshold was set to .95.} Empirically, almost 70\% of cases are terminated early.

% \renee{This seems consistent with a threshold-based definition, see if you can work this in through out.}

\section{Empirical Evaluation}\label{sec:experiments}
% \rs{update:} We introduce our evaluation methodology (\cref{sec:expeval}), based on which, we select the output explanation (\cref{sec:choosing}). Then, we compare its performance to baselines (\cref{sec:res1}) and analyze its components using an ablation study (\cref{sec:ablation}).

We now compare our performance to baselines (\cref{sec:res1}) and analyze the components using an ablation study (\cref{sec:ablation}).

\subsection{Experimental Setup}\label{sec:setup}
We now detail our benchmarks, implementation, and baselines. The benchmark and code are available in our repository~\cite{gitURL}.

% \subsection{Benchmarks}\label{sec:benchmark}
\subsubsection{Benchmarks}\label{sec:benchmark}
%As semantic data versioning is novel, we design a new benchmark, termed Semantic Data Versioning Benchmark (SDVB), composed of five \emph{version-sets}. We also adopt a publicly available dataset designed by Yang et al.~\cite{yang2021auto} for a similar task of synthesizing data pipelines. 

\revision{We design a new benchmark for the novel task of semantic data versioning, termed Semantic Data Versioning Benchmark (SDVB), composed of five \emph{version-sets}.} We also adopt a publicly available dataset designed by Yang et al.~\cite{yang2021auto} for a similar task of synthesizing data pipelines.

% \subsubsection{Semantic Data Versioning Benchmark (SDVB)}
% The SDVB is composed of five version-sets each revolving around a seed dataset of a different topic. The examined versions are originated in well-known datasets detailed in Table~\ref{tab:SDVB}. We provide a sample of the benchmark in an anonymous repository.~\cite{gitURL} The benchmark and its generation code will become publicly available upon acceptance.
% \vspace{.1in}
\noindent\textbf{Semantic Data Versioning Benchmark (SDVB):}
\revision{SDVB contains a total of 342 dataset versions (136 version pairs) over five different topics, ranging in length (number of tuples) and width (number of attributes).\footnote{Not all versions use all original attributes.} Each topic represents a \emph{version-set} that was derived from a well-known seed dataset detailed in Table~\ref{tab:SDVB}, which includes smaller datasets (e.g., IRIS) along side bigger datasets (e.g., WINE).}

\noindent\textbf{Version Generation:} \revision{Given a seed dataset, we revise it to generate a version of it by first selecting a subset of change dimensions (e.g., \rhda and \lhdr).} Then, based on the dimension, we perform a set of transformations (some sampled and some manually created). %footnote{The full SDVB and its generation code will become available upon acceptance.}
We assure that each \revision{of the five version-sets cover all change dimensions.} Prior to version generating, each dataset is split into $T$ and $\tilde{T}$ (80\%-20\%), where the latter is a hold-out to compute generalizability. Following Section~\ref{sec:valngen}, the same changes applied to $T$ to generate $T^{\prime}$ are applied to $\tilde{T}$ to generate $\tilde{T}^{\prime}$.\footnote{The numbers reported in Table~\ref{tab:SDVB} include the hold-outs.}

\revision{Finally, note that a version may be created using more than one change and, in practice, the aforementioned number of versions is actually \textbf{composed of 1,702 changes}. For example, to create the 72 WINE dataset versions, a total of 681 changes were applied over the original dataset and its versions.} 

% The SDVB is composed of five version-sets 
% derived from well-known seed datasets detailed in Table~\ref{tab:SDVB}. We provide a sample of the benchmark and its generation notebook in our repository~\cite{gitURL}.\footnote{The full SDVB and its generation code will become available upon acceptance.} SDVB contains smaller datasets (e.g., IRIS) along side bigger datasets (e.g., WINE). In total SDVB contains 342 dataset versions (136 version pairs) over five different topics, ranging in length (number of tuples) and width (number of attributes).\footnote{Not all versions use all original attributes.} 

% Note that a version may be created using more than one change and, in practice, the aforementioned number of versions is actually \textbf{composed of 1,702 changes}. For example, to create the 72 WINE dataset versions, a total of 681 changes were applied over the original dataset and its versions. 

\begin{table}[t]
	\caption{Semantic Data Versioning Benchmark Details.}\label{tab:SDVB}
% 	\vspace{-.1in}
	\scalebox{0.65}{\begin{tabular}{|l|c|c|c|c|} 
		\bottomrule
		Topic (Name) & \# of Original & \# of Original & \# of & \# of \\
		& Tuples  & Attributes & Versions & Version-pairs \\\toprule
		Movies and TV shows~\cite{IMDB} (IMDB) & 1,000 & 6 & 72 & 29\\\midrule
		NBA Players~\cite{NBA} (NBA) & 11,700 & 9 & 68 & 27\\\midrule
		Wines Reviews~\cite{WINE} (WINE) & 129,971 & 6 & 72 & 29\\\midrule
		Iris Flowers~\cite{IRIS} (IRIS) & 150 & 5 & 58 & 22\\\midrule
		Titanic Passengers~\cite{TITANIC} (TITANIC) & 891 & 6 & 72 & 29\\\bottomrule
	\end{tabular}}
% 	\vspace{-.15in}
\end{table}

\noindent\textbf{Auto-Pipeline Benchmark}~\cite{AutoPipelineRepo}: This benchmark contains real data pipelines extracted from Github notebooks. As we focus on dataset versions, we filter out pipelines that include more than one table (e.g., 
%rjm performing 
those that use a join). %Specifically, f
Following %\rjm{
Yang et al.%}
%the evaluation process reported in~
~\cite{yang2021auto}, we consider the ``test'' table as $T$ and the ``target'' table as $T^{\prime}$. For a fair comparison, we run
\texttt{Explain-Da-V} and all baselines on all the data and do not consider generalizability for this benchmark.

\begin{table*}
	\caption{\revision{%Foofah, Foofah+, Auto-pipeline*, and \texttt{Explain-Da-V} 
	Performance in terms of Validity (Val), Generalizability (Gen) and average number of explanations the method chooses from (\# $\mathcal{E}$). For \texttt{Explain-Da-V}, we also report (in parenthesis) the proportion of explanations with Val/Gen score of 1}.}\label{tab:res}
% 	\vspace{-.1in}
	\scalebox{0.725}{\revision{\begin{tabular}{|l|cc|c||cc|c||cc|c||cc|c||cc|c||cc|c||} 
		\bottomrule
		\quad\quad\textbf{Dataset}$\rightarrow$ & \multicolumn{3}{c|}{\textbf{IMDB}} & \multicolumn{3}{c|}{\textbf{NBA}} & \multicolumn{3}{c|}{\textbf{WINE}} & \multicolumn{3}{c|}{\textbf{IRIS}} & \multicolumn{3}{c|}{\textbf{TITANIC}} & \multicolumn{3}{c|}{\textbf{Auto-pipeline}}\\\
		\hspace{-.05in}$\downarrow$\textbf{Method} & \textbf{Val} & \textbf{Gen} & \textbf{\# $\mathcal{E}$} & \textbf{Val} & \textbf{Gen} & \textbf{\# $\mathcal{E}$} & \textbf{Val} & \textbf{Gen} & \textbf{\# $\mathcal{E}$} & \textbf{Val} & \textbf{Gen} & \textbf{\# $\mathcal{E}$}& \textbf{Val} & \textbf{Gen} & \textbf{\# $\mathcal{E}$} & \textbf{Val} & \textbf{Gen} & \textbf{\# $\mathcal{E}$} \\\toprule
        Foofah &.42&.42&3.7&.28&.28&4.2&.29&.29&3.9&.23&.23&3.1&.29&.29&4.1&.55&-&3.3\\
        Foofah+ &.44&.44&3.7&.29&.29&4.2&.34&.34&3.9&.25&.25&3.1&.37&.37&4.1&.55&-&3.3\\
        Auto-pipeline* &.44&.44&3.7&.30&.30&4.2&.33&.33&3.9&.26&.26&3.1&.37&.37&4.1&.78&-&3.3\\\midrule
        \texttt{Explain-Da-V} & \textbf{.73} (.64) &\textbf{.60} (.56) &6.4&\textbf{.90} (.89) &\textbf{.79} (.69) &7.3&\textbf{.87} (.76) &\textbf{.81} (.59) &6.8&\textbf{.93} (.88) &\textbf{.83} (.76) &8.9&\textbf{.88} (.79) &\textbf{.77} (.68) &7.2&\textbf{.82} (.78) & -&5.7\\
         + over baseline& +65\% & +36\%& & +202\% & +167\% && +156\% & +138\% && +254\% & +217\%&& +140\%& +109\%&& +5\% & -&\\\bottomrule
	\end{tabular}}}
% 	\vspace{-.125in}
\end{table*}
\ifdefined\TechReport
\begin{table*}
	\caption{Foofah, Foofah+, Auto-pipeline*, and \texttt{Explain-Da-V} performance in terms of Validity (Val.) and Generalizability (Gen.) for Numeric goals}\label{tab:res}
% 	\vspace{-.1in}
	\scalebox{0.85}{\begin{tabular}{|l|cc|cc|cc|cc|cc|cc|} 
		\bottomrule
		 \quad\quad\textbf{Dataset}$\rightarrow$ & \multicolumn{2}{c|}{\textbf{IMDB}} & \multicolumn{2}{c|}{\textbf{NBA}} & \multicolumn{2}{c|}{\textbf{WINE}} & \multicolumn{2}{c|}{\textbf{IRIS}} & \multicolumn{2}{c|}{\textbf{TITANIC}}\\\
		\hspace{-.05in}$\downarrow$\textbf{Method} & \textbf{Val.} & \textbf{Gen.} & \textbf{Val.} & \textbf{Gen.} & \textbf{Val.} & \textbf{Gen.} & \textbf{Val.} & \textbf{Gen.} & \textbf{Val.} & \textbf{Gen.} \\\toprule
        Foofah  &.20&.20&.11&.11&.14&.14&.18&.18&.12&.12\\
        Foofah+ &.20&.20&.15&.15&.16&.16&.21&.21&.24&.24\\
        Auto-pipeline* &.22&.22&.16&.16&.18&.18&.22&.22&.26&.26\\\midrule
        \texttt{Explain-Da-V} & \textbf{.97} (+340\%)&\textbf{.85}(+286\%)&\textbf{.72} (+350\%) &\textbf{.66} (+313\%) &\textbf{.87} (+383\%) &\textbf{.75} (+316\%) &\textbf{.99} (+350\%) &\textbf{.85} (+286\%)&\textbf{.92} (+253\%)&\textbf{.84} (+223\%)\\\bottomrule
	\end{tabular}}
% 	\vspace{-.1in}
\end{table*}
\begin{table*}
	\caption{Foofah, Foofah+, Auto-pipeline*, and \texttt{Explain-Da-V} performance in terms of Validity (Val.) and Generalizability (Gen.) for Categorical goals}\label{tab:res}
% 	\vspace{-.1in}
	\scalebox{0.85}{\begin{tabular}{|l|cc|cc|cc|cc|cc|cc|} 
		\bottomrule
		 \quad\quad\textbf{Dataset}$\rightarrow$ & \multicolumn{2}{c|}{\textbf{IMDB}} & \multicolumn{2}{c|}{\textbf{NBA}} & \multicolumn{2}{c|}{\textbf{WINE}} & \multicolumn{2}{c|}{\textbf{IRIS}} & \multicolumn{2}{c|}{\textbf{TITANIC}} \\\
		\hspace{-.05in}$\downarrow$\textbf{Method} & \textbf{Val.} & \textbf{Gen.} & \textbf{Val.} & \textbf{Gen.} & \textbf{Val.} & \textbf{Gen.} & \textbf{Val.} & \textbf{Gen.} & \textbf{Val.} & \textbf{Gen.}\\\toprule
        Foofah &.15&.15&.09&.09&.16&.16&.23&.23&.22&.22\\
        Foofah+ &.25&.25&.12&.12&.18&.18&.25&.25&.22&.22\\
        Auto-pipeline* &.27&.27&.12&.12&.17&.17&.26&.26&.22&.22\\\midrule
        \texttt{Explain-Da-V} & \textbf{.86} (+218\%)&\textbf{.86}(+218\%)&\textbf{.99} (+725\%) &\textbf{.87} (+625\%) &\textbf{.89} (+423\%)&\textbf{.78} (+358\%) &\textbf{.98} (+277\%) &\textbf{.91} (+250\%)&\textbf{.88} (+300\%)&\textbf{.83} (+277\%)\\\bottomrule
	\end{tabular}}
% 	\vspace{-.1in}
\end{table*}
\begin{table*}
	\caption{Foofah, Foofah+, Auto-pipeline*, and \texttt{Explain-Da-V} performance in terms of Validity (Val.) and Generalizability (Gen.) for Textual goals}\label{tab:res}
% 	\vspace{-.1in}
	\scalebox{0.85}{\begin{tabular}{|l|cc|cc|cc|cc|cc|cc|} 
		\bottomrule
		 \quad\quad\textbf{Dataset}$\rightarrow$ & \multicolumn{2}{c|}{\textbf{IMDB}} & \multicolumn{2}{c|}{\textbf{NBA}} & \multicolumn{2}{c|}{\textbf{WINE}} & \multicolumn{2}{c|}{\textbf{IRIS}} & \multicolumn{2}{c|}{\textbf{TITANIC}} \\\
		\hspace{-.05in}$\downarrow$\textbf{Method} & \textbf{Val.} & \textbf{Gen.} & \textbf{Val.} & \textbf{Gen.} & \textbf{Val.} & \textbf{Gen.} & \textbf{Val.} & \textbf{Gen.} & \textbf{Val.} & \textbf{Gen.} \\\toprule
        Foofah &.50&.50&.48&.48&.44&.44&.33&.33&.47&.47\\
        Foofah+ &.52&.52&.50&.50&.53&.53&.42&.42&.48&.48\\
        Auto-pipeline* &.52&.52&.56&.56&.53&.53&.38&.38&.50&.50\\\midrule
        \texttt{Explain-Da-V} & \textbf{.62} (+19\%)&\textbf{.57}(+10\%)&\textbf{.88} (+57\%) &\textbf{.67} (+20\%) &\textbf{.58} (+9\%)&\textbf{.56} (+6\%) &\textbf{.82} (+115\%) &\textbf{.72} (+89\%)&\textbf{.79} (+58\%)&\textbf{.69} (+38\%)\\\bottomrule
	\end{tabular}}
% 	\vspace{-.1in}
\end{table*}
\else
\fi
\subsubsection{Implementation} \texttt{Explain-Da-V} was implemented in python, following Sections~\ref{section:vertical} and~\ref{section:horizontal}. Main parts of the code are provided in our repository~\cite{gitURL}. Linear regression with Lasso~\cite{Lasso} and Rigde~\cite{Rigde},\footnote{We first tried applying Lasso and if failed we applied Rigde.} regularization and decision trees~\cite{DecisionTree} were implemented with Scikit-learn. We extended Foofah's python publicly available implementation~\cite{FoofahCode}. We use the Featuretools~\cite{Featuretools} framework to generate aggregated and group by features (see Section~\ref{sec:col_addition}).

%Scikit-learn was used to implement the machine learning algorithms, namely, linear regression with Lasso~\cite{Lasso} and Rigde~\cite{Rigde},\footnote{we first tried applying Lasso and if failed we applied Rigde.} regularization and decision trees~\cite{DecisionTree}. We extended Foofah's python publicly available implementation~\cite{FoofahCode}. We use the Featuretools~\cite{Featuretools} framework to generate aggregated and group by features (see Section~\ref{sec:col_addition}). %To resemble a data lake scenario, the attribute names are disregarded from the analysis. 
% \renee{Hmm, what is the significance of this?  Do others require names to do the attribute matching?  Aren't we assuming a matching is given?} 

\subsubsection{Baselines}\label{sec:baselines}
\textbf{Foofah}~\cite{foofah} is used as a PBE baseline (see Section~\ref{sec:related3}). % using the same implementation as above.
%We also use 
\textbf{Foofah+} %to denote 
denotes Foofah with our novel extensions (e.g., textual-to-numeric, see Section~\ref{sec:tex_methods}). %Finally, a
As Auto-pipeline's implementation is not publicly available, we reproduced its search methodology\footnote{Reinforcement learning requires training data, which we assume unavailable.} using Foofah's framework by implementing the operators provided by Yang et al. (\textbf{Auto-pipeline*}) ~\cite{yang2021auto}. Search has an exponential worst case time complexity, so we apply a 60 second timeout for all methods following the default in Foofah~\cite{foofah}.\footnote{We note that Auto-pipeline default timeout limit is an hour} 

We also ran AutoPandas~\cite{autopandas} using their %publicly 
available implementation~\cite{AutoPandasCode}. AutoPandas creates a search space based on pandas~\cite{Pandas} operations and prunes the space of programs using deep learning. Similar to the reported performance in Auto-pipeline~\cite{yang2021auto}, AutoPandas performance was inferior and thus not reported. We also experimented with SQUARES~\cite{orvalho2020squares}, a recent query reverse engineering framework, and, similarly, do not report its inferior results. Since SQUARES was designed to synthesize traditional SQL queries it can sometimes resolve selection predicates; yet, it fails to cope with other changes %dimensions 
such as attributes added using transformations.

% As another PBE baseline, we also ran AutoPandas~\cite{autopandas} using their publicly available implementation~\cite{AutoPandasCode}. AutoPandas creates a search space based on pandas~\cite{Pandas} operations and prunes the space of programs using deep learning. Similar to the reported performance in Auto-pipeline~\cite{yang2021auto}, AutoPandas performance was inferior and thus not reported. We also experimented with SQUARES~\cite{orvalho2020squares}, a recent query reverse engineering framework, and, similarly, do not report its inferior results. Since SQUARES was designed to synthesize traditional SQL queries it can sometimes resolve selection predicates; yet, it fails to cope with other change dimensions such as attributes added using transformations. 
% \renee{describe why squares is worse - e.g., SQUARES is not a full version manager, but rather aims at learning a selection predicat...}

% Finally, to allow a fair comparison, we ``find the origin'' (see Section~\ref{sec:origin}) for each of the baselines as well. \renee{do baselines assume origin is given?  or do they use full relation as origin?  Clarify. }
% Additionally, horizontal explanations are solved iteratively (each attribute at a time). For vertical addition explanations the tuples of $T$ are used as input and the tuples of $T^{\prime}$ as output (similarly for vertical removal with $T^{\prime}$ as input and $T$ as output).

Finally, a na\"ive implementation of the baselines would use all of $T$ and $T^{\prime}$ as input-output examples. However, to allow a fair comparison, we ``find the origin'' (see Section~\ref{sec:origin}) for each of the baselines %as well \renee{do baselines assume origin is given?  or do they use full relation as origin?  Clarify. }
%Additionally, 
and vertical explanations are solved iteratively (each attribute at a time). For horizontal addition explanations the tuples of $T$ are used as input and the tuples of $T^{\prime}$ as output (similarly for horizontal removal with $T^{\prime}$ as input and $T$ as output).

% \rs{if we have time: \textbf{Baseline-original}, \textbf{Baseline-iterative}, \textbf{Baseline-with-find-origin}. If not only \textbf{Baseline-original}}

% \begin{table*}
% 	\caption{Foofah, Foofah+, Auto-pipeline*, and \texttt{Explain-Da-V} performance in terms of Validity (Val.) and Generalizability (Gen.)}\label{tab:res}
% 	\scalebox{0.65}{\begin{tabular}{|l|cc|cc|cc|cc|cc|cc|} 
% 		\bottomrule
% 		\quad\quad\textbf{Dataset}$\rightarrow$ & \multicolumn{2}{c|}{\textbf{IMDB}} & \multicolumn{2}{c|}{\textbf{NBA}} & \multicolumn{2}{c|}{\textbf{WINE}} & \multicolumn{2}{c|}{\textbf{IRIS}} & \multicolumn{2}{c|}{\textbf{TITANIC}} & \multicolumn{2}{c|}{\textbf{Auto-pipeline}}\\\
% 		\hspace{-.05in}$\downarrow$\textbf{Method} & \textbf{Val.} & \textbf{Gen.} & \textbf{Val.} & \textbf{Gen.} & \textbf{Val.} & \textbf{Gen.} & \textbf{Val.} & \textbf{Gen.} & \textbf{Val.} & \textbf{Gen.} & \textbf{Val.} & \textbf{Gen.}\\\toprule
%         Foofah &.415&.415&.280&.280&.291&.291&.230&.230&.290&.290&.545&.545\\
%         Foofah+ &.443&.443&.292&.292&.341&.341&.251&.251&.366&.366&.552&.552\\
%         Auto-pipeline* &.437&.437&.297&.297&.330&.327&.262&.262&.366&.366&.776&.776\\\midrule
%         \texttt{Explain-Da-V} & \textbf{.732} (+65\%)&\textbf{.602}(+36\%)&\textbf{.899} (+202\%) &\textbf{.793} (+167\%) &\textbf{.873} (+156\%)&\textbf{.811} (+138\%) &\textbf{.977} (+273\%) &\textbf{.831} (+217\%)&\textbf{.879} (+140\%)&\textbf{.768} (+109\%)&\textbf{.816} (+5\%) &\textbf{.816} (+5\%)\\\bottomrule
% 	\end{tabular}}
% \end{table*}

\subsubsection{\revision{Evaluation Measures}} \label{sec:measures}\revision{The explanations provided by our baselines are of a single type (programs, not regressors or decision trees), thus, in \cref{sec:res1}, we compare the Validity (Val.) and Generalizability (Gen.) of \texttt{Explain-Da-V} to the baselines. Since \texttt{Explain-Da-V} can return explanations that do not have a validity/generalizability score of 1.0, we also report the proportion of such explanations out of all output explanations. We further report the average number of explanations (\# $\mathcal{E}$) from which the method selects the most explainable valid (see \cref{sec:search}). We also compare and report runtimes. \cref{sec:ablation} also uses explainability (conciseness and concentration).}

\subsection{\texttt{Explain-Da-V} Compared to Baselines}\label{sec:res1}
% \rs{here we present the main results compared to the baseline over SDVB and APB}
% We now compare the results of \texttt{Explain-Da-V} to the baselines (Section~\ref{sec:baselines}) over the benchmarks described in Section~\ref{sec:benchmark}. %Since the explanations provided by our baselines are of a single type (programs, not regressors or decision trees), we only compare Validity (Val.) and Generalizability (Gen.) and analyze explainability in Section~\ref{sec:ablation}. 
% The results are reported in Table~\ref{tab:res}. 

\revision{The comparison between \texttt{Explain-Da-V} and the baselines (\cref{sec:baselines}) over the benchmarks (\cref{sec:benchmark}) is reported in Table~\ref{tab:res}.}

\texttt{Explain-Da-V} performs much better than the baselines %. This out-performance is 
mainly due to its ability to cope with varying data types (numeric and categorical in addition to textual). %Note that our baselines usually have the same validity and generalizability. 
The adapted Auto-pipeline benchmark is an exception where \texttt{Explain-Da-V} only performs slightly better than Auto-pipeline*. %in terms of validity and generalizability 
%as it was
%which is not designed for data versioning. 
\revision{Also, even if we zoom-in only on textual transformations (provided in a technical report~\cite{technical_report}), %\renee{be sure this exists and also summarize actual results in cover letter} \rs{DONE.} 
\texttt{Explain-Da-V} still out-performs all baselines. Even when we evaluate only the 100\% valid/generalizable explanations returned by \texttt{Explain-Da-V} (denoted in parenthesis in \cref{tab:res}), we observe a significant improvement.} Across baselines, we observe that extending Foofah (Foofah+) provides an average validity and generalizability boost of 9.5\%% in terms of validity and generalizability
, showing the benefit of the extended search space. \revision{All methods select among multiple explanations (\# $\mathcal{E}$, see \cref{sec:measures}) based on multiple origins (see \cref{sec:baselines}). \texttt{Explain-Da-V} considers almost twice as many explanations since it generates expanded origins for numeric explanations (see \cref{sec:num_methods}). }
%Comparing across baselines, we observe that extending Foofah (Foofah+) provides an average boost of 9.5\% in terms of validity and generalizability, which shows the benefit of extending the search space. \revision{All methods select among multiple explanations (\# $\mathcal{E}$, see \cref{sec:measures}) based on multiple origins (see \cref{sec:baselines}). \texttt{Explain-Da-V} considers almost twice as much explanations since it generates expanded origins for numeric explanations (see \cref{sec:num_methods}). }

Comparing among version-sets, we observe that in the IRIS dataset, \texttt{Explain-Da-V} obtained the best performance (.927 Val. and .831 Gen.) and the highest improvement. For the IMDB version-set, \texttt{Explain-Da-V} obtained the 
%rjm least 
worst performance (.732 Val. and .602 Gen.) and lowest improvement (among the newly suggested benchmark version-sets). IRIS is mostly composed of numeric attributes (4 out of 5) which are solved using our \revision{numeric change explanations (\cref{sec:num_methods}) }%numeric-to-numeric methodology 
and are not dealt with by the baselines. \revision{Note that accordingly, \texttt{Explain-Da-V} considers almost three times as %\rjm{
many %} 
explanations. Yet, the numeric extensions  %\renee{extensions to what?  - have about "innovations"} 
introduced in \cref{sec:num_methods} and the fact that we find the origin helps to home in on a valid solution quite quickly (see runtime below).} The IMDB version-set, on the other hand, contains more textual attributes (5 out of 6) and involves changes that \texttt{Explain-Da-V} fails to solve. %\rs{maybe add the XX min to XX change example}.
For example, one of the IMDB version-sets involves a transformation that adds an attribute containing the count of the number of genres from a \texttt{Genre} attribute.
%rjm runon sentence
In the \texttt{Genre} attribute, the genres are separated by a comma (e.g., \texttt{Drama, Romance}). A correct transformation would, for example, count the number of commas and add 1. While finding a transformation that counts the number of commas is a practical task for \texttt{Explain-Da-V} (which includes textual transformations and aggregations), such a composition is 
%rjm far mote challenging. 
not currently possible.  
%rj In practice,
Instead, the explanation \texttt{Explain-Da-V} chose (most valid, see \cref{sec:search}) uses an \texttt{IMDB Rating} attribute to determine the number of genres using a decision tree with a validity of 0.65.\footnote{The explanation is available in the repository~\cite{expexp}.}
% \renee{add example?  makes benchmark sound more real?} 

Finally, we note that the performance varies with respect to the different change dimensions. Interestingly, if we only look at vertical removals (Section~\ref{sec:col_removal}), 
% \renee{4.2 is on vertical (attribute) removals, not horizontal}
all three baselines have a validity and generalizability score of 1. The reason for that is their ability to discover projections (in their terms applying a \texttt{drop} operation over an attribute). Although it successfully finds these transformations, it lacks the ability to explain the semantics of the attribute removal. \texttt{Explain-Da-V}, although not perfectly valid and generalizable (.95), is more expressive in term of explaining the removal. For example, explaining that an attribute was removed because it contains duplicated information (see Section~\ref{sec:col_removal}). When looking at tuple removal, \texttt{Explain-Da-V} performs much better than the baselines. Since Auto-pipeline does ``not consider row-level filtering''~\cite{yang2021auto}, we recall the comparison against SQUARES (see Section~\ref{sec:baselines}). %\rjm{Despite its focus on learning a selection predicate,}
Despite its focus on learning a selection predicate, SQUARES is able to resolve cases where a predicate was applied with 0.6 validity (\texttt{Explain-Da-V} obtains 0.73 over these changes). This is because SQUARES was not able to resolve removing tuples containing NaN values and duplicate tuples (two cases in the benchmark). \ifdefined\TechReport
\else
\footnote{\revision{Additional cases from the experiments are provided in a technical report~\cite{technical_report}.}}
\fi%\rjm{This is because SQUARES was}  not able to resolve removing tuples containing NaN values and duplicate tuples \rjm{(two cases in the benchmark)}.

\noindent{\textbf{Runtime}:} In these experiments, excluding timeouts (see Section~\ref{sec:baselines}), finding an explanation using foofah took an average of 4.9 seconds, foofah+ 12.4 seconds, Auto-pipeline* 8.1 seconds, and \texttt{Explain-Da-V} 2.4 seconds. A reason for that difference is that fitting a regressor (linear time complexity) and learning a decision tree (quadratic complexity) are more efficient than search (exponential).

% \rs{complete renee: “in these experiments, for IMDB, foofah took around W time units, foofah+ around X, Auto-pripeline around Y. and Explain-Da-V around Z” }

\subsection{\texttt{Explain-Da-V} Ablation Study}\label{sec:ablation}
\begin{figure}[t]
     \centering
     \begin{subfigure}[b]{0.2\textwidth}
         \centering
         \includegraphics[width=\textwidth,trim=0 0 30 30]{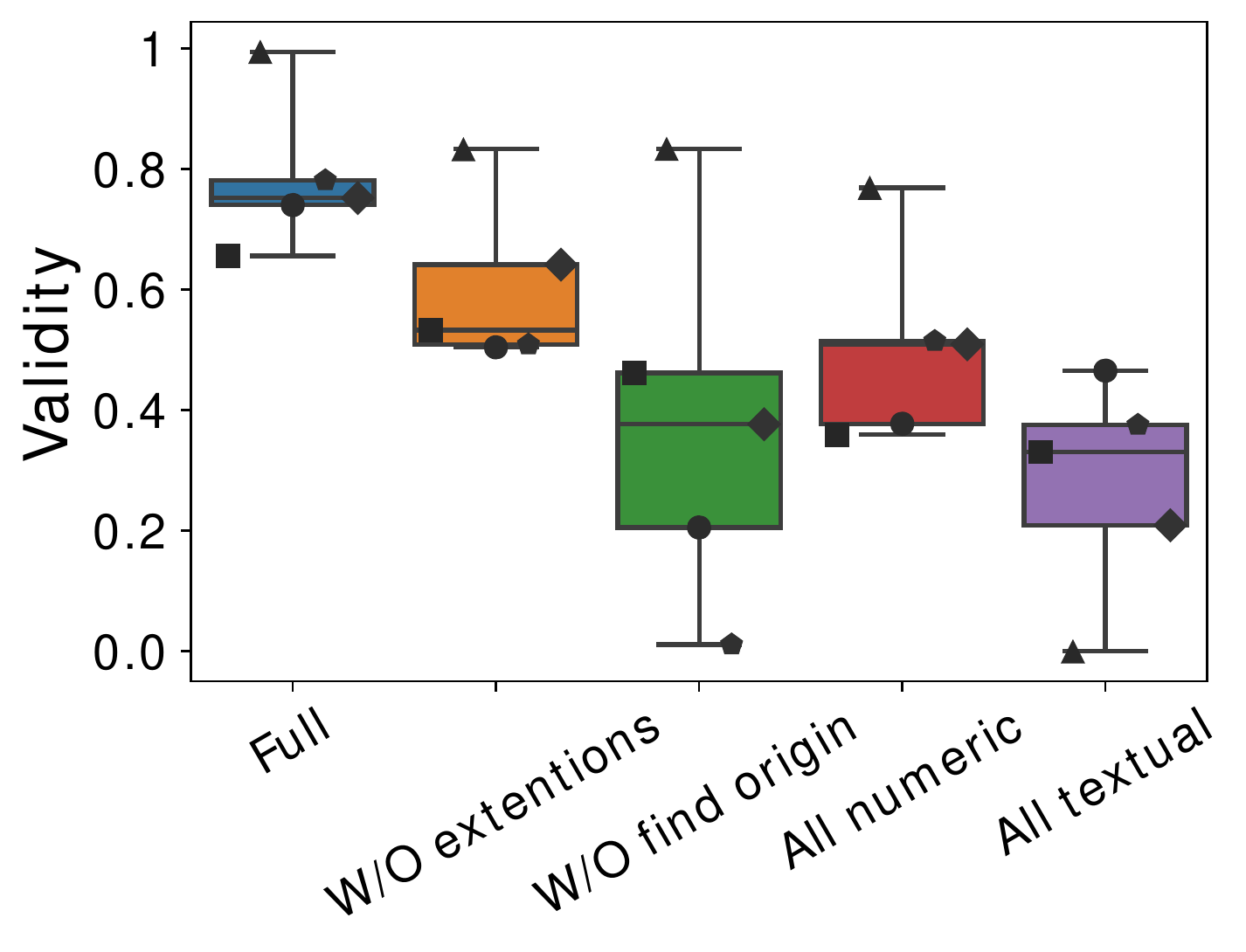}
         \caption{Validity}
         \label{fig:validity}
     \end{subfigure}
     \hfill
     \begin{subfigure}[b]{0.065\textwidth}
         \centering
         \includegraphics[width=\textwidth,trim=0 0 30 30]{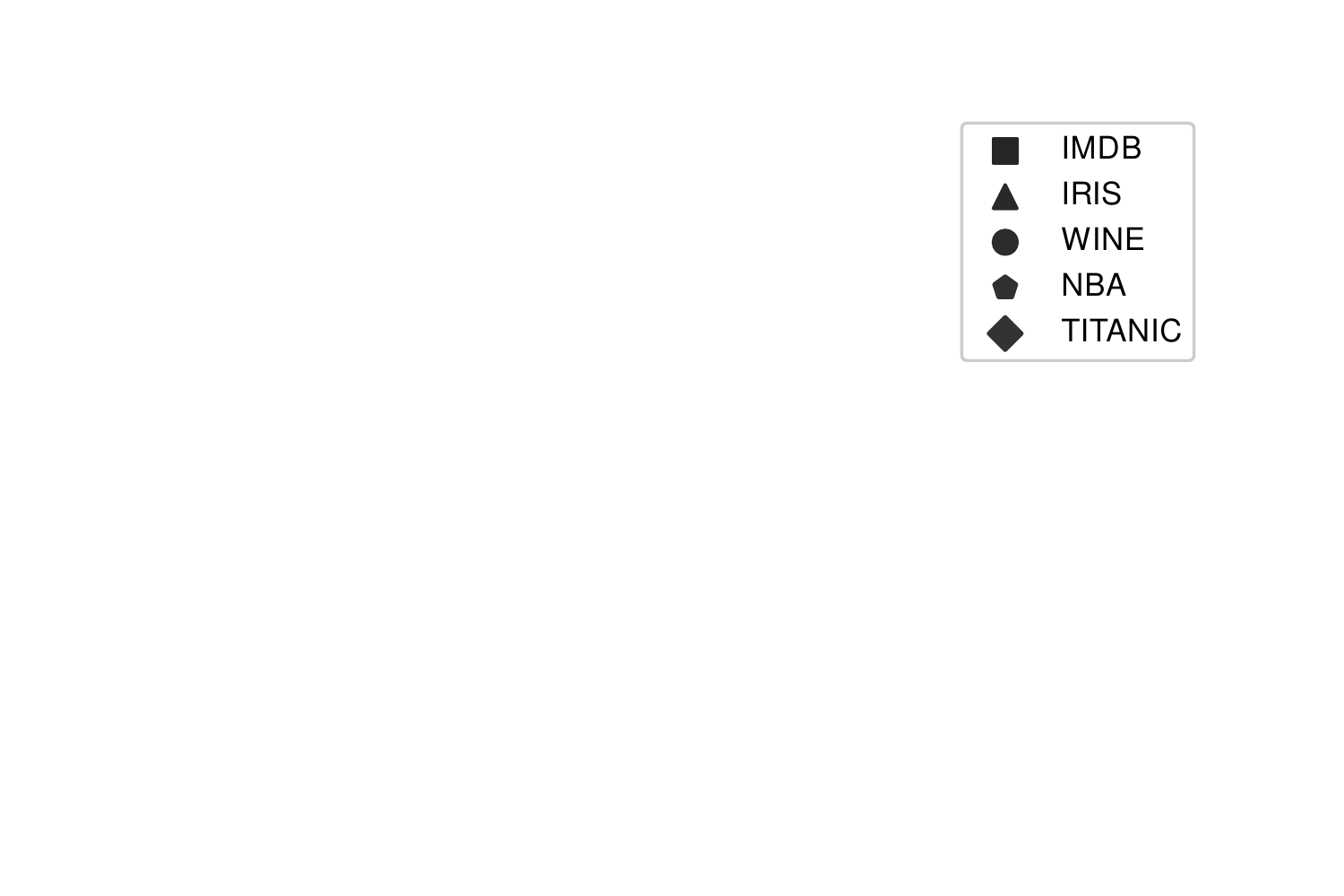}
     \end{subfigure}
     \begin{subfigure}[b]{0.2\textwidth}
         \centering
         \includegraphics[width=\textwidth,trim=0 0 30 30]{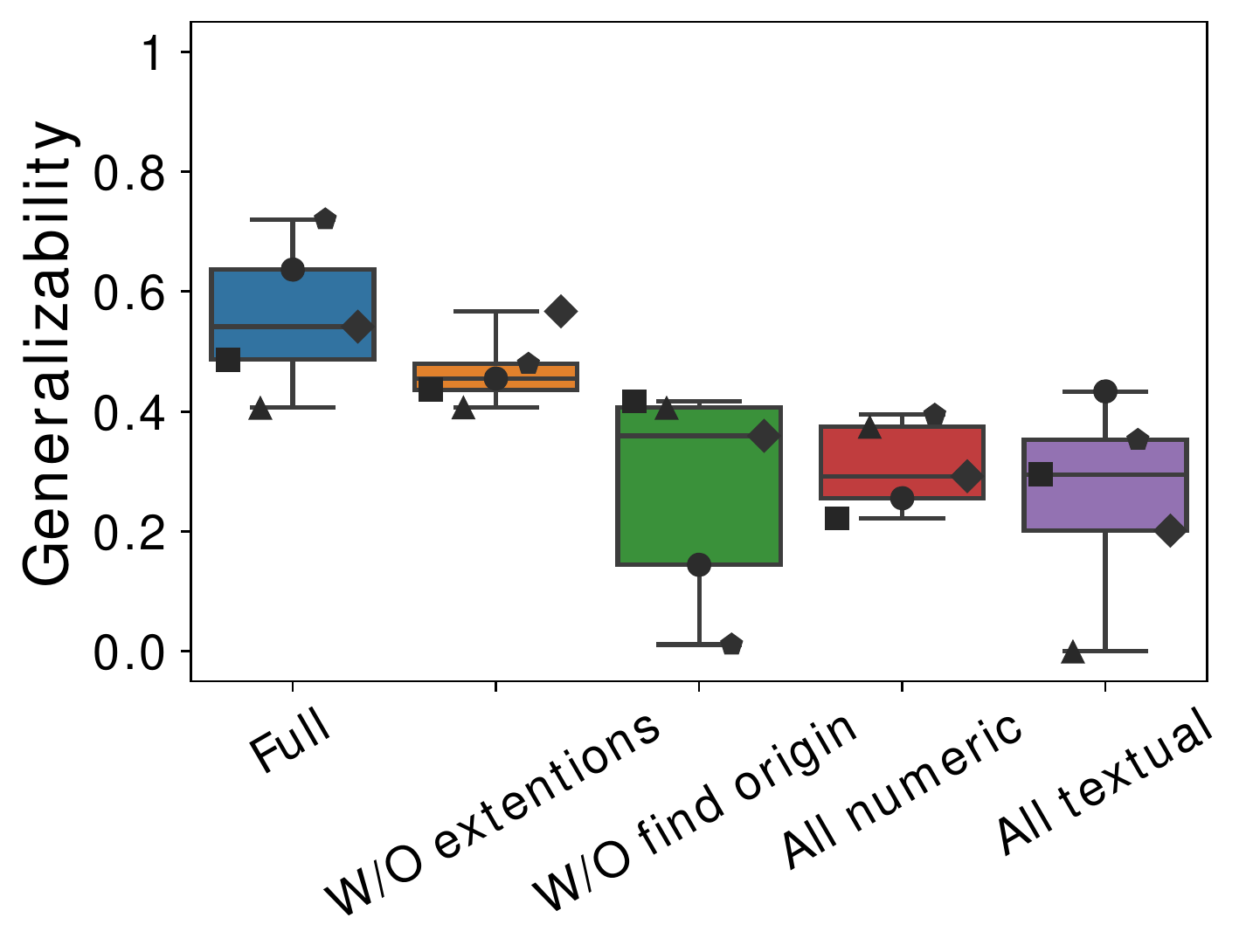}
         \caption{Generalizability}
         \label{fig:generalizability}
     \end{subfigure}
     \hfill
     \begin{subfigure}[b]{0.2\textwidth}
         \centering
         \includegraphics[width=\textwidth,trim=0 0 30 30]{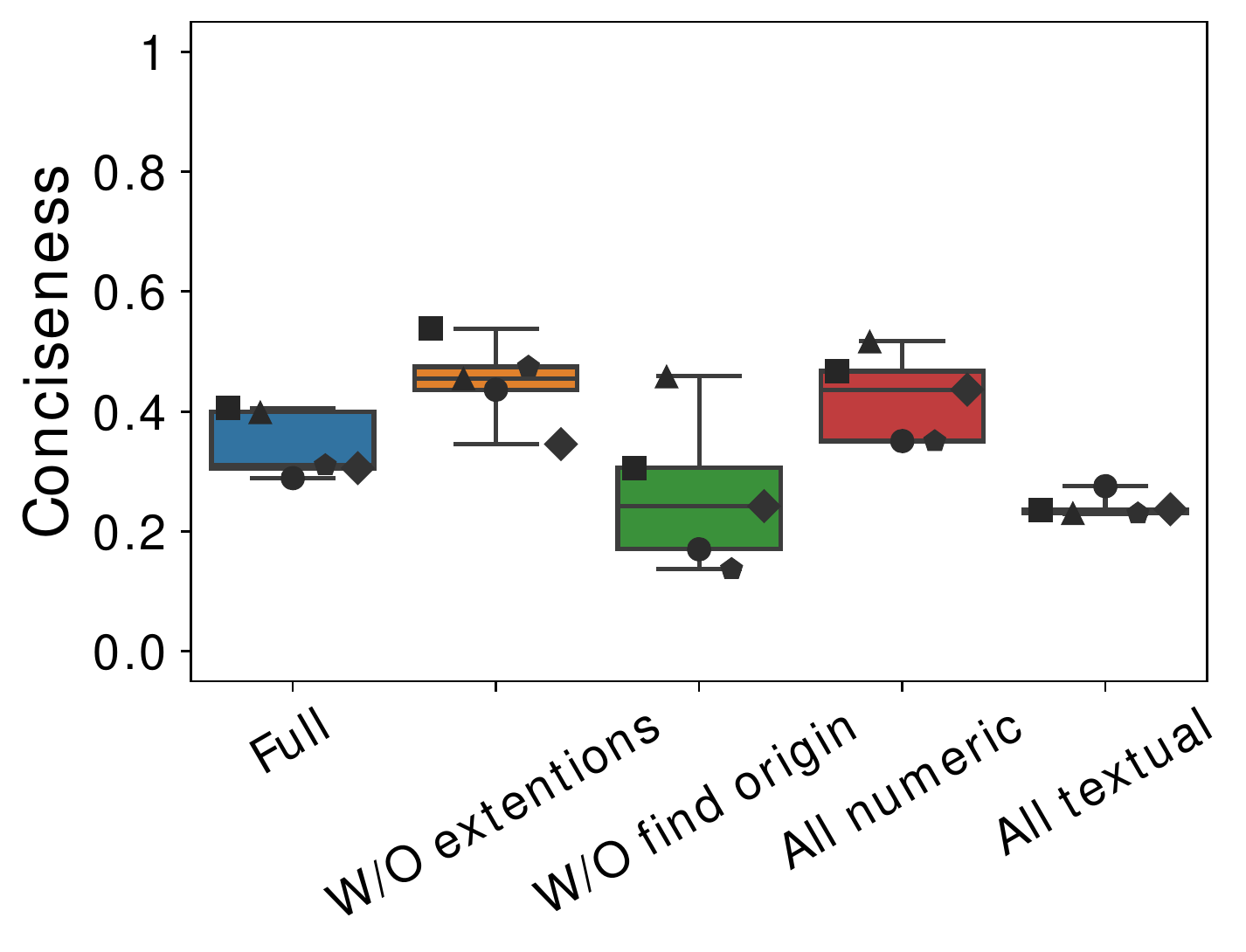}
         \caption{Conciseness}
         \label{fig:breadth}
     \end{subfigure}
     \hfill
     \begin{subfigure}[b]{0.2\textwidth}
         \centering
         \includegraphics[width=\textwidth,trim=0 0 30 30]{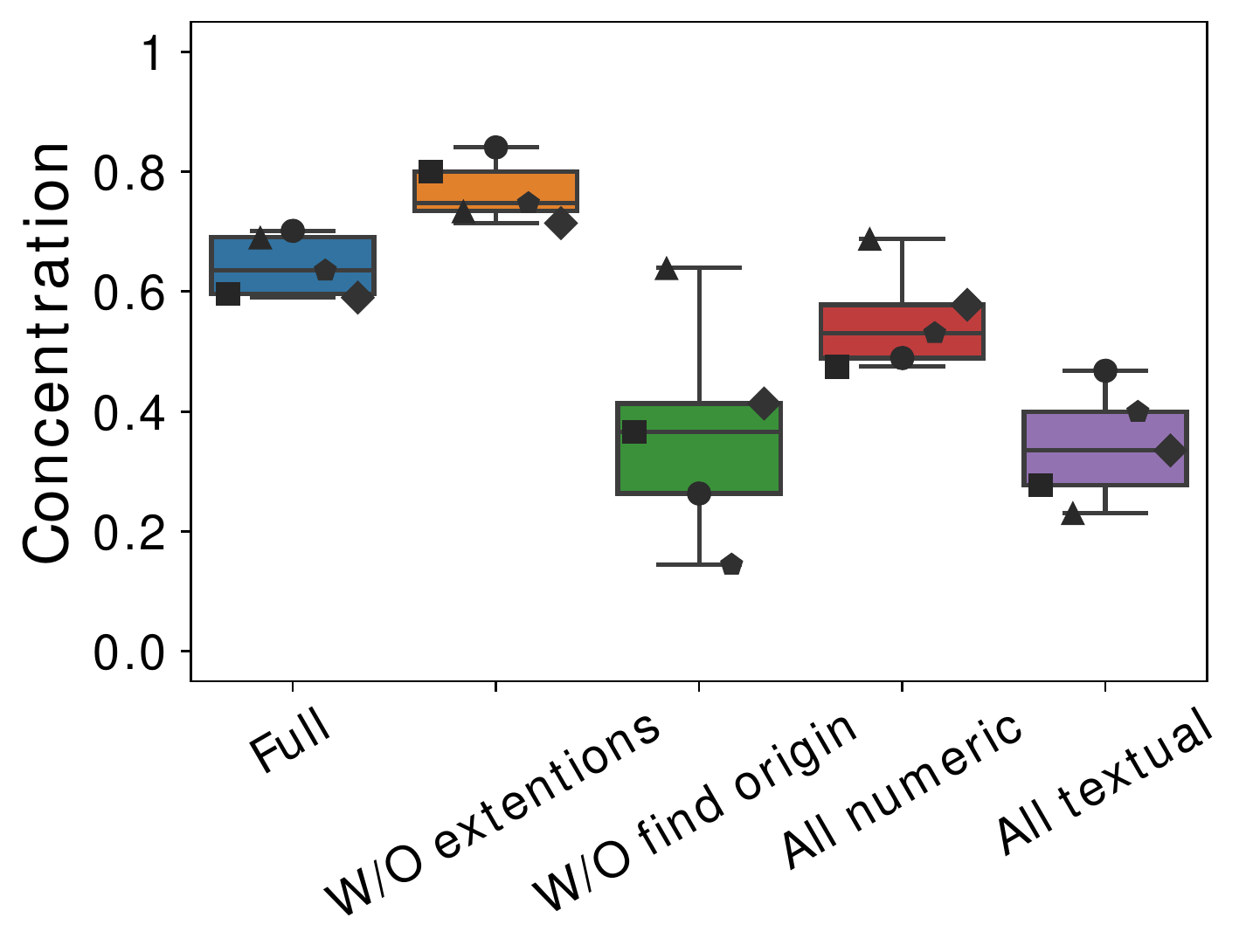}
         \caption{Concentration}
         \label{fig:concentration}
     \end{subfigure}
        \caption{Ablation Study over SDVB datasets.}
        \label{fig:ablation}
        % \vspace{-.25in}
\end{figure}

Figure~\ref{fig:ablation} provides an ablation study of \texttt{Explain-Da-V}. We focus on vertical addition explanations and analyze \texttt{Explain-Da-V} performance without finding the origin (W/O find origin), i.e., using $T$ as a whole to explain a given goal and without the extensions for numeric-to-numeric transformations (W/O extensions). %\renee{So in W/O origin case, the origin is assumed to be full relation?} \rs{Yes. Added a note.} 
We also analyze the resolved data types by applying \texttt{Explain-Da-V} assuming all types are numeric (All numeric) or all textual (All textual).    
% \renee{the fonts in figure are too small to read and a nasty reviewer could reject for violating the style guidelines.}\rs{Done}

As illustrated in Figures~\ref{fig:validity} and~\ref{fig:generalizability}, the full \texttt{Explain-Da-V} provides the most valid and generalizable performance. Adding extensions and finding the origin provide an average performance boost of 30\% and 107\%, respectively, in terms of validity, while addressing all attribute types as numeric and textual decreases the validity by 35\% and 64\%, respectively. The NBA version-set demonstrates an interesting case. Since it contains diverse attributes of varying types, without finding origin, \texttt{Explain-Da-V} obtains very low validity and generalizability. Similarly, as mentioned above, since the IRIS version-set mainly consists of numeric attributes, treating all attributes as textual results in very low performance.  

Examining the explainability % of solutions 
(Figures~\ref{fig:breadth}-\ref{fig:concentration}), we observe that while less valid and generalizable, explanations without extensions are more concise and concentrated. If an origin is not found, the explanations are usually less concise and much less concentrated. Finally, numeric explanations are more explainable than textual explanations especially in terms of conciseness. %\rs{REMEMBER TO ADD BACK}Note that understanding textual explanations also requires domain knowledge in programming while numeric explanations only require basic math. 
\ifdefined\TechReport
% \vspace{.1cm}
\noindent\textbf{Limitations via Interesting Cases:}
Consider the NBA version-set. One vertical addition in this dataset creates a binary attribute representing an indicator for a ``double double'' performance.\footnote{In basketball, a double-double is when a player accumulates ten or more in two of statistical categories \url{https://en.wikipedia.org/wiki/Double-double}.} A valid and generalizable transformation generated by \texttt{Explain-Da-V} returns $1$ if $A_{ast}>10$ (more than 10 assists) simply because for each tuple $r_i\in T$ such that $\pi_{A_{ast}}[r_i]>10$ it also happens that  $\pi_{A_{pts}}[r_i]>10$ (more than 10 points). While correct in the %version-set versioning 
specific scenario, it does not capture the 
%rjm do we want to mention intent
%true intention of the user that created this attribute. 
semantics of a double double.

Another interesting case considers a square-root transformation over movie rating ($A_{rating}$) in the IMDB version-set. \texttt{Explain-Da-V} generated two valid and generalizable explanations one using the transformation $sqrt(A_{rating})$ (using the $math$ extensions) and one using a polynomial transformation $1.074 + 0.262\cdot A_{rating} -0.005\cdot A_{rating}^{2}$. Obviously, these two transformations are different; however, if we zoom in on the range $x\in[6,10]$ (typical ratings in IMDB), we observe that the transformations behave almost identically. For example, a rating of 9 would be transformed to 3 and 3.02, respectively. An illustration is given in our repository~\cite{sqrtEx}.
\else
% The interested reader is referred to the technical report~\cite{technical_report}, where we provide some additional interesting cases from the experiments.
\fi
\section{Conclusion}
This work laid the groundwork for explaining semantic changes in data versioning. 
%rjm shortening We introduced the notation of explanation and formally defined change dimensions to be explained. Our method, 
\texttt{Explain-Da-V}, uses different types of techniques to resolve and explain changes between a pair of dataset versions. We introduced measures to evaluate explanations and show that \texttt{Explain-Da-V} performs better than multiple baselines over an existing adapted benchmark and a newly introduced data versioning benchmark. In future work, we intend to extend \texttt{Explain-Da-V} to address additional data types, e.g., dates, and address changes that are triggered by external data such as performing joins and unions. \shepherd{An additional future challenge is to formulate the version explanation problem as a multi-objective optimization problem that collectively optimizes validity, generalizability, and explainability.} 

% \added{In this work we cover ``internal'' additions over textual, numeric and categorical data.} 
% /In future work, we intend to extend \texttt{Explain-Da-V} to address additional data types, e.g., dates, and address changes that are triggered by external data such as performing joins and unions. 

\bibliographystyle{ACM-Reference-Format.bst}
\bibliography{sample}

\end{document}